\documentclass{aa}  
\usepackage{natbib}
\usepackage{orcidlink}
\usepackage{threeparttable}

\bibpunct{(}{)}{;}{a}{}{,} 

\usepackage{graphicx}

\usepackage{placeins}

\usepackage{txfonts,textcomp}
\newcommand{\tred}[1]{{\textcolor{black}{#1}}} 

\begin{document}

   \title{A novel optimal transport-based approach for interpolating spectral time series}

   \subtitle{ Paving the way for photometric classification of supernovae}

   \author{M. Ramirez \inst{1,2}\orcidlink{0009-0001-3293-7741}
          \and
          G. Pignata\inst{3,2}\orcidlink{0000-0003-0006-0188}
          \and
          Francisco F\"orster\inst{2,4,5,6}\orcidlink{0000-0003-3459-2270}
          \and
           Santiago Gonz\'alez-Gait\'an\inst{7}\orcidlink{0000-0001-9541-0317}
           \and
            Claudia P. Guti\'errez \inst{10,11}\orcidlink{0000-0003-2375-2064}
            \and
          B. Ayala\inst{1,2} \orcidlink{0009-0002-2564-1069}
          \and
          Guillermo Cabrera-Vives \inst{2,12,13}\orcidlink{0000-0002-2720-7218}
          \and
            M\'arcio Catelan\inst{2,8,9}\orcidlink{0000-0001-6003-8877}
          \and
          A. M. Mu\~noz Arancibia\inst{2,5}\orcidlink{0000-0002-8722-516X}
          \and
        J.~Pineda-García\inst{1,2}\orcidlink{0000-0003-0737-8463}
          }
    
   \institute{Instituto de Astrofisica, Departamento de Fisica, Facultad de Ciencias Exactas, Universidad Andres Bello, Fernandez Concha 700, Las Condes, Santiago RM, Chile.\\
              \email{m.ramirezvalenzuela@uandresbello.edu}
         \and
             Millennium Institute of Astrophysics, Nuncio Monse\~{n}or Sotero Sanz 100, Of. 104, Providencia, Santiago, Chile.
        \and
            Instituto de Alta Investigaci\'on, Universidad de Tarapac\'a, Casilla 7D, Arica, Chile.
        \and
        Data and Artificial Intelligence Initiative (IDIA), Faculty of Physical and Mathematical Sciences, Universidad de Chile, Chile.
        \and 
        Center for Mathematical Modeling, Universidad de Chile, Beauchef 851, Santiago 8370456, Chile
        \and 
        Departamento de Astronom\'{\i}a, Universidad de Chile, Chile
        \and
        CENTRA, Instituto Superior T\'ecnico, Universidade de Lisboa, Av. Rovisco Pais 1, 1049-001 Lisboa, Portugal.
        \and
        Instituto de Astrof\'{\i}sica, Pontificia Universidad Cat\'olica de Chile, Av. Vicu\~{n}a Mackenna 4860, 7820436 Macul, Santiago, Chile.
        \and
        Centro de Astroingenier{\'{\i}}a, Pontificia Universidad Cat{\'{o}}lica de Chile, Av. Vicu\~{n}a Mackenna 4860, 7820436 Macul, Santiago, Chile.
        \and
        Institut d’Estudis Espacials de Catalunya (IEEC), Gran Capit\`a, 2-4, Edifici Nexus, Desp. 201, E-08034 Barcelona, Spain. 
        \and
        Institute of Space Sciences (ICE, CSIC), Campus UAB, Carrer de Can Magrans, s/n, E-08193 Barcelona, Spain.
        \and 
        Department of Computer Science, Universidad de Concepción, Chile.
        \and 
        Data Science Unit, Universidad de Concepción, Edmundo Larenas 310, Concepción, Chile.
            }

   \date{Received January 8, 2024; accepted September 12, 2024}

 
\abstract
   {The \textit{Vera C. Rubin} Observatory is set to discover 1 million supernovae (SNe) within its first operational year.  Given the impracticality
of spectroscopic classification at such scales, it is mandatory to develop a reliable photometric classification framework.}
    {\tred{This paper introduces a novel method for creating spectral time series that can be used not only to generate synthetic light curves for photometric classification, but also in applications such as K-corrections and bolometric corrections. This approach is particularly valuable in the era of large astronomical surveys, where it can significantly enhance the analysis and understanding of an increasing number of SNe, even in the absence of extensive spectroscopic data}.}
   {By employing interpolations based on optimal transport theory, starting from a spectroscopic sequence, we derive weighted average spectra with high cadence. The weights incorporate an uncertainty factor for penalizing interpolations between spectra that show significant epoch differences and lead to a poor match between the synthetic and observed photometry.}
   {Our analysis reveals that \tred{even with a phase difference of up to 40 days between pairs of spectra, optical transport can generate interpolated spectral time series that closely resemble the original ones.} Synthetic photometry extracted from these spectral time series aligns well with observed photometry. The best results are achieved in the \textit{V} band, with relative residuals of less than 10\%  for 87\% and 84\% of the data for type Ia and II, respectively. For the \textit{B}, \textit{g}, \textit{R,} and \textit{r} bands, the relative residuals are between 65\% and  87\% within the previously mentioned 10\% threshold for both classes. The worse results correspond to the  \textit{i} and \textit{I} bands, where, in the case of SN~Ia, the values drop to 53\% and 42\%, respectively.}
   {We introduce a new method for constructing spectral time series for individual SNe starting from a sparse spectroscopic sequence, and demonstrate its capability to produce reliable light curves that can be used for photometric classification.}


   \maketitle
%

\section{Introduction}
\label{sec:introduction}

Supernovae (SNe) are transient astronomical events that occur during the terminal phases of stellar evolution. SNe play an important role in galactic evolution, influencing both chemical evolution and energy dynamics across galaxies \citep[e.g.,][]{Ceverino2009}. Additionally, due to their luminous and homogeneous intensity, type Ia SNe serve as robust distance indicators, enhancing our understanding of cosmic scales \citep[e.g.,][]{Riess1998,Perlmutter1999,Abbott2019}.

The current classification scheme of SNe is mainly based on differences in their spectra, but initially they were classified into Type I and Type II based on the presence or absence of hydrogen lines in their spectra \citep{minkowski}. Furthermore, a variety of subtypes emerge depending on the manifestation and/or strength of other chemical elements \citep{wheler,Elias1985,Harkness1987,W&H1990}.
On one hand, Type I has been divided into Ia, Ib, and Ic, where SNe~Ia show strong signatures of Si~II {near their peak}, while SNe~Ib are defined by the presence of strong He~I features. SNe~Ic do not show strong lines of  either He or Si~II. 
On the other hand, a subtype of type II SNe is SN~IIb, where spectra at early phases are dominated by strong H~I lines, but weaken with time while He~I features strengthen.
Finally, the spectra of SN~IIn are dominated by prominent narrow emission lines of the Balmer series at all phases.
From the explosion mechanism point of view, SNe can be broadly classified into thermonuclear and core collapse (CC) explosions. Thermonuclear SN coincides with the SN~Ia observational class and is believed to be the result of the thermonuclear incineration of a carbon-oxygen white dwarf in a binary system \citep[e.g,][]{1960ApJ...132..565H}. On the other hand, CC explosions include all the other SN observational types mentioned above (II, IIb, IIn, Ib and Ic) and are expected to be the result of the collapse of the iron-degenerate core of a massive star (M $\gtrapprox$ 8 $M_{\odot}$)  \citep[for a review see][]{2009ARA&A..47...63S}

The Legacy Survey of Space and Time (LSST) that will be carried out by the \textit{Vera C. Rubin} Observatory with the Simonyi Survey Telescope is projected to discover more than $10^7$ SNe spanning
a considerable redshift range during its ten years of operations  \citep{LSST}. Given the time investment that would be required to carry the spectroscopic classification described above on the LSST sources, the development of reliable photometric classification algorithms is fundamental. This will enable the complete realization of the enormous potential of the photometric dataset produced by current and future transient surveys.
 The performance of a photometric classification based on machine learning techniques \citep[e.g.,][]{Ishida2013,Lochner2016,Charnock2017,Boone2019,villar_2019,Moller2020}, regardless of the specific algorithm used, strongly depends on the dataset employed for training \citep[e.g.,][]{Richards2012,Karpenka2013,Millard2015}.
 In this respect, spectro-photometric time series offer a valuable resource, as they enable the construction of synthetic light curves at various redshifts, which are ideal for use as training sets.

A key aspect of training sets is that they must be representative of the diversity of the classes that are the targets for classification.  The first spectral time series were constructed by combining spectra from multiple SNe belonging to the same class. These spectral "templates" are representative of the entire class of the objects that were employed in their construction. Synthetic light curves for individual objects are then generated by warping the spectra to match their observed color.  Spectral templates have been generated for Type Ia SNe \citep[e.g.,][]{Nugent2002,Hsiao2007,Lu2023} and CC~SNe \citep[e.g.,][]{Kessler2010,Kessler2019Plastic}. Nevertheless, this approach can potentially reduce the intrinsic diversity between the members of a given class, introducing biases if the spectral templates are used to generate training sets for photometric classification. This bias is particularly important for CC~SNe, because of the large heterogeneity that they display within their classes. Building up spectral time series for individual SNe belonging to a given family preserves the diversity, making them particularly suitable for generating synthetic light curves for training sets. \citet{Vincenzi2019} compiled a set of 67 spectral time-series across various SN types (II, IIn, IIb, Ib, Ic, Ic-BL), integrating photometric and spectroscopic data from the literature with \tred{Gaussian processes (GPs)}.

In this context, we introduce a novel method for constructing spectral time series of SNe based on optimal transport (OT) and the Wasserstein barycenter. OT theory has found large applications in a variety of scientific fields, from economics to biology, physics, data science, and machine learning. OT has also been applied in the field of astronomy. \citet{Uriel_2002}, for example, demonstrated that the reconstruction of the early density fluctuations of the Universe is effectively an optimization problem, leveraging optimal mass transportation techniques. Similarly, \citet{Levy_2021} advanced this field by developing a fast semi-discrete OT algorithm, providing a unique and efficient approach to modeling these early cosmic structures. \citet{Nikakhtar_2022,Nikakhtar_2023} apply OT for reconstructing biased tracers in redshift space and baryon acoustic oscillations,  enhancing our understanding of the structure of the Universe. \citet{Rawson_2022} employ OT to interpolate high-resolution images from low-resolution data, selecting the best match through a process that optimizes for the smallest Wasserstein distance, effectively refining the interpolation into a more precise reconstruction. Here, our method utilizes well-calibrated spectro-photometric data from individual events, and is therefore a versatile, data-driven approach applicable to various types of SNe.

This paper is organized as follows: In Sect. \ref{sec:Data_sample} we detail the data sample and discuss our selection criteria. Section \ref{sec:Methodology} is dedicated to our methodology, where we elucidate the foundations of OT. Section \ref{sec:Analysis OT} presents the tests conducted with models and the recipe we use for the production of the spectral time series. In addition, we validate our approach with observed photometry. Finally, we summarize our findings in Sect. \ref{sec:Conclusion}.

\section{Data sample}\label{sec:Data_sample}

The spectra and light curves used for this work were retrieved from the open supernova catalog \citep[][]{Guillochon_2017}, from The Weizmann Interactive Supernova Data Repository (\mbox{ WISeREP} \footnote{\url{https://www.wiserep.org/}}) \citep[][]{2012PASP..124..668Y}, and from the literature. The data cleaning process is described in Appendix \ref{sec:Data_Cleaning}.

From the whole sample, we selected SNe of type Ia and II for which photometry is available in at least two bands within the Johnson-Cousins (JC) or Sloan (SDSS) photometric systems and for which a minimum of three spectra have been obtained (at least one before maximum light in the case of type Ia SNe). These selection criteria yielded an initial sample of 458 type Ia and  138 type II SNe.
We assess the calibration of the spectra by comparing the flux from synthetic photometry $F_{syn}$ with that from observed photometry $F_{obs}$ across as many bands as possible through the following ratio:

\begin{equation}
    \frac{F_{syn}}{F_{obs}} = K_x,
    \label{eq:K}
\end{equation}

\noindent where the  $K_x$ corresponds to a given band $x$. $F_{syn}$ is calculated using the following equation: 

\begin{equation}
    F_{syn}=F_0 \frac{\int f(\lambda) R(\lambda) d\lambda }{\int R(\lambda) d\lambda},   
    \label{eq:Syn_photometry}
\end{equation}

\noindent where $f(\lambda)$ represents the flux density of the spectrum and $R(\lambda)$ is the band response. Both of these variables are functions of the wavelength $\lambda$. $F_0$ is the zero point (see Appendix \ref{sec:Standars_stars}). We only compute $F_{syn}$ when at least 95\% of the response band is covered by the spectrum.

$F_{obs}$ is obtained by interpolating the observed light curve using the Automatic Loess Regression (ALR) technique as outlined in \citet{Osmar_2019} at the corresponding spectrum date.

For each spectrum, we compute the $K_x$ values for as many bands as possible. This allows us to rescale the spectrum against the photometry by dividing its flux by the median of the $K_x$ values ($\tilde{K}$) and also to evaluate the quality of its calibration, computing the median absolute deviation (MAD; hereafter $MAD(K)$) of these values. As the aim of this work is to assess the performance of OT, all the results presented in this paper are based on spectra with a relative error ($MAD(K)/\tilde{K}$) of less than 10\%, which make up our "golden sample." The final golden sample consists of 110 SNe~Ia and 31 SNe~II, which are reported in Table \ref{tab:SN included}.

\section{Methodology}\label{sec:Methodology}

 In this work, we employ OT to interpolate between SN spectra. Introduced by \citet{Monge1784}, OT is a mathematical framework designed to help find the most efficient way to move "mass" between different distributions, understood in this context as an abstract representation of resources, probabilities, and distributed data, which in our case are spectra. For a more detailed explanation of the computational foundations, please refer to  \citet{Villani2009}, \citet{Peyre2020}, and \citet{Zhang2021}.
 
\begin{figure}[!htb]
    \centering
    \includegraphics[width=1\columnwidth]{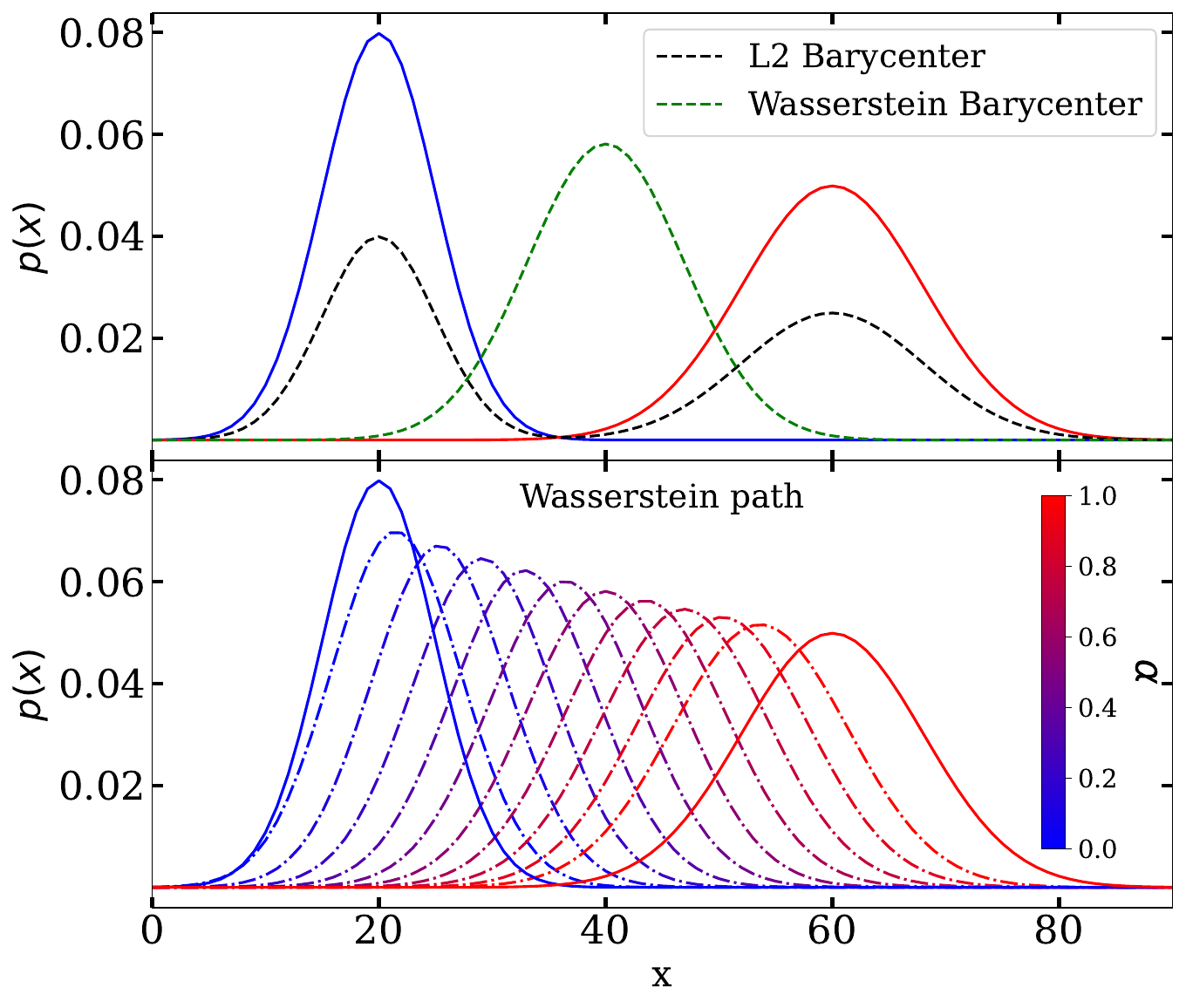}
    \caption{Comparison between the $L2$ barycenter and the Wasserstein barycenter. The starting and final distributions are in blue and red, respectively. The black dashed distribution represents the barycenter calculated using the $L2$ distance, whereas the green dashed distribution is computed using the Wasserstein distance. The bottom panel illustrates the interpolation path  of the Wasserstein barycenter for different $\alpha$ values, highlighting its transition from the starting distribution to the final one.}
    \label{fig:OT_example}
\end{figure}

Moving to the mathematical formalism, we first consider the simpler, 1D case, where we move something from position $(x_1,x_2,x_3,\ldots,x_N)$ to a new location $(y_1,y_2,y_3,\ldots,y_N)$. The problem is to find the optimal transport plan $T(x_i)=y_i$ that minimizes the total transportation costs $C_T$. The total cost is given by

\begin{equation}
    C_T=\sum_i^N{c(x_i,T(x_i))},
    \label{eq:CT}
\end{equation}

\noindent where $c$ is the cost function of moving from one point to another. Equation \ref{eq:CT} establishes that, for each transport  $(x) \xrightarrow[]{} (y),$ the unit cost of the transport depends on the quantity to be transported (the transportation plan), given by $T,$ and the starting point. By adding over all possible origins and destinations, we get the total cost. With the OT plan found by minimizing Eq. \ref{eq:CT}, it is possible to construct the Wasserstein distance as in \citet{Kolouri_2017}.
Wasserstein distance is a metric for quantifying the distance between two probability distributions, and this distance is a way of measuring how much work it takes to transform one distribution into another, also referred to as the cost of moving. The Wasserstein barycenter is then the distribution that results from minimizing the total sum of these distances (costs) to all other distributions. It is like finding a middle point, not in terms of physical distance, but in terms of how much you would have to change each distribution to reach this middle point. A parameter $\alpha$, ranging from 0 to 1, is defined to control the interpolation between the two distributions. An $\alpha$ value of 0 interpolates entirely to the first distribution, while a value of 1 interpolates to the second one. We show a practical illustration of how OT works in  Fig. \ref{fig:OT_example}. In the top panel, the starting and final distributions  are shown in blue
and red, while the  Wasserstein barycenter and $L2$ are shown  in green and black, respectively. The barycenter $L2$, also known as the Euclidean barycenter, is calculated as the average of each corresponding pair of points of the two distributions, that is, it provides a linear interpolation. As visible in the plot, the $L2$ barycenter results in a bimodal distribution, while the Wasserstein barycenter produces a distribution that is transitional between the initial ones. The lower panel of Fig. \ref{fig:OT_example} shows the interpolation path of the Wasserstein barycenter for different weights, that is, different values of $\alpha$, illustrating the transition of the interpolations from one distribution to the other. 

\begin{figure}[!htb]
    \centering

    \includegraphics[width=1\columnwidth]{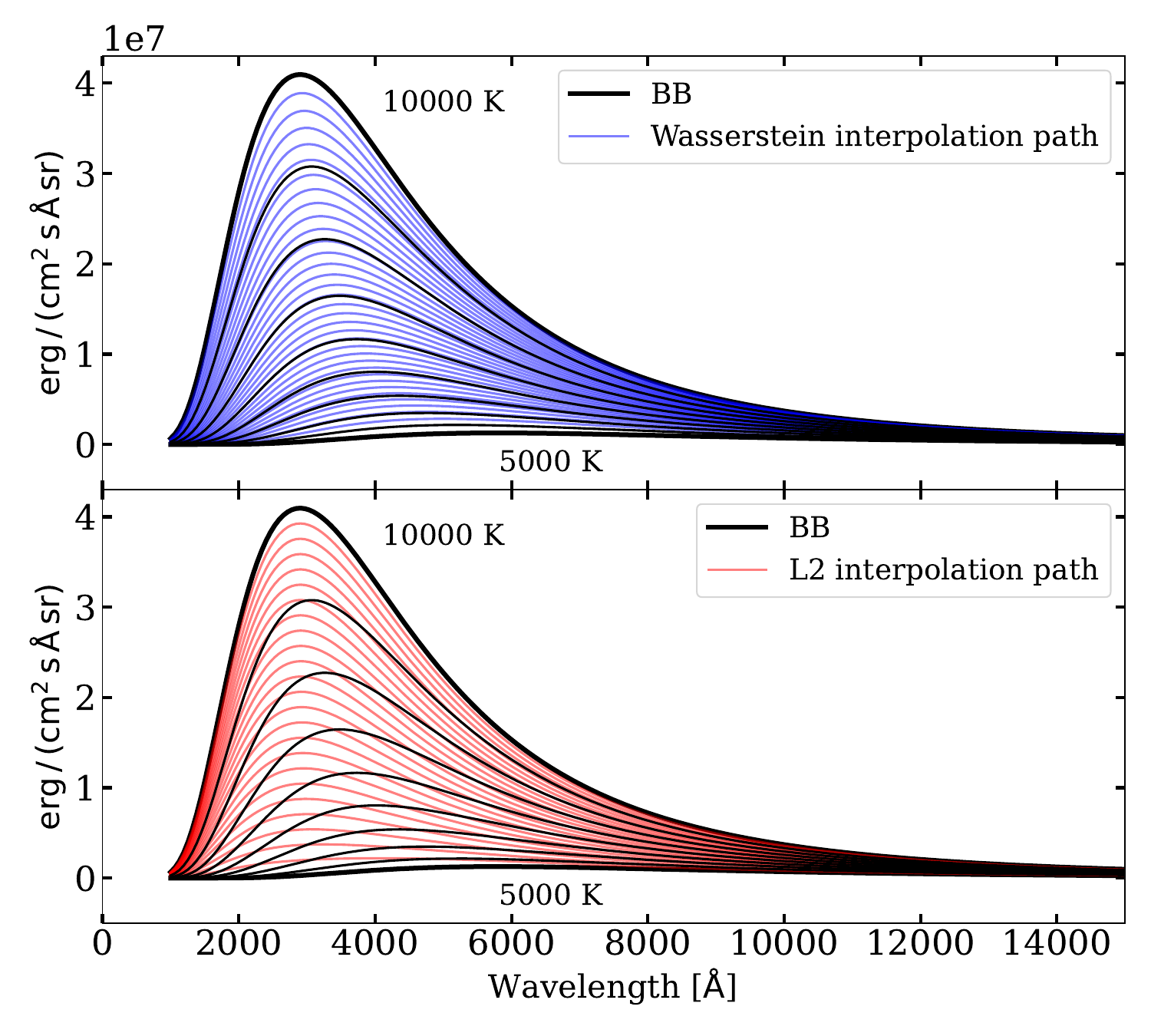}
    \caption{Examples of black body interpolations using Wasserstein and $L2$ barycenters. In both panels, black lines show black body radiation at different temperatures. In the top panel, OT interpolations are represented in blue, while in the bottom panel, linear interpolations are presented in red.}
    \label{fig:ot_bb}
\end{figure}

Moving closer to the subject of this paper, in Fig. \ref{fig:ot_bb} we apply $L2$ and OT methods to compute interpolations among black bodies. As visible in the top panel, the OT method effectively replicates the thermal evolution of black body radiation. On the contrary, the $L2$ barycenter leads to interpolations that do not accurately reflect the physical changes expected in a real black body as its temperature varies (see bottom panel).

Encouraged by these results, in this work we use the Python Optimal Transport (POT) library from \citet{flamary2021pot} to perform the interpolations in time between spectra of SNe. We specifically employ the barycenter computation function \footnote{\url{https://pythonot.github.io/gen_modules/ot.bregman}} for interpolating between two spectra, which takes a matrix $A$ that contains the distributions, in our case the spectra, a loss matrix ($M$), the regularization term ($reg$), and the weights of each distribution. The normalized loss matrix $M$ of size $n\times n $ was calculated with the utils function provided by the POT library \footnote{\url{https://pythonot.github.io/gen_modules/ot.utils.html}}.

In this case, an $\alpha$ value of 0 interpolates entirely to the spectrum of the first date, while a value of 1 interpolates to the spectrum of the second date. Consequently, a value of 0.5 results in a spectrum interpolated exactly halfway between the two dates. Considering the scenario where the initial and final phases are given by $ph_a = 2$ and $ph_b = 6$, respectively, to interpolate to intermediate phases $ph_c = [3, 4, 5]$,  $\alpha$ values of [0.25, 0.5, 0.75] must be employed. These $\alpha$ values correspond to the proportional distances of $ph_c$ within the interval defined by $ph_a$ and $ph_b$. This parameter shapes the weight array input for the {\fontfamily{lmtt}\selectfont ot.bregman.barycenter()} function using the form [$1-\alpha$,$\alpha$] for the two distributions.

\section{Analysis of the spectral interpolation performance}\label{sec:Analysis OT}

To assess the performance of OT for producing spectral time series, we conducted three sets of tests.
To avoid spurious residuals introduced by the noise and miscalibration that naturally affect the observed spectra, for the first two sets of tests, we use publicly available model spectra from \citet{Dessart_2014} and \citet{Dessart_2013}, for Type Ia and II SNe, respectively. For consistency with the tests performed on observed data, we calculated synthetic photometry for the JC and Sloan bands from the spectral time series models; this photometry is used to compute $\tilde{K}$ and $MAD(K)$ for each interpolated
spectrum. A phase to each spectrum is assigned, taking as a reference the epoch of the maximum light in the $V$ band obtained by fitting a second-order polynomial around the peak for SNe~Ia, while for SNe~II we considered the midpoint of the transition phase ($t_{PT}$) as defined in  \citet{Olivares2010}.

\subsection{Optimal transport on model spectra}

For our purposes, the most basic form of interpolation involves pairing two spectra. Therefore, our initial set of tests focused on this procedure. The test is conducted in the following way: Let us assume we have spectra at phases $ph_1$, $ph_2$, and $ph_3$ with $ph_1 < ph_2 < ph_3$. We first interpolate a spectrum at $ph_2$ from the spectra at $ph_1$ and $ph_3$; this spectrum is then rescaled against the photometry, as detailed in Sect. \ref{sec:Data_sample}, and compared to the actual $ph_2$ spectrum, computing the mean relative spectral residual, $\epsilon$, which is defined as follows:

\begin{equation}\label{eq:epsilon}
 \epsilon = \frac{1}{n} \sum_{\lambda=\lambda_0}^{n} \left( \frac{|f_{M_{\lambda}}-f_{I_{\lambda}}|} {f_{M_{\lambda}}} \right),
\end{equation}

\noindent where $f_{M}$ and $f_{I}$ represent the flux of the model and the flux of the interpolated spectra, respectively. 
This procedure is applied across all spectral pair combinations, utilizing a moving grid and progressively increasing the time interval between them.

\begin{figure} [!htb]
    \centering
    \includegraphics[width=1\columnwidth]{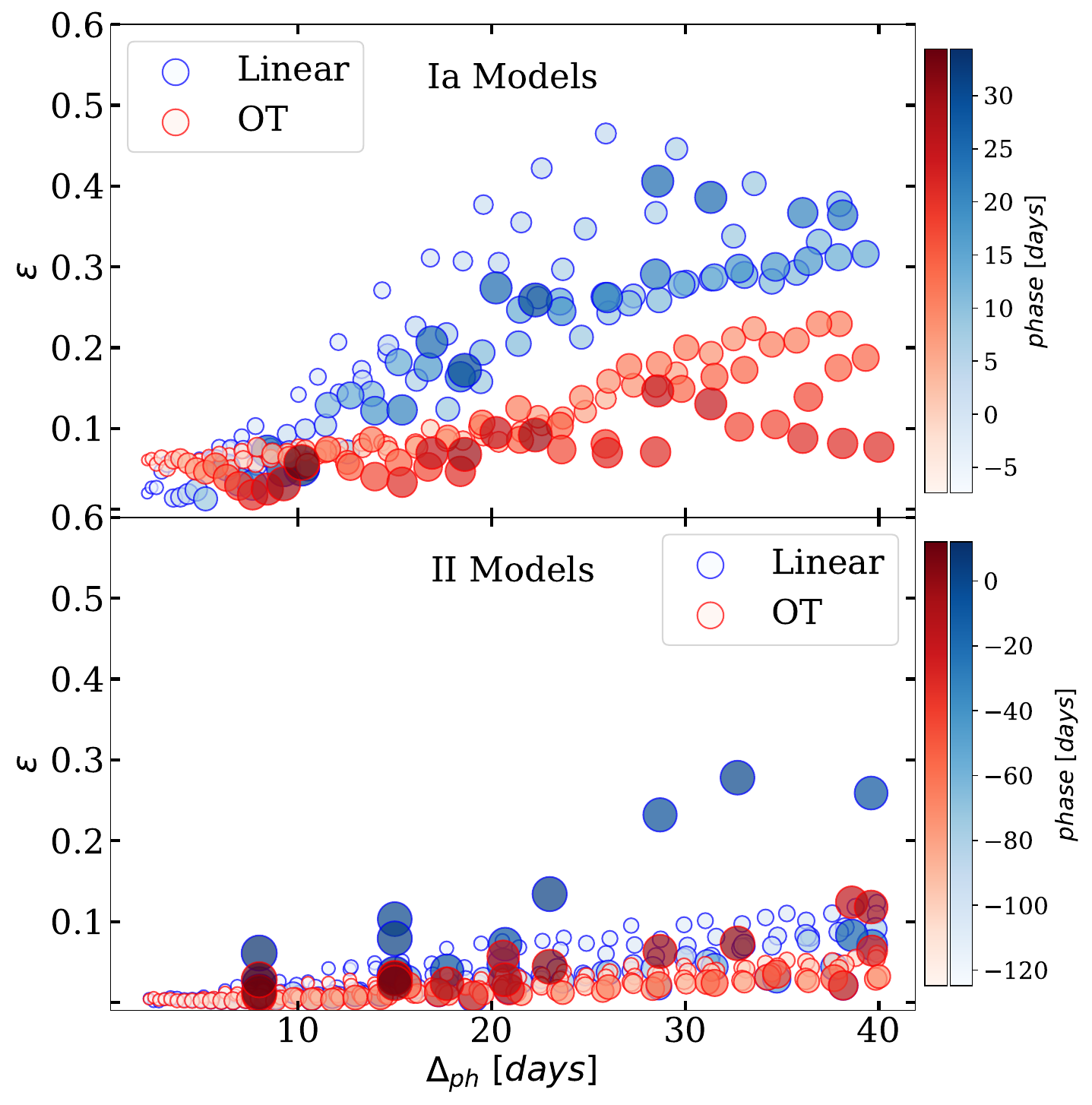}
    \caption{Relative spectral residuals $\epsilon$ as a function of $\Delta_{ph}$. Blue circles correspond to the linear interpolation and the red ones to the OT. The size and darkness of the circles increase with the phase of the interpolated spectrum.}
    \label{fig:Lineal_OT}
\end{figure}

As mentioned in Sect.~\ref{sec:Methodology}, OT employs an $\alpha$ value to define the position of the interpolation, which in this study corresponds to the phase of the spectrum we aim to compare with. It is worth mentioning that we only test interpolations with an $\alpha$ value ranging from 0.45 to 0.55. This is because an $\alpha$ of 0.5 is the most challenging case to interpolate, given that it represents the largest phase distance from the two spectra.
For comparison, we also compute the same interpolations between the same pairs of spectra but employ standard linear interpolation.
The results of this first set of tests can be seen in Fig.~\ref{fig:Lineal_OT}, where we indicate with  $\Delta_{ph}$ the phase difference between the pair of spectra. As expected, for small $\Delta_{ph}$, the relative spectral residual is low for both interpolation methods. OT shows a slower increase in $\epsilon$, which remains below 10\% even for phase gaps of 40 days for type II and 25 days for type Ia SNe. In contrast, linear interpolation sees a faster increase in $\epsilon$, exceeding 40\% as phase gaps become larger.
To account for the increment of $\epsilon$ introduced by an increasing $\Delta_{ph}$ between spectra, we fit a plane in the $\Delta_{ph}-ph-\epsilon$ space for both  SNe~Ia and SNe~II.
This plane allows us to assign an uncertainty $\Sigma(\epsilon)$, which we use to penalize interpolation between pairs of spectra with large $\Delta_{ph}$.

\begin{figure*}[!htb]
\centering
\includegraphics[width=1\textwidth]{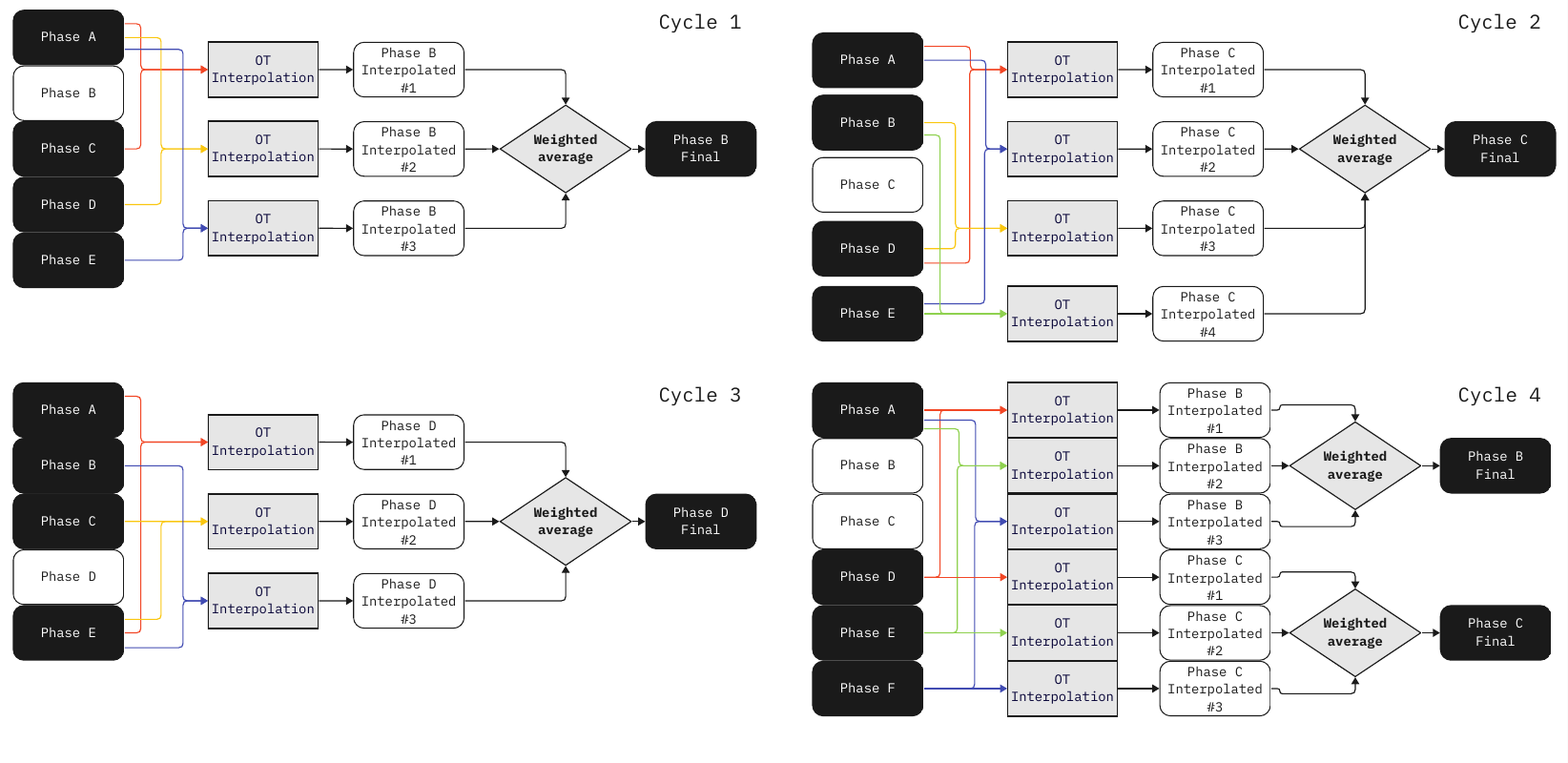}
\caption{Averaging scheme for the second test: Black boxes represent the spectra used for interpolation, and white boxes indicate the target phase for interpolation. Colored lines link pairs of spectra, which are used to create the interpolated spectrum at the position marked by the white box. To compute the final spectrum at a given phase, a weighted average is calculated across all interpolated spectra.}
\label{fig:scheme_combinations}

\end{figure*}

In the second set of tests, to include as much information as possible in generating a given interpolated spectrum, we consider not just one pair of spectra, but all possible combinations of pairs between four spectra. This approach is applied to both the linear and the OT method, ensuring that the same information is used in both cases. While a larger number of spectra could enhance the interpolation by providing more information, this scenario is not often realistic, given that not all SNe have an extensive number of available and well-calibrated spectra. The iterative procedure is illustrated in Fig. \ref{fig:scheme_combinations}, where the black boxes represent the four spectra used in the interpolation, and the white boxes indicate the position of the spectrum being interpolated. Colored lines connect the pairs of spectra used to compute the interpolated one. As in the previous test, following interpolation, we rescaled the spectra against the photometry. If there are three connections, this means that the weighted average spectrum is computed with these three interpolated spectra. As illustrated, once all possible interpolations among the existing spectrum pairs are completed, the grid moves to include an additional spectrum in the interpolation process, a concept exemplified in cycle 4.

The flux of the final spectrum $\overline{f(\lambda)}$ is computed as a weighted average over all the interpolated spectra, as follows:

\begin{equation}
\overline{f(\lambda)}=\frac{\sum_{i=1}^n{\frac{f_i(\lambda)}{(MAD(K)_i^2+\Sigma(\epsilon)_i^2)}}}{\sum_{i=1}^n{\frac{1}{(MAD(K)_i^2+\Sigma(\epsilon)_i^2)}}},
    \label{eq:mean_flux}
\end{equation}

\noindent where $f_i(\lambda)$ is the flux for the different spectra for a given $\lambda$ and $MAD(K)$ has the same meaning as in Sect. \ref{sec:Data_sample}.

\begin{figure}[!htb]
    \centering
    \includegraphics[width=\columnwidth]{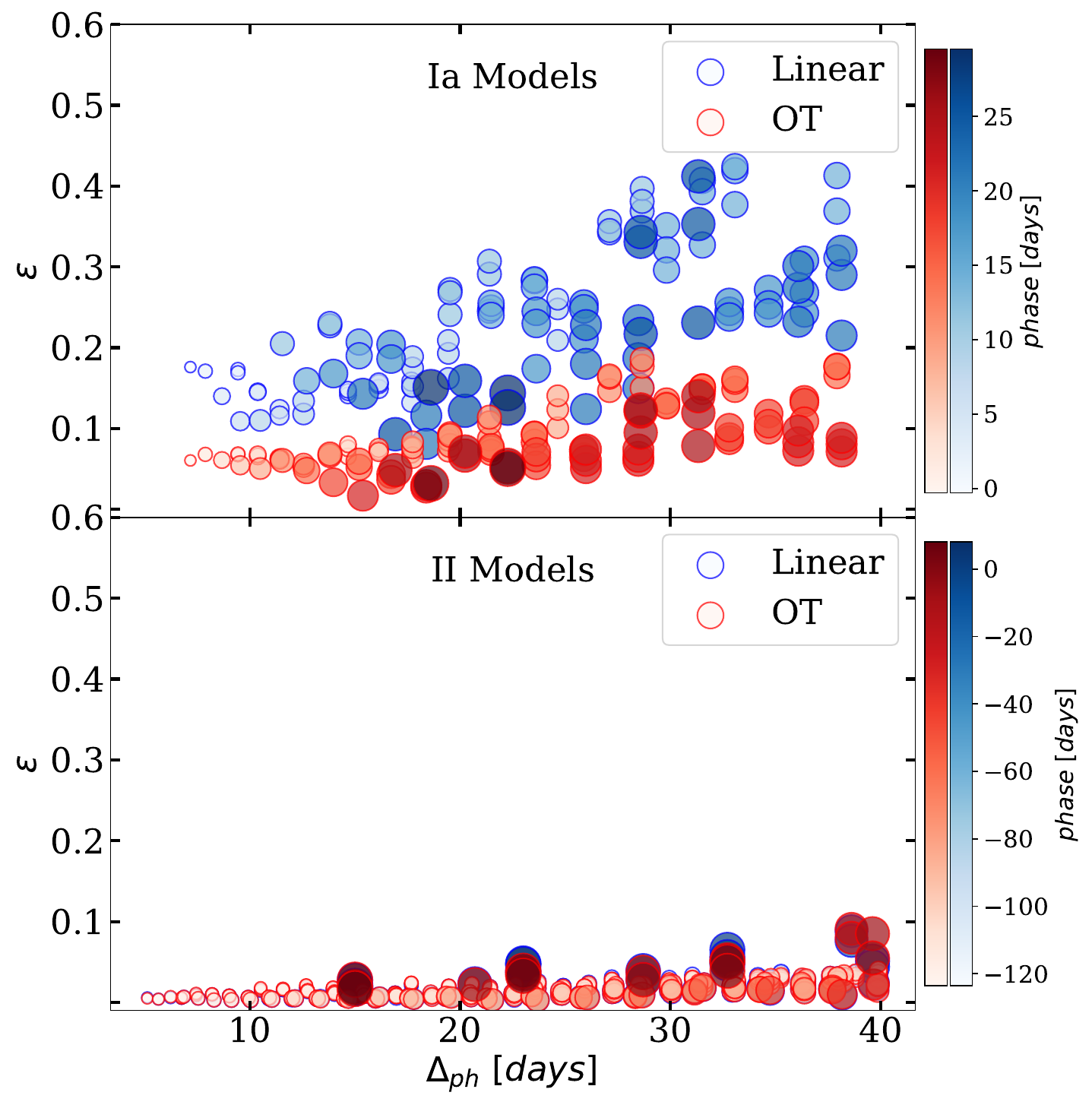}
    \caption{Same as Fig.~\ref{fig:Lineal_OT}, but in this case Eq.~\ref{eq:mean_flux} has been used to compute the flux of the spectra. We note that in this figure a given $\Delta_{ph}$ refers to the largest phase interval of the interpolated spectral pairs that enter in the weighted mean.} 
    \label{fig:GP_OT}
\end{figure}

 The results of this second set of tests are displayed in Fig. \ref{fig:GP_OT}, where the red circles represent the relative spectral residuals for OT interpolation and the blue circles represent those for the linear interpolation. As in Fig.~\ref{fig:Lineal_OT}, the size of the circles increases with phase. As expected, the residuals of the weighted spectra are larger than in  Fig.~\ref{fig:Lineal_OT} at a given $\Delta_{ph}$. This is because the weighted average includes spectra with larger $\Delta_{ph}$  than in Fig.~\ref{fig:Lineal_OT}. However, the inclusion of more than one pair of spectra in computing the interpolated spectrum is crucial in the case of observed spectra because it reduces the effect of noise and miscalibration.

\subsection{Optimal transport on observed spectra}
\label{subsec:OT_observed_spectra}

To evaluate the performance of OT on observed data, we generated a set of time series from the golden sample of spectra outlined in Sect. \ref{sec:Data_sample}. For each SN, we calculated interpolated spectra with a daily cadence from all the potential combinations of paired spectra within the sample. This implies that for a specific phase, we generate as many interpolated spectra as the number of combinations that include this phase. The process is the same as illustrated in Fig. \ref{fig:scheme_combinations}; however, in this instance, we did not limit ourselves to using only four spectra. Our goal is to use as much data as possible, and so we included all spectra with $\Delta_{ph}$ shorter than 40 days. For a given epoch, for each interpolated spectrum, we compute  $MAD(K)$, $\Sigma(\epsilon)$, and then rescale the spectrum using the $\tilde{K}$ value.  As the edges of spectra are usually poorly calibrated, we incorporate a weight term, $C(\lambda)$, which decreases linearly from 1 to 0 within a  $50 \, \AA$ window located at the boundaries of the interpolated spectrum.  The flux of the combined spectrum is computed as follows:

\begin{equation}
        \overline{f(\lambda)}=\frac{{\sum_{i=1}^n \frac{f_i(\lambda) C_i(\lambda)}{(MAD(K)_i^2+\Sigma(\epsilon)_i^2)}}}{{\sum_{i=1}^n \frac{C_i(\lambda)}{(MAD(K)_i^2+\Sigma(\epsilon)_i^2)}}}.
        \label{eq:mean_flux_2}
\end{equation}

\begin{figure*}[!htb]
    \centering
    \includegraphics[width=\textwidth,height=14cm]{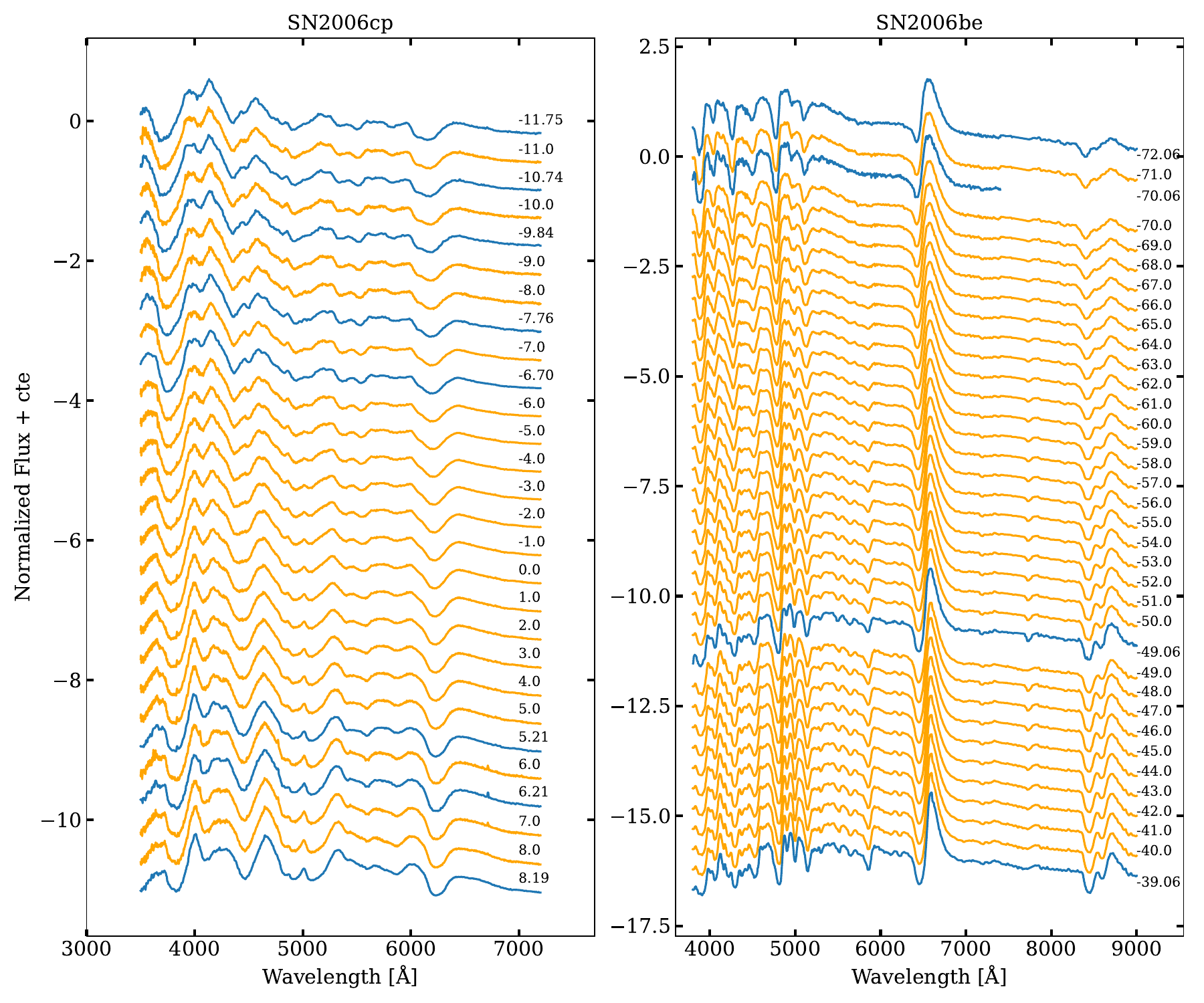}
    \caption{Spectral time series for Type Ia SN2006cp and Type II SN2006be. The observed spectra are shown in blue, while our spectral time series, calculated with the weighted average OT, are displayed in orange. The black numbers represent the phase with respect to the maximum flux and the $t_{pt}$, respectively.}
    \label{fig:template}
\end{figure*}

\begin{figure*}[!htb]
    \centering
    \includegraphics[width=\textwidth,height=10cm]{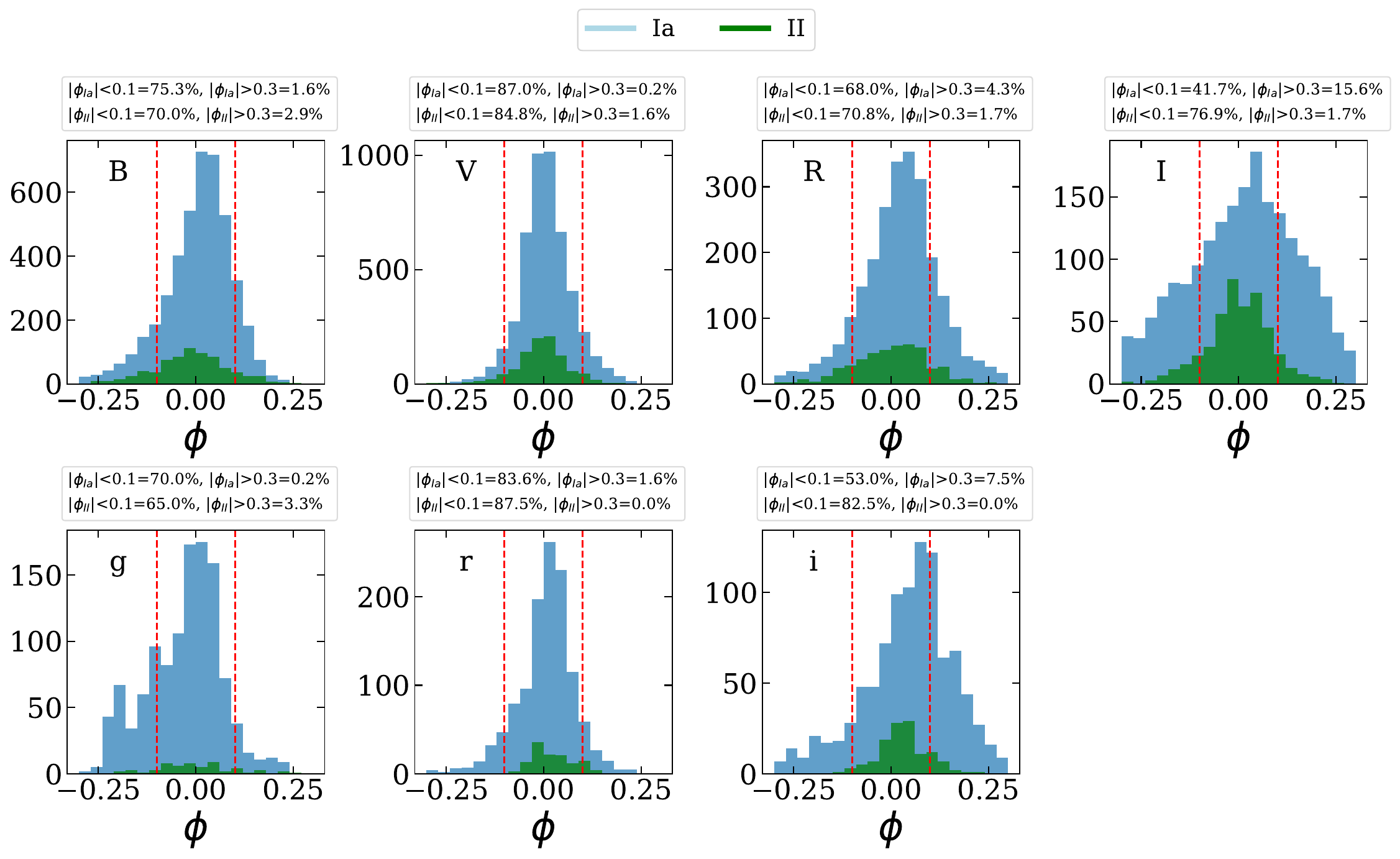}
    \caption{Relative photometric residuals $\phi$ for each of the \textit{BVRIgri} filters. The red dotted line denotes the zones where $|\phi|<0.1$. \tred{The legend also includes the fraction of relative residuals that fall outside the range shown in the histogram ($|\phi| > 0.3$).}}
    \label{fig:no-weighted_Ia}
\end{figure*}

Having produced our spectral time series, examples of which are reported in Fig. \ref{fig:template}, we proceed to compute the synthetic photometry using Eq.~\ref{eq:Syn_photometry} and compare it with the observed one. This is done for those bands for which at least 95\% of the total response is covered by the spectrum. \tred{L}inear interpolation is applied to the resulting synthetic light curves, enabling us to evaluate $F_{syn}$ for the corresponding $F_{obs}$ dates. We then compute  the relative photometric residuals $\phi$ and $\Phi$ as follows:

\begin{eqnarray}
  \phi =  \frac{F_{syn}-F_{obs}}{F_{obs}}, \\
  \Phi =  \frac{F_{syn}-F_{obs}}{\sigma_{obs}},
\end{eqnarray}

\noindent where $\sigma_{obs}$  is the error associated with the observed flux.
Figures \ref{fig:no-weighted_Ia} and \ref{fig:weighted_Ia} present these results for $\phi$ and $\Phi$, respectively.

\begin{figure*}[!htb]
    \centering
    \includegraphics[width=\textwidth,height=10cm]{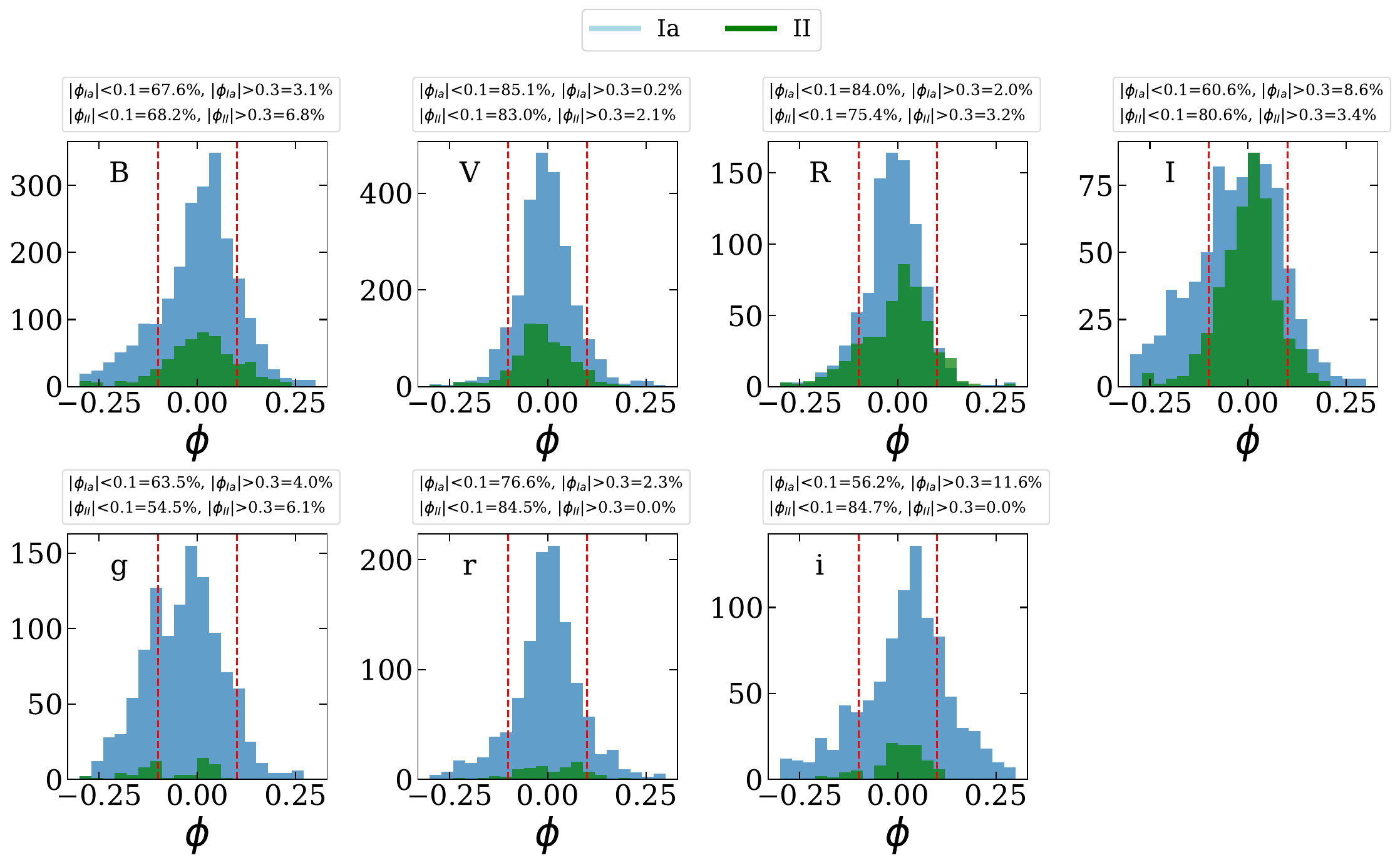}
    \caption{\tred{Same as Fig.~\ref{fig:no-weighted_Ia}, but with the relative residuals computed  using the leave-one-out cross-validation test.}}
    \label{fig:no-weighted_Ia_oneout}
\end{figure*}

We observe that the relative photometric residuals are generally below 10\%, and we do not see significant differences between SN~Ia and SN~II types, with the only exception being the $I$ and $i$ bands, where residuals are much lower in the latter than the former. We believe that these larger residuals are mostly due to the difference in natural bands of the instruments with which the observed photometry was obtained. Both $I$ and $i$ bands include the Ca~II NIR triplet feature and, especially for the $I$ band, the red cutoff can vary significantly between imagers \citep[e.g.,][]{2008MNRAS.388..971P}, including a different fraction of the P-Cygni profile. In SN~Ia, the Ca~II NIR triplet feature is much stronger than in SN~II, which can explain the larger residuals visible in the plots. In the case of $\Phi$, we observe similar trends. In most bands, the difference between synthetic and observed photometry falls within three times the error of the observed photometry.
\tred{In the previous test, although the spectra are only scaled by a constant factor $\tilde{K}$, this factor still contains information from all the available photometric observations. To ensure that this information is not entering into the estimation of the relative residual for any given band, we conducted an additional test using a leave-one-out cross-validation technique. For this test, we selected only SNe with spectra covered by at least four bands. This approach allows us to compute $\tilde{K}$ using at least three bands, even after leaving one out. For example, if we are computing the relative residual between the synthetic and observed photometry for the \textit{B} band, and the available bands are \textit{BVRI}, the scaling factor $\tilde{K}$ is computed using only the \textit{VRI} bands.
The results of this test are shown in Figs. \ref{fig:no-weighted_Ia_oneout} and \ref{fig:weighted_Ia_oneout}. As can be seen, for some bands, the distribution of residuals became slightly wider (\textit{BVgr}) or slightly narrower (\textit{Ri}) with respect Figs. \ref{fig:no-weighted_Ia} and \ref{fig:weighted_Ia}, respectively. The only case where the distribution became significantly narrower is in the case of SN Ia for the \textit{I} band, where the fraction of residuals below 0.1 magnitudes increased from 41.7\% to 60.6\%. The latter supports the hypothesis that a significant portion of the residuals in the $I$ band are due to differences in the natural bands of the instruments used to observe the SNe, rather than to a decrease in the performance of the interpolation method within the wavelength range covered by this band.} These results demonstrate that the spectral time series we generate with OT can produce accurate and reliable synthetic light curves that closely resemble the observed photometry.
The performance across these different SN types is consistent, showing the versatility of our method.

\section{Conclusions}
\label{sec:Conclusion}

In this study, we assessed the performance of the OT interpolation in producing spectral time series. Using SN models from \citet{Dessart_2013} and \citet{Dessart_2014}, we first tested the OT interpolation between pairs of spectra, finding that even with phase differences of 40 days, the relative spectral residuals ($\epsilon$) stay below 20\% and 10\% for SNe~Ia and SNe~II, respectively (Fig. \ref{fig:Lineal_OT}). To include more information in the generation of a given interpolation, we included all possible combinations between four synthetic spectra. Again the relative spectral residuals ($\epsilon$) stay below 20\% and 10\% for SN~Ia and SN~II, respectively (Fig. \ref{fig:GP_OT}). Our findings indicate that the error associated with the OT method increases at a slower rate with the phase difference compared to the linear method. This means that OT demonstrates a superior capability in preserving the spectral shape as the phase gap increases.

Finally, using the observed spectra of our golden sample of SNe described in Sect. \ref{sec:Data_sample}, we computed spectral time series, from which we constructed synthetic light curves in the \textit{BVRIgri} bands. We find that a significant portion of the relative photometric residuals ($\phi$) for both SN types generally fall below 10\% error (Fig. \ref{fig:no-weighted_Ia}), indicating a good match between the synthetic and observed light curves. For Type Ia SNe, this is particularly evident in the \textit{B}, \textit{V}, \textit{r}, and \textit{g} bands, while the \textit{I} and \textit{i} bands show more variability. \tred{Nevertheless, we find evidence that, for the \textit{I} band, differences between the natural bands of the instruments used to observe the SNe contribute at least in part to the residuals.} In the case of Type~II SNe, the residuals also show a good match across all bands, with notably high percentages of residuals within acceptable error margins. When examining the residuals weighted by observational error ($\Phi$), we observe that the majority of the differences between synthetic and observed photometry for both types of SNe are within three times the error of the observed photometry, although some specific bands for Type Ia SNe show larger deviations (Fig. \ref{fig:weighted_Ia}). 

In conclusion, the OT interpolation method emerges as a robust and innovative approach for creating spectral time series; it effectively performs accurate interpolations even in scenarios with substantial phase gaps between spectra, demonstrating its capability to produce high-quality synthetic light curves.  These \tred{spectral} time series \tred{are} highly suitable for generating training sets, which are essential for photometric classification algorithms. \tred{This aspect is particularly important in large astronomical surveys where extensive spectroscopic data may not be available. Additionally, the series may also be useful in performing K-corrections and bolometric corrections.}


\begin{acknowledgements}
The authors acknowledge support from National Agency for Research and Development (ANID) grants ANID-PFCHA/Doctorado Nacional/2020-21202606 (MR), ANID-PFCHA/Doctorado Nacional/2022-21221964 (BA). Support from the Chilean Ministry of Economy, Development, and Tourism's Millennium Science Initiative through grant ICN\textunderscore 12009, awarded to the Millennium Institute of Astrophysics (GP, MC, AMMA, FF); by FONDECYT Regular grant 1231637(MC), FONDECYT Regular 1200710 (FF) and by ANID's Basal project FB210003(MC,FF). BASAL project FB210005 (AMMA), BASAL Center of Mathematical Modeling Grant PAI AFB-170001 (FF). SGG acknowledges support by FCT under Project CRISP PTDC/FIS-AST-31546/2017 and Project~No.~UIDB/00099/2020. CPG acknowledges financial support from the Secretary of Universities and Research (Government of Catalonia) and by the Horizon 2020 Research and Innovation Programme of the European Union under the Marie Sk\l{}odowska-Curie and the Beatriu de Pin\'os 2021 BP 00168 programme, from the Spanish Ministerio de Ciencia e Innovaci\'on (MCIN) and the Agencia Estatal de Investigaci\'on (AEI) 10.13039/501100011033 under the PID2020-115253GA-I00 HOSTFLOWS project, and the program Unidad de Excelencia Mar\'ia de Maeztu CEX2020-001058-M.
\end{acknowledgements}

%
\bibliographystyle{aa} 
\bibliography{references} 

\begin{thebibliography}{245}
\expandafter\ifx\csname natexlab\endcsname\relax\def\natexlab#1{#1}\fi

\bibitem[{199(1997)}]{1997ASIC..486.....R}
 1997, NATO Advanced Study Institute (ASI) Series C, Vol. 486, {Thermonuclear supernovae}

\bibitem[{{Abbott} {et~al.}(2019){Abbott}, {Allam}, {Andersen}, {Angus}, {Asorey}, {Avelino}, {Avila}, {Bassett}, {Bechtol}, {Bernstein}, {Bertin}, {Brooks}, {Brout}, {Brown}, {Burke}, {Calcino}, {Carnero Rosell}, {Carollo}, {Carrasco Kind}, {Carretero}, {Casas}, {Castander}, {Cawthon}, {Challis}, {Childress}, {Clocchiatti}, {Cunha}, {D'Andrea}, {da Costa}, {Davis}, {Davis}, {De Vicente}, {DePoy}, {Desai}, {Diehl}, {Doel}, {Drlica-Wagner}, {Eifler}, {Evrard}, {Fernandez}, {Filippenko}, {Finley}, {Flaugher}, {Foley}, {Fosalba}, {Frieman}, {Galbany}, {Garc{\'\i}a-Bellido}, {Gaztanaga}, {Giannantonio}, {Glazebrook}, {Goldstein}, {Gonz{\'a}lez-Gait{\'a}n}, {Gruen}, {Gruendl}, {Gschwend}, {Gupta}, {Gutierrez}, {Hartley}, {Hinton}, {Hollowood}, {Honscheid}, {Hoormann}, {Hoyle}, {James}, {Jeltema}, {Johnson}, {Johnson}, {Kasai}, {Kent}, {Kessler}, {Kim}, {Kirshner}, {Kovacs}, {Krause}, {Kron}, {Kuehn}, {Kuhlmann}, {Kuropatkin}, {Lahav}, {Lasker}, {Lewis}, {Li}, {Lidman}, {Lima}, {Lin}, {Macaulay}, {Maia}, {Mandel},
  {March}, {Marriner}, {Marshall}, {Martini}, {Menanteau}, {Miller}, {Miquel}, {Miranda}, {Mohr}, {Morganson}, {Muthukrishna}, {M{\"o}ller}, {Neilsen}, {Nichol}, {Nord}, {Nugent}, {Ogando}, {Palmese}, {Pan}, {Plazas}, {Pursiainen}, {Romer}, {Roodman}, {Rozo}, {Rykoff}, {Sako}, {Sanchez}, {Scarpine}, {Schindler}, {Schubnell}, {Scolnic}, {Serrano}, {Sevilla-Noarbe}, {Sharp}, {Smith}, {Soares-Santos}, {Sobreira}, {Sommer}, {Spinka}, {Suchyta}, {Sullivan}, {Swann}, {Tarle}, {Thomas}, {Thomas}, {Troxel}, {Tucker}, {Uddin}, {Walker}, {Wester}, {Wiseman}, {Wolf}, {Yanny}, {Zhang}, {Zhang}, \& {DES Collaboration}}]{Abbott2019}
{Abbott}, T.~M.~C., {Allam}, S., {Andersen}, P., {et~al.} 2019, \apjl, 872, L30

\bibitem[{{Aldering} {et~al.}(2002){Aldering}, {Adam}, {Antilogus}, {Astier}, {Bacon}, {Bongard}, {Bonnaud}, {Copin}, {Hardin}, {Henault}, {Howell}, {Lemonnier}, {Levy}, {Loken}, {Nugent}, {Pain}, {Pecontal}, {Pecontal}, {Perlmutter}, {Quimby}, {Schahmaneche}, {Smadja}, \& {Wood-Vasey}}]{2002SPIE.4836...61A}
{Aldering}, G., {Adam}, G., {Antilogus}, P., {et~al.} 2002, in Society of Photo-Optical Instrumentation Engineers (SPIE) Conference Series, Vol. 4836, Survey and Other Telescope Technologies and Discoveries, ed. J.~A. {Tyson} \& S.~{Wolff}, 61--72

\bibitem[{{Altavilla} {et~al.}(2004){Altavilla}, {Fiorentino}, {Marconi}, {Musella}, {Cappellaro}, {Barbon}, {Benetti}, {Pastorello}, {Riello}, {Turatto}, \& {Zampieri}}]{2004MNRAS.349.1344A}
{Altavilla}, G., {Fiorentino}, G., {Marconi}, M., {et~al.} 2004, \mnras, 349, 1344

\bibitem[{{Altavilla} {et~al.}(2007){Altavilla}, {Stehle}, {Ruiz-Lapuente}, {Mazzali}, {Pignata}, {Balastegui}, {Benetti}, {Blanc}, {Canal}, {Elias-Rosa}, {Goobar}, {Harutyunyan}, {Pastorello}, {Patat}, {Rich}, {Salvo}, {Schmidt}, {Stanishev}, {Taubenberger}, {Turatto}, \& {Hillebrandt}}]{2007A&A...475..585A}
{Altavilla}, G., {Stehle}, M., {Ruiz-Lapuente}, P., {et~al.} 2007, \aap, 475, 585

\bibitem[{{Anderson} {et~al.}(2014){Anderson}, {Gonz{\'a}lez-Gait{\'a}n}, {Hamuy}, {Guti{\'e}rrez}, {Stritzinger}, {Olivares E.}, {Phillips}, {Schulze}, {Antezana}, {Bolt}, {Campillay}, {Castell{\'o}n}, {Contreras}, {de Jaeger}, {Folatelli}, {F{\"o}rster}, {Freedman}, {Gonz{\'a}lez}, {Hsiao}, {Krzemi{\'n}ski}, {Krisciunas}, {Maza}, {McCarthy}, {Morrell}, {Persson}, {Roth}, {Salgado}, {Suntzeff}, \& {Thomas-Osip}}]{2014ApJ...786...67A}
{Anderson}, J.~P., {Gonz{\'a}lez-Gait{\'a}n}, S., {Hamuy}, M., {et~al.} 2014, \apj, 786, 67

\bibitem[{{Andrews} {et~al.}(2010){Andrews}, {Gallagher}, {Clayton}, {Sugerman}, {Chatelain}, {Clem}, {Welch}, {Barlow}, {Ercolano}, {Fabbri}, {Wesson}, \& {Meixner}}]{2010ApJ...715..541A}
{Andrews}, J.~E., {Gallagher}, J.~S., {Clayton}, G.~C., {et~al.} 2010, \apj, 715, 541

\bibitem[{{Argo} {et~al.}(2004){Argo}, {Muxlow}, {Beswick}, {Pedlar}, \& {Marcaide}}]{2004IAUC.8399....3A}
{Argo}, M.~K., {Muxlow}, T.~W.~B., {Beswick}, R.~J., {Pedlar}, A., \& {Marcaide}, J.~M. 2004, \iaucirc, 8399, 3

\bibitem[{{Argyle} {et~al.}(1994){Argyle}, {Morrison}, {Knudsen}, {Einicke}, {Helmer}, {Nicolet}, {Mikuz}, \& {Green}}]{1994IAUC.5976....3A}
{Argyle}, R.~W., {Morrison}, L.~V., {Knudsen}, T., {et~al.} 1994, \iaucirc, 5976, 3

\bibitem[{{Behrend} {et~al.}(2004){Behrend}, {Roy}, {Rinner}, {Antonini}, {Pravec}, {Harris}, {Sposetti}, {Durkee}, \& {Klotz}}]{2004IAUC.8265....2B}
{Behrend}, R., {Roy}, R., {Rinner}, C., {et~al.} 2004, \iaucirc, 8265, 2

\bibitem[{{Benetti} {et~al.}(1994){Benetti}, {Cappellaro}, {Turatto}, {della Valle}, {Mazzali}, \& {Gouiffes}}]{1994A&A...285..147B}
{Benetti}, S., {Cappellaro}, E., {Turatto}, M., {et~al.} 1994, \aap, 285, 147

\bibitem[{{Bessell} {et~al.}(1998){Bessell}, {Castelli}, \& {Plez}}]{Bessell_1998}
{Bessell}, M.~S., {Castelli}, F., \& {Plez}, B. 1998, \aap, 333, 231

\bibitem[{{Blondin} {et~al.}(2007{\natexlab{a}}){Blondin}, {Kirshner}, {Challis}, \& {Berlind}}]{2007CBET.1062....2B}
{Blondin}, S., {Kirshner}, R., {Challis}, P., \& {Berlind}, P. 2007{\natexlab{a}}, Central Bureau Electronic Telegrams, 1062, 2

\bibitem[{{Blondin} {et~al.}(2007{\natexlab{b}}){Blondin}, {Kirshner}, {Challis}, \& {Calkins}}]{2007CBET.1048....2B}
{Blondin}, S., {Kirshner}, R., {Challis}, P., \& {Calkins}, M. 2007{\natexlab{b}}, Central Bureau Electronic Telegrams, 1048, 2

\bibitem[{{Blondin} {et~al.}(2012){Blondin}, {Matheson}, {Kirshner}, {Mandel}, {Berlind}, {Calkins}, {Challis}, {Garnavich}, {Jha}, {Modjaz}, {Riess}, \& {Schmidt}}]{2012AJ....143..126B}
{Blondin}, S., {Matheson}, T., {Kirshner}, R.~P., {et~al.} 2012, \aj, 143, 126

\bibitem[{{Boone}(2019)}]{Boone2019}
{Boone}, K. 2019, \aj, 158, 257

\bibitem[{{Bouma} {et~al.}(1998){Bouma}, {Lehky}, \& {Carvajal}}]{1998IAUC.6993....3B}
{Bouma}, R.~J., {Lehky}, M., \& {Carvajal}, J. 1998, \iaucirc, 6993, 3

\bibitem[{{Branch} {et~al.}(2003){Branch}, {Garnavich}, {Matheson}, {Baron}, {Thomas}, {Hatano}, {Challis}, {Jha}, \& {Kirshner}}]{2003AJ....126.1489B}
{Branch}, D., {Garnavich}, P., {Matheson}, T., {et~al.} 2003, \aj, 126, 1489

\bibitem[{{Brown} {et~al.}(2014){Brown}, {Breeveld}, {Holland}, {Kuin}, \& {Pritchard}}]{2014Ap&SS.354...89B}
{Brown}, P.~J., {Breeveld}, A.~A., {Holland}, S., {Kuin}, P., \& {Pritchard}, T. 2014, \apss, 354, 89

\bibitem[{{Brown} {et~al.}(2012){Brown}, {Dawson}, {Harris}, {Olmstead}, {Milne}, \& {Roming}}]{2012ApJ...749...18B}
{Brown}, P.~J., {Dawson}, K.~S., {Harris}, D.~W., {et~al.} 2012, \apj, 749, 18

\bibitem[{{Bufano} {et~al.}(2009){Bufano}, {Immler}, {Turatto}, {Landsman}, {Brown}, {Benetti}, {Cappellaro}, {Holland}, {Mazzali}, {Milne}, {Panagia}, {Pian}, {Roming}, {Zampieri}, {Breeveld}, \& {Gehrels}}]{2009ApJ...700.1456B}
{Bufano}, F., {Immler}, S., {Turatto}, M., {et~al.} 2009, \apj, 700, 1456

\bibitem[{{Burns} {et~al.}(2018){Burns}, {Parent}, {Phillips}, {Stritzinger}, {Krisciunas}, {Suntzeff}, {Hsiao}, {Contreras}, {Anais}, {Boldt}, {Busta}, {Campillay}, {Castell{\'o}n}, {Folatelli}, {Freedman}, {Gonz{\'a}lez}, {Hamuy}, {Heoflich}, {Krzeminski}, {Madore}, {Morrell}, {Persson}, {Roth}, {Salgado}, {Ser{\'o}n}, \& {Torres}}]{2018arXiv180906381B}
{Burns}, C.~R., {Parent}, E., {Phillips}, M.~M., {et~al.} 2018, \apj, 869, 56

\bibitem[{{Cellier-Holzem} {et~al.}(2012){Cellier-Holzem}, {Canto}, {Antilogus}, {Bongard}, {Pain}, {Copin}, {Gangler}, {Pereira}, {Rigault}, {Smadja}, {Aldering}, {Birchall}, {Childress}, {Fakhouri}, {Kim}, {Nordin}, {Nugent}, {Perlmutter}, {Runge}, {Saunders}, {Suzuki}, {Thomas}, {Pecontal}, {Buton}, {Feindt}, {Kerschhaggl}, {Kowalski}, {Benitez}, {Hillebrandt}, {Kromer}, {Sasdelli}, {Sternberg}, {Taubenberger}, {Baugh}, {Chen}, {Chotard}, {Tao}, {Fouchez}, {Tilquin}, {Hadjiyska}, {Rabinowitz}, {Baltay}, {Ellman}, {McKinnon}, \& {Effron}}]{2012ATel.4566....1C}
{Cellier-Holzem}, F., {Canto}, A., {Antilogus}, P., {et~al.} 2012, The Astronomer's Telegram, 4566, 1

\bibitem[{{Ceverino} \& {Klypin}(2009)}]{Ceverino2009}
{Ceverino}, D. \& {Klypin}, A. 2009, \apj, 695, 292

\bibitem[{{Chakraborti} {et~al.}(2016){Chakraborti}, {Ray}, {Smith}, {Margutti}, {Pooley}, {Bose}, {Sutaria}, {Chandra}, {Dwarkadas}, {Ryder}, \& {Maeda}}]{2016ApJ...817...22C}
{Chakraborti}, S., {Ray}, A., {Smith}, R., {et~al.} 2016, \apj, 817, 22

\bibitem[{{Charnock} \& {Moss}(2017)}]{Charnock2017}
{Charnock}, T. \& {Moss}, A. 2017, \apjl, 837, L28

\bibitem[{{Childress} {et~al.}(2015){Childress}, {Hillier}, {Seitenzahl}, {Sullivan}, {Maguire}, {Taubenberger}, {Scalzo}, {Ruiter}, {Blagorodnova}, {Camacho}, {Castillo}, {Elias-Rosa}, {Fraser}, {Gal-Yam}, {Graham}, {Howell}, {Inserra}, {Jha}, {Kumar}, {Mazzali}, {McCully}, {Morales-Garoffolo}, {Pandya}, {Polshaw}, {Schmidt}, {Smartt}, {Smith}, {Sollerman}, {Spyromilio}, {Tucker}, {Valenti}, {Walton}, {Wolf}, {Yaron}, {Young}, {Yuan}, \& {Zhang}}]{2015arXiv150702501C}
{Childress}, M.~J., {Hillier}, D.~J., {Seitenzahl}, I., {et~al.} 2015, \mnras, 454, 3816

\bibitem[{{Childress} {et~al.}(2013){Childress}, {Scalzo}, {Sim}, {Tucker}, {Yuan}, {Schmidt}, {Cenko}, {Silverman}, {Contreras}, {Hsiao}, {Phillips}, {Morrell}, {Jha}, {McCully}, {Filippenko}, {Anderson}, {Benetti}, {Bufano}, {de Jaeger}, {Forster}, {Gal-Yam}, {Le Guillou}, {Maguire}, {Maund}, {Mazzali}, {Pignata}, {Smartt}, {Spyromilio}, {Sullivan}, {Taddia}, {Valenti}, {Bayliss}, {Bessell}, {Blanc}, {Carson}, {Clubb}, {de Burgh-Day}, {Desjardins}, {Fang}, {Fox}, {Gates}, {Ho}, {Keller}, {Kelly}, {Lidman}, {Loaring}, {Mould}, {Owers}, {Ozbilgen}, {Pei}, {Pickering}, {Pracy}, {Rich}, {Schaefer}, {Scott}, {Stritzinger}, {Vogt}, \& {Zhou}}]{2013ApJ...770...29C}
{Childress}, M.~J., {Scalzo}, R.~A., {Sim}, S.~A., {et~al.} 2013, \apj, 770, 29

\bibitem[{{Childress} {et~al.}(2016){Childress}, {Tucker}, {Yuan}, {Scalzo}, {Ruiter}, {Seitenzahl}, {Zhang}, {Schmidt}, {Anguiano}, {Aniyan}, {Bayliss}, {Bento}, {Bessell}, {Bian}, {Davies}, {Dopita}, {Fogarty}, {Fraser-McKelvie}, {Freeman}, {Kuruwita}, {Medling}, {Murphy}, {Murphy}, {Owers}, {Panther}, {Sweet}, {Thomas}, \& {Zhou}}]{2016arXiv160708526C}
{Childress}, M.~J., {Tucker}, B.~E., {Yuan}, F., {et~al.} 2016, \pasa, 33, e055

\bibitem[{{Chomiuk} {et~al.}(2016){Chomiuk}, {Soderberg}, {Chevalier}, {Bruzewski}, {Foley}, {Parrent}, {Strader}, {Badenes}, {Fransson}, {Kamble}, {Margutti}, {Rupen}, \& {Simon}}]{2016ApJ...821..119C}
{Chomiuk}, L., {Soderberg}, A.~M., {Chevalier}, R.~A., {et~al.} 2016, \apj, 821, 119

\bibitem[{{Chomiuk} {et~al.}(2012){Chomiuk}, {Soderberg}, {Moe}, {Chevalier}, {Rupen}, {Badenes}, {Margutti}, {Fransson}, {Fong}, \& {Dittmann}}]{2012ApJ...750..164C}
{Chomiuk}, L., {Soderberg}, A.~M., {Moe}, M., {et~al.} 2012, \apj, 750, 164

\bibitem[{{Christensen} {et~al.}(2003){Christensen}, {Becker}, {Jahnke}, {Kelz}, {Roth}, {S{\'a}nchez}, \& {Wisotzki}}]{2003A&A...401..479C}
{Christensen}, L., {Becker}, T., {Jahnke}, K., {et~al.} 2003, \aap, 401, 479

\bibitem[{{Contreras} {et~al.}(2010){Contreras}, {Hamuy}, {Phillips}, {Folatelli}, {Suntzeff}, {Persson}, {Stritzinger}, {Boldt}, {Gonz{\'a}lez}, {Krzeminski}, {Morrell}, {Roth}, {Salgado}, {Maureira}, {Burns}, {Freedman}, {Madore}, {Murphy}, {Wyatt}, {Li}, \& {Filippenko}}]{2010AJ....139..519C}
{Contreras}, C., {Hamuy}, M., {Phillips}, M.~M., {et~al.} 2010, \aj, 139, 519

\bibitem[{{Corelli} {et~al.}(2008){Corelli}, {Yamaoka}, \& {Itagaki}}]{2008CBET.1228....2C}
{Corelli}, P., {Yamaoka}, H., \& {Itagaki}, K. 2008, Central Bureau Electronic Telegrams, 1228, 2

\bibitem[{{Cousins}(1980)}]{Cousins_1980}
{Cousins}, A.~W.~J. 1980, South African Astronomical Observatory Circular, 1, 234

\bibitem[{{Cousins}(1984)}]{Cousins_1984}
{Cousins}, A.~W.~J. 1984, South African Astronomical Observatory Circular, 8, 69

\bibitem[{{de Jaeger} {et~al.}(2019){de Jaeger}, {Zheng}, {Stahl}, {Filippenko}, {Brink}, {Bigley}, {Blanchard}, {Blanchard}, {Bradley}, {Cargill}, {Casper}, {Cenko}, {Channa}, {Choi}, {Clubb}, {Cobb}, {Cohen}, {de Kouchkovsky}, {Ellison}, {Falcon}, {Fox}, {Fuller}, {Ganeshalingam}, {Gould}, {Graham}, {Halevi}, {Hayakawa}, {Hestenes}, {Hyland}, {Jeffers}, {Joubert}, {Kandrashoff}, {Kelly}, {Kim}, {Kim}, {Kumar}, {Leonard}, {Li}, {Lowe}, {Lu}, {Mason}, {McAllister}, {Mauerhan}, {Modjaz}, {Molloy}, {Perley}, {Pina}, {Poznanski}, {Ross}, {Shivvers}, {Silverman}, {Soler}, {Stegman}, {Taylor}, {Tang}, {Wilkins}, {Wang}, {Wang}, {Yuk}, {Yunus}, \& {Zhang}}]{2019MNRAS.490.2799D}
{de Jaeger}, T., {Zheng}, W., {Stahl}, B.~E., {et~al.} 2019, \mnras, 490, 2799

\bibitem[{{Dessart} {et~al.}(2008){Dessart}, {Blondin}, {Brown}, {Hicken}, {Hillier}, {Holland}, {Immler}, {Kirshner}, {Milne}, {Modjaz}, \& {Roming}}]{2008ApJ...675..644D}
{Dessart}, L., {Blondin}, S., {Brown}, P.~J., {et~al.} 2008, \apj, 675, 644

\bibitem[{{Dessart} {et~al.}(2014){Dessart}, {Blondin}, {Hillier}, \& {Khokhlov}}]{Dessart_2014}
{Dessart}, L., {Blondin}, S., {Hillier}, D.~J., \& {Khokhlov}, A. 2014, \mnras, 441, 532

\bibitem[{Dessart {et~al.}(2013)Dessart, Hillier, Waldman, \& Livne}]{Dessart_2013}
Dessart, L., Hillier, D.~J., Waldman, R., \& Livne, E. 2013, Monthly Notices of the Royal Astronomical Society, 433, 1745

\bibitem[{{Dhungana} {et~al.}(2016){Dhungana}, {Kehoe}, {Vinko}, {Silverman}, {Wheeler}, {Zheng}, {Marion}, {Fox}, {Akerlof}, {Biro}, {Borkovits}, {Cenko}, {Clubb}, {Filippenko}, {Ferrante}, {Gibson}, {Graham}, {Hegedus}, {Kelly}, {Kelemen}, {Lee}, {Marschalko}, {Moln{\'a}r}, {Nagy}, {Ordasi}, {Pal}, {Sarneczky}, {Shivvers}, {Szakats}, {Szalai}, {Szegedi-Elek}, {Sz{\'e}kely}, {Szing}, {Tak{\'a}ts}, \& {Vida}}]{2016ApJ...822....6D}
{Dhungana}, G., {Kehoe}, R., {Vinko}, J., {et~al.} 2016, \apj, 822, 6

\bibitem[{{Diepvens} {et~al.}(2006){Diepvens}, {Gonzalez}, {Bouma}, \& {King}}]{2006IAUC.8766....3D}
{Diepvens}, A., {Gonzalez}, J.~J., {Bouma}, R.~J., \& {King}, B. 2006, \iaucirc, 8766, 3

\bibitem[{{Doi} {et~al.}(2007){Doi}, {Nakano}, {Itagaki}, {Naito}, \& {Iizuka}}]{2007CBET..848....1D}
{Doi}, T., {Nakano}, S., {Itagaki}, K., {Naito}, H., \& {Iizuka}, R. 2007, Central Bureau Electronic Telegrams, 848, 1

\bibitem[{{Drake} {et~al.}(2009){Drake}, {Djorgovski}, {Mahabal}, {Beshore}, {Larson}, {Graham}, {Williams}, {Christensen}, {Catelan}, {Boattini}, {Gibbs}, {Hill}, \& {Kowalski}}]{2009ApJ...696..870D}
{Drake}, A.~J., {Djorgovski}, S.~G., {Mahabal}, A., {et~al.} 2009, \apj, 696, 870

\bibitem[{{Elias} {et~al.}(1985){Elias}, {Matthews}, {Neugebauer}, \& {Persson}}]{Elias1985}
{Elias}, J.~H., {Matthews}, K., {Neugebauer}, G., \& {Persson}, S.~E. 1985, \apj, 296, 379

\bibitem[{{Elias-Rosa} {et~al.}(2006){Elias-Rosa}, {Benetti}, {Cappellaro}, {Turatto}, {Mazzali}, {Patat}, {Meikle}, {Stehle}, {Pastorello}, {Pignata}, {Kotak}, {Harutyunyan}, {Altavilla}, {Navasardyan}, {Qiu}, {Salvo}, \& {Hillebrandt}}]{2006MNRAS.369.1880E}
{Elias-Rosa}, N., {Benetti}, S., {Cappellaro}, E., {et~al.} 2006, \mnras, 369, 1880

\bibitem[{{Elmhamdi} {et~al.}(2003){Elmhamdi}, {Danziger}, {Chugai}, {Pastorello}, {Turatto}, {Cappellaro}, {Altavilla}, {Benetti}, {Patat}, \& {Salvo}}]{2003MNRAS.338..939E}
{Elmhamdi}, A., {Danziger}, I.~J., {Chugai}, N., {et~al.} 2003, \mnras, 338, 939

\bibitem[{{Fabbri} {et~al.}(2011){Fabbri}, {Otsuka}, {Barlow}, {Gallagher}, {Wesson}, {Sugerman}, {Clayton}, {Meixner}, {Andrews}, {Welch}, \& {Ercolano}}]{2011MNRAS.418.1285F}
{Fabbri}, J., {Otsuka}, M., {Barlow}, M.~J., {et~al.} 2011, \mnras, 418, 1285

\bibitem[{{Faran} {et~al.}(2014){Faran}, {Poznanski}, {Filippenko}, {Chornock}, {Foley}, {Ganeshalingam}, {Leonard}, {Li}, {Modjaz}, {Nakar}, {Serduke}, \& {Silverman}}]{2014MNRAS.442..844F}
{Faran}, T., {Poznanski}, D., {Filippenko}, A.~V., {et~al.} 2014, \mnras, 442, 844

\bibitem[{{Filippenko} \& {Foley}(2004)}]{2004IAUC.8453....3F}
{Filippenko}, A.~V. \& {Foley}, R.~J. 2004, \iaucirc, 8453, 3

\bibitem[{Flamary {et~al.}(2021)Flamary, Courty, Gramfort, Alaya, Boisbunon, Chambon, Chapel, Corenflos, Fatras, Fournier, Gautheron, Gayraud, Janati, Rakotomamonjy, Redko, Rolet, Schutz, Seguy, Sutherland, Tavenard, Tong, \& Vayer}]{flamary2021pot}
Flamary, R., Courty, N., Gramfort, A., {et~al.} 2021, Journal of Machine Learning Research, 22, 1

\bibitem[{{Folatelli} {et~al.}(2013){Folatelli}, {Morrell}, {Phillips}, {Hsiao}, {Campillay}, {Contreras}, {Castell{\'o}n}, {Hamuy}, {Krzeminski}, {Roth}, {Stritzinger}, {Burns}, {Freedman}, {Madore}, {Murphy}, {Persson}, {Prieto}, {Suntzeff}, {Krisciunas}, {Anderson}, {F{\"o}rster}, {Maza}, {Pignata}, {Rojas}, {Boldt}, {Salgado}, {Wyatt}, {Olivares E.}, {Gal-Yam}, \& {Sako}}]{2013ApJ...773...53F}
{Folatelli}, G., {Morrell}, N., {Phillips}, M.~M., {et~al.} 2013, \apj, 773, 53

\bibitem[{{Foley} {et~al.}(2012){Foley}, {Challis}, {Filippenko}, {Ganeshalingam}, {Landsman}, {Li}, {Marion}, {Silverman}, {Beaton}, {Bennert}, {Cenko}, {Childress}, {Guhathakurta}, {Jiang}, {Kalirai}, {Kirshner}, {Stockton}, {Tollerud}, {Vink{\'o}}, {Wheeler}, \& {Woo}}]{2012ApJ...744...38F}
{Foley}, R.~J., {Challis}, P.~J., {Filippenko}, A.~V., {et~al.} 2012, \apj, 744, 38

\bibitem[{{Foley} {et~al.}(2008){Foley}, {Filippenko}, \& {Jha}}]{2008ApJ...686..117F}
{Foley}, R.~J., {Filippenko}, A.~V., \& {Jha}, S.~W. 2008, \apj, 686, 117

\bibitem[{{Foley} {et~al.}(2014){Foley}, {Fox}, {McCully}, {Phillips}, {Sand}, {Zheng}, {Challis}, {Filippenko}, {Folatelli}, {Hillebrandt}, {Hsiao}, {Jha}, {Kirshner}, {Kromer}, {Marion}, {Nelson}, {Pakmor}, {Pignata}, {R{\"o}pke}, {Seitenzahl}, {Silverman}, {Skrutskie}, \& {Stritzinger}}]{2014MNRAS.443.2887F}
{Foley}, R.~J., {Fox}, O.~D., {McCully}, C., {et~al.} 2014, \mnras, 443, 2887

\bibitem[{{Foley} {et~al.}(2018){Foley}, {Scolnic}, {Rest}, {Jha}, {Pan}, {Riess}, {Challis}, {Chambers}, {Coulter}, {Dettman}, {Foley}, {Fox}, {Huber}, {Jones}, {Kilpatrick}, {Kirshner}, {Schultz}, {Siebert}, {Flewelling}, {Gibson}, {Magnier}, {Miller}, {Primak}, {Smartt}, {Smith}, {Wainscoat}, {Waters}, \& {Willman}}]{2017arXiv171102474F}
{Foley}, R.~J., {Scolnic}, D., {Rest}, A., {et~al.} 2018, \mnras, 475, 193

\bibitem[{{Fox} {et~al.}(2010){Fox}, {Chevalier}, \& {Skrutskie}}]{2010ATel.2665....1F}
{Fox}, O.~D., {Chevalier}, R.~A., \& {Skrutskie}, M.~F. 2010, The Astronomer's Telegram, 2665, 1

\bibitem[{{Fox} {et~al.}(2013){Fox}, {Filippenko}, {Skrutskie}, {Silverman}, {Ganeshalingam}, {Cenko}, \& {Clubb}}]{2013AJ....146....2F}
{Fox}, O.~D., {Filippenko}, A.~V., {Skrutskie}, M.~F., {et~al.} 2013, \aj, 146, 2

\bibitem[{{Friedman} {et~al.}(2015){Friedman}, {Wood-Vasey}, {Marion}, {Challis}, {Mandel}, {Bloom}, {Modjaz}, {Narayan}, {Hicken}, {Foley}, {Klein}, {Starr}, {Morgan}, {Rest}, {Blake}, {Miller}, {Falco}, {Wyatt}, {Mink}, {Skrutskie}, \& {Kirshner}}]{2015ApJS..220....9F}
{Friedman}, A.~S., {Wood-Vasey}, W.~M., {Marion}, G.~H., {et~al.} 2015, \apjs, 220, 9

\bibitem[{{Frieman} {et~al.}(2006){Frieman}, {Prasad}, {Li}, {Itagaki}, {Nakano}, {Quimby}, {Mondol}, {Puckett}, {Pelloni}, \& {Winslow}}]{2006IAUC.8754....1F}
{Frieman}, J., {Prasad}, R.~R., {Li}, W., {et~al.} 2006, \iaucirc, 8754, 1

\bibitem[{{Frieman} \& {Sloan Digital Sky Survey Collaboration}(2006)}]{2006IAUC.8770....2F}
{Frieman}, J. \& {Sloan Digital Sky Survey Collaboration}, I. 2006, \iaucirc, 8770, 2

\bibitem[{{Frisch} {et~al.}(2002){Frisch}, {Matarrese}, {Mohayaee}, \& {Sobolevski}}]{Uriel_2002}
{Frisch}, U., {Matarrese}, S., {Mohayaee}, R., \& {Sobolevski}, A. 2002, \nat, 417, 260

\bibitem[{{Fukugita} {et~al.}(1996){Fukugita}, {Ichikawa}, {Gunn}, {Doi}, {Shimasaku}, \& {Schneider}}]{Fukugita_1996}
{Fukugita}, M., {Ichikawa}, T., {Gunn}, J.~E., {et~al.} 1996, \aj, 111, 1748

\bibitem[{{Gagliano} {et~al.}(2009){Gagliano}, {Puckett}, \& {Orff}}]{2009CBET.2005....1G}
{Gagliano}, R., {Puckett}, T., \& {Orff}, T. 2009, Central Bureau Electronic Telegrams, 2005, 1

\bibitem[{{Gal-Yam} {et~al.}(2008){Gal-Yam}, {Bufano}, {Barlow}, {Baron}, {Benetti}, {Cappellaro}, {Challis}, {Ellis}, {Filippenko}, {Foley}, {Fox}, {Hicken}, {Kirshner}, {Leonard}, {Li}, {Maoz}, {Matheson}, {Mazzali}, {Modjaz}, {Nomoto}, {Ofek}, {Simon}, {Small}, {Smith}, {Turatto}, {Van Dyk}, \& {Zampieri}}]{2008ApJ...685L.117G}
{Gal-Yam}, A., {Bufano}, F., {Barlow}, T.~A., {et~al.} 2008, \apjl, 685, L117

\bibitem[{{Galbany} {et~al.}(2016{\natexlab{a}}){Galbany}, {Hamuy}, {Phillips}, {Suntzeff}, {Maza}, {de Jaeger}, {Moraga}, {Gonz{\'a}lez-Gait{\'a}n}, {Krisciunas}, {Morrell}, {Thomas-Osip}, {Krzeminski}, {Gonz{\'a}lez}, {Antezana}, {Wishnjewski}, {McCarthy}, {Anderson}, {Guti{\'e}rrez}, {Stritzinger}, {Folatelli}, {Anguita}, {Galaz}, {Green}, {Impey}, {Kim}, {Kirhakos}, {Malkan}, {Mulchaey}, {Phillips}, {Pizzella}, {Prosser}, {Schmidt}, {Schommer}, {Sherry}, {Strolger}, {Wells}, \& {Williger}}]{2016AJ....151...33G}
{Galbany}, L., {Hamuy}, M., {Phillips}, M.~M., {et~al.} 2016{\natexlab{a}}, \aj, 151, 33

\bibitem[{{Galbany} {et~al.}(2016{\natexlab{b}}){Galbany}, {Moreno-Raya}, {Ruiz-Lapuente}, {Gonz{\'a}lez Hern{\'a}ndez}, {M{\'e}ndez}, {Vallely}, {Baron}, {Dom{\'\i}nguez}, {Hamuy}, {L{\'o}pez-S{\'a}nchez}, {Moll{\'a}}, {Catal{\'a}n}, {Cooke}, {Fari{\~n}a}, {G{\'e}nova-Santos}, {Karjalainen}, {Lietzen}, {McCormac}, {Riddick}, {Rubi{\~n}o-Mart{\'\i}n}, {Skillen}, {Tudor}, \& {Vaduvescu}}]{2015arXiv151006596G}
{Galbany}, L., {Moreno-Raya}, M.~E., {Ruiz-Lapuente}, P., {et~al.} 2016{\natexlab{b}}, \mnras, 457, 525

\bibitem[{{Gandhi} {et~al.}(2013){Gandhi}, {Yamanaka}, {Tanaka}, {Nozawa}, {Kawabata}, {Saviane}, {Maeda}, {Moriya}, {Hattori}, {Sasada}, \& {Itoh}}]{2013ApJ...767..166G}
{Gandhi}, P., {Yamanaka}, M., {Tanaka}, M., {et~al.} 2013, \apj, 767, 166

\bibitem[{{Ganeshalingam} {et~al.}(2010){Ganeshalingam}, {Li}, {Filippenko}, {Anderson}, {Foster}, {Gates}, {Griffith}, {Grigsby}, {Joubert}, {Leja}, {Lowe}, {Macomber}, {Pritchard}, {Thrasher}, \& {Winslow}}]{2010ApJS..190..418G}
{Ganeshalingam}, M., {Li}, W., {Filippenko}, A.~V., {et~al.} 2010, \apjs, 190, 418

\bibitem[{{G{\'o}mez} \& {L{\'o}pez}(1998)}]{1998AJ....115.1096G}
{G{\'o}mez}, G. \& {L{\'o}pez}, R. 1998, \aj, 115, 1096

\bibitem[{{G{\'o}mez} \& {L{\'o}pez}(2000)}]{2000AJ....120..367G}
{G{\'o}mez}, G. \& {L{\'o}pez}, R. 2000, \aj, 120, 367

\bibitem[{{Gonzalez}(2005)}]{2005IAUC.8470....3G}
{Gonzalez}, J.~J. 2005, \iaucirc, 8470, 3

\bibitem[{{Goobar} {et~al.}(2014){Goobar}, {Johansson}, {Amanullah}, {Cao}, {Perley}, {Kasliwal}, {Ferretti}, {Nugent}, {Harris}, {Gal-Yam}, {Ofek}, {Tendulkar}, {Dennefeld}, {Valenti}, {Arcavi}, {Banerjee}, {Venkataraman}, {Joshi}, {Ashok}, {Cenko}, {Diaz}, {Fremling}, {Horesh}, {Howell}, {Kulkarni}, {Papadogiannakis}, {Petrushevska}, {Sand}, {Sollerman}, {Stanishev}, {Bloom}, {Surace}, {Dupuy}, \& {Liu}}]{2014ApJ...784L..12G}
{Goobar}, A., {Johansson}, J., {Amanullah}, R., {et~al.} 2014, \apjl, 784, L12

\bibitem[{{Graham} {et~al.}(2017){Graham}, {Kumar}, {Hosseinzadeh}, {Hiramatsu}, {Arcavi}, {Howell}, {Valenti}, {Sand}, {Parrent}, {McCully}, \& {Filippenko}}]{2017MNRAS.472.3437G}
{Graham}, M.~L., {Kumar}, S., {Hosseinzadeh}, G., {et~al.} 2017, \mnras, 472, 3437

\bibitem[{{Graur} {et~al.}(2016){Graur}, {Zurek}, {Shara}, {Riess}, {Seitenzahl}, \& {Rest}}]{2016ApJ...819...31G}
{Graur}, O., {Zurek}, D., {Shara}, M.~M., {et~al.} 2016, \apj, 819, 31

\bibitem[{{Green}(2005)}]{2005IAUC.8604....3G}
{Green}, D.~W.~E. 2005, \iaucirc, 8604, 3

\bibitem[{{Guillochon} {et~al.}(2017){Guillochon}, {Parrent}, {Kelley}, \& {Margutti}}]{Guillochon_2017}
{Guillochon}, J., {Parrent}, J., {Kelley}, L.~Z., \& {Margutti}, R. 2017, \apj, 835, 64

\bibitem[{{Guti{\'e}rrez} {et~al.}(2017){Guti{\'e}rrez}, {Anderson}, {Hamuy}, {Morrell}, {Gonz{\'a}lez-Gaitan}, {Stritzinger}, {Phillips}, {Galbany}, {Folatelli}, {Dessart}, {Contreras}, {Della Valle}, {Freedman}, {Hsiao}, {Krisciunas}, {Madore}, {Maza}, {Suntzeff}, {Prieto}, {Gonz{\'a}lez}, {Cappellaro}, {Navarrete}, {Pizzella}, {Ruiz}, {Smith}, \& {Turatto}}]{2017ApJ...850...89G}
{Guti{\'e}rrez}, C.~P., {Anderson}, J.~P., {Hamuy}, M., {et~al.} 2017, \apj, 850, 89

\bibitem[{{Hamuy} {et~al.}(2001){Hamuy}, {Pinto}, {Maza}, {Suntzeff}, {Phillips}, {Eastman}, {Smith}, {Corbally}, {Burstein}, {Li}, {Ivanov}, {Moro-Martin}, {Strolger}, {de Souza}, {dos Anjos}, {Green}, {Pickering}, {Gonz{\'a}lez}, {Antezana}, {Wischnjewsky}, {Galaz}, {Roth}, {Persson}, \& {Schommer}}]{2001ApJ...558..615H}
{Hamuy}, M., {Pinto}, P.~A., {Maza}, J., {et~al.} 2001, \apj, 558, 615

\bibitem[{{Hanzl}(1998)}]{1998IAUC.6978....4H}
{Hanzl}, D. 1998, \iaucirc, 6978, 4

\bibitem[{{Harkness} {et~al.}(1987){Harkness}, {Wheeler}, {Margon}, {Downes}, {Kirshner}, {Uomoto}, {Barker}, {Cochran}, {Dinerstein}, {Garnett}, \& {Levreault}}]{Harkness1987}
{Harkness}, R.~P., {Wheeler}, J.~C., {Margon}, B., {et~al.} 1987, \apj, 317, 355

\bibitem[{{Harutyunyan} {et~al.}(2008){Harutyunyan}, {Pfahler}, {Pastorello}, {Taubenberger}, {Turatto}, {Cappellaro}, {Benetti}, {Elias-Rosa}, {Navasardyan}, {Valenti}, {Stanishev}, {Patat}, {Riello}, {Pignata}, \& {Hillebrandt}}]{2008A&A...488..383H}
{Harutyunyan}, A.~H., {Pfahler}, P., {Pastorello}, A., {et~al.} 2008, \aap, 488, 383

\bibitem[{{Hendry} {et~al.}(2006){Hendry}, {Smartt}, {Crockett}, {Maund}, {Gal-Yam}, {Moon}, {Cenko}, {Fox}, {Kudritzki}, {Benn}, \& {{\O}stensen}}]{2006MNRAS.369.1303H}
{Hendry}, M.~A., {Smartt}, S.~J., {Crockett}, R.~M., {et~al.} 2006, \mnras, 369, 1303

\bibitem[{{Hendry} {et~al.}(2005){Hendry}, {Smartt}, {Maund}, {Pastorello}, {Zampieri}, {Benetti}, {Turatto}, {Cappellaro}, {Meikle}, {Kotak}, {Irwin}, {Jonker}, {Vermaas}, {Peletier}, {van Woerden}, {Exter}, {Pollacco}, {Leon}, {Verley}, {Benn}, \& {Pignata}}]{2005MNRAS.359..906H}
{Hendry}, M.~A., {Smartt}, S.~J., {Maund}, J.~R., {et~al.} 2005, \mnras, 359, 906

\bibitem[{{Heraudeau} {et~al.}(1994){Heraudeau}, {Prugniel}, \& {Taupenas}}]{1994IAUC.5952....3H}
{Heraudeau}, P., {Prugniel}, P., \& {Taupenas}, J. 1994, \iaucirc, 5952, 3

\bibitem[{{Hicken} {et~al.}(2009){Hicken}, {Challis}, {Jha}, {Kirshner}, {Matheson}, {Modjaz}, {Rest}, {Wood-Vasey}, {Bakos}, {Barton}, {Berlind}, {Bragg}, {Brice{\~n}o}, {Brown}, {Caldwell}, {Calkins}, {Cho}, {Ciupik}, {Contreras}, {Dendy}, {Dosaj}, {Durham}, {Eriksen}, {Esquerdo}, {Everett}, {Falco}, {Fernandez}, {Gaba}, {Garnavich}, {Graves}, {Green}, {Groner}, {Hergenrother}, {Holman}, {Hradecky}, {Huchra}, {Hutchison}, {Jerius}, {Jordan}, {Kilgard}, {Krauss}, {Luhman}, {Macri}, {Marrone}, {McDowell}, {McIntosh}, {McNamara}, {Megeath}, {Mochejska}, {Munoz}, {Muzerolle}, {Naranjo}, {Narayan}, {Pahre}, {Peters}, {Peterson}, {Rines}, {Ripman}, {Roussanova}, {Schild}, {Sicilia-Aguilar}, {Sokoloski}, {Smalley}, {Smith}, {Spahr}, {Stanek}, {Barmby}, {Blondin}, {Stubbs}, {Szentgyorgyi}, {Torres}, {Vaz}, {Vikhlinin}, {Wang}, {Westover}, {Woods}, \& {Zhao}}]{2009ApJ...700..331H}
{Hicken}, M., {Challis}, P., {Jha}, S., {et~al.} 2009, \apj, 700, 331

\bibitem[{{Hicken} {et~al.}(2012){Hicken}, {Challis}, {Kirshner}, {Rest}, {Cramer}, {Wood-Vasey}, {Bakos}, {Berlind}, {Brown}, {Caldwell}, {Calkins}, {Currie}, {de Kleer}, {Esquerdo}, {Everett}, {Falco}, {Fernandez}, {Friedman}, {Groner}, {Hartman}, {Holman}, {Hutchins}, {Keys}, {Kipping}, {Latham}, {Marion}, {Narayan}, {Pahre}, {Pal}, {Peters}, {Perumpilly}, {Ripman}, {Sipocz}, {Szentgyorgyi}, {Tang}, {Torres}, {Vaz}, {Wolk}, \& {Zezas}}]{2012ApJS..200...12H}
{Hicken}, M., {Challis}, P., {Kirshner}, R.~P., {et~al.} 2012, \apjs, 200, 12

\bibitem[{{Hicken} {et~al.}(2017){Hicken}, {Friedman}, {Blondin}, {Challis}, {Berlind}, {Calkins}, {Esquerdo}, {Matheson}, {Modjaz}, {Rest}, \& {Kirshner}}]{2017ApJS..233....6H}
{Hicken}, M., {Friedman}, A.~S., {Blondin}, S., {et~al.} 2017, \apjs, 233, 6

\bibitem[{{Hicken} {et~al.}(2007){Hicken}, {Garnavich}, {Prieto}, {Blondin}, {DePoy}, {Kirshner}, \& {Parrent}}]{2007ApJ...669L..17H}
{Hicken}, M., {Garnavich}, P.~M., {Prieto}, J.~L., {et~al.} 2007, \apjl, 669, L17

\bibitem[{Hoaglin(2013)}]{Hoaglin2013Volume1H}
Hoaglin, D.~C. 2013, in Volume 16: How to Detect and Handle Outliers

\bibitem[{{Holtzman} {et~al.}(2008){Holtzman}, {Marriner}, {Kessler}, {Sako}, {Dilday}, {Frieman}, {Schneider}, {Bassett}, {Becker}, {Cinabro}, {DeJongh}, {Depoy}, {Doi}, {Garnavich}, {Hogan}, {Jha}, {Konishi}, {Lampeitl}, {Marshall}, {McGinnis}, {Miknaitis}, {Nichol}, {Prieto}, {Riess}, {Richmond}, {Romani}, {Smith}, {Takanashi}, {Tokita}, {van der Heyden}, {Yasuda}, \& {Zheng}}]{2008AJ....136.2306H}
{Holtzman}, J.~A., {Marriner}, J., {Kessler}, R., {et~al.} 2008, \aj, 136, 2306

\bibitem[{{Hoyle} \& {Fowler}(1960)}]{1960ApJ...132..565H}
{Hoyle}, F. \& {Fowler}, W.~A. 1960, \apj, 132, 565

\bibitem[{{Hsiao} {et~al.}(2007){Hsiao}, {Conley}, {Howell}, {Sullivan}, {Pritchet}, {Carlberg}, {Nugent}, \& {Phillips}}]{Hsiao2007}
{Hsiao}, E.~Y., {Conley}, A., {Howell}, D.~A., {et~al.} 2007, \apj, 663, 1187

\bibitem[{{Inserra} {et~al.}(2013){Inserra}, {Pastorello}, {Turatto}, {Pumo}, {Benetti}, {Cappellaro}, {Botticella}, {Bufano}, {Elias-Rosa}, {Harutyunyan}, {Taubenberger}, {Valenti}, \& {Zampieri}}]{2013A&A...555A.142I}
{Inserra}, C., {Pastorello}, A., {Turatto}, M., {et~al.} 2013, \aap, 555, A142

\bibitem[{{Inserra} {et~al.}(2011){Inserra}, {Turatto}, {Pastorello}, {Benetti}, {Cappellaro}, {Pumo}, {Zampieri}, {Agnoletto}, {Bufano}, {Botticella}, {Della Valle}, {Elias Rosa}, {Iijima}, {Spiro}, \& {Valenti}}]{2011MNRAS.417..261I}
{Inserra}, C., {Turatto}, M., {Pastorello}, A., {et~al.} 2011, \mnras, 417, 261

\bibitem[{{Ishida} \& {de Souza}(2013)}]{Ishida2013}
{Ishida}, E.~E.~O. \& {de Souza}, R.~S. 2013, \mnras, 430, 509

\bibitem[{{Jeffery} {et~al.}(1992){Jeffery}, {Leibundgut}, {Kirshner}, {Benetti}, {Branch}, \& {Sonneborn}}]{1992ApJ...397..304J}
{Jeffery}, D.~J., {Leibundgut}, B., {Kirshner}, R.~P., {et~al.} 1992, \apj, 397, 304

\bibitem[{{Jha} {et~al.}(2006){Jha}, {Kirshner}, {Challis}, {Garnavich}, {Matheson}, {Soderberg}, {Graves}, {Hicken}, {Alves}, {Arce}, {Balog}, {Barmby}, {Barton}, {Berlind}, {Bragg}, {Brice{\~n}o}, {Brown}, {Buckley}, {Caldwell}, {Calkins}, {Carter}, {Concannon}, {Donnelly}, {Eriksen}, {Fabricant}, {Falco}, {Fiore}, {Garcia}, {G{\'o}mez}, {Grogin}, {Groner}, {Groot}, {Haisch}, {Hartmann}, {Hergenrother}, {Holman}, {Huchra}, {Jayawardhana}, {Jerius}, {Kannappan}, {Kim}, {Kleyna}, {Kochanek}, {Koranyi}, {Krockenberger}, {Lada}, {Luhman}, {Luu}, {Macri}, {Mader}, {Mahdavi}, {Marengo}, {Marsden}, {McLeod}, {McNamara}, {Megeath}, {Moraru}, {Mossman}, {Muench}, {Mu{\~n}oz}, {Muzerolle}, {Naranjo}, {Nelson-Patel}, {Pahre}, {Patten}, {Peters}, {Peters}, {Raymond}, {Rines}, {Schild}, {Sobczak}, {Spahr}, {Stauffer}, {Stefanik}, {Szentgyorgyi}, {Tollestrup}, {V{\"a}is{\"a}nen}, {Vikhlinin}, {Wang}, {Willner}, {Wolk}, {Zajac}, {Zhao}, \& {Stanek}}]{2006AJ....131..527J}
{Jha}, S., {Kirshner}, R.~P., {Challis}, P., {et~al.} 2006, \aj, 131, 527

\bibitem[{{Johansson} {et~al.}(2017){Johansson}, {Goobar}, {Kasliwal}, {Helou}, {Masci}, {Tinyanont}, {Jencson}, {Cao}, {Fox}, {Kromer}, {Amanullah}, {Banerjee}, {Joshi}, {Jerkstrand}, {Kankare}, \& {Prince}}]{2017MNRAS.466.3442J}
{Johansson}, J., {Goobar}, A., {Kasliwal}, M.~M., {et~al.} 2017, \mnras, 466, 3442

\bibitem[{{Karpenka} {et~al.}(2013){Karpenka}, {Feroz}, \& {Hobson}}]{Karpenka2013}
{Karpenka}, N.~V., {Feroz}, F., \& {Hobson}, M.~P. 2013, \mnras, 429, 1278

\bibitem[{{Kessler} {et~al.}(2010){Kessler}, {Bassett}, {Belov}, {Bhatnagar}, {Campbell}, {Conley}, {Frieman}, {Glazov}, {Gonz{\'a}lez-Gait{\'a}n}, {Hlozek}, {Jha}, {Kuhlmann}, {Kunz}, {Lampeitl}, {Mahabal}, {Newling}, {Nichol}, {Parkinson}, {Sajeeth Philip}, {Poznanski}, {Richards}, {Rodney}, {Sako}, {Schneider}, {Smith}, {Stritzinger}, \& {Varughese}}]{Kessler2010}
{Kessler}, R., {Bassett}, B., {Belov}, P., {et~al.} 2010, \pasp, 122, 1415

\bibitem[{{Kessler} {et~al.}(2019){Kessler}, {Narayan}, {Avelino}, {Bachelet}, {Biswas}, {Brown}, {Chernoff}, {Connolly}, {Dai}, {Daniel}, {Di Stefano}, {Drout}, {Galbany}, {Gonz{\'a}lez-Gait{\'a}n}, {Graham}, {Hlo{\v{z}}ek}, {Ishida}, {Guillochon}, {Jha}, {Jones}, {Mandel}, {Muthukrishna}, {O'Grady}, {Peters}, {Pierel}, {Ponder}, {Pr{\v{s}}a}, {Rodney}, {Villar}, {LSST Dark Energy Science Collaboration}, \& {Transient and Variable Stars Science Collaboration}}]{Kessler2019Plastic}
{Kessler}, R., {Narayan}, G., {Avelino}, A., {et~al.} 2019, \pasp, 131, 094501

\bibitem[{Kolouri {et~al.}(2017)Kolouri, Park, Thorpe, Slepcev, \& Rohde}]{Kolouri_2017}
Kolouri, S., Park, S.~R., Thorpe, M., Slepcev, D., \& Rohde, G.~K. 2017, IEEE Signal Processing Magazine, 34, 43

\bibitem[{{Kotak} {et~al.}(2009){Kotak}, {Meikle}, {Farrah}, {Gerardy}, {Foley}, {Van Dyk}, {Fransson}, {Lundqvist}, {Sollerman}, {Fesen}, {Filippenko}, {Mattila}, {Silverman}, {Andersen}, {H{\"o}flich}, {Pozzo}, \& {Wheeler}}]{2009ApJ...704..306K}
{Kotak}, R., {Meikle}, W.~P.~S., {Farrah}, D., {et~al.} 2009, \apj, 704, 306

\bibitem[{{Kotak} {et~al.}(2005){Kotak}, {Meikle}, {Pignata}, {Stehle}, {Smartt}, {Benetti}, {Hillebrandt}, {Lennon}, {Mazzali}, {Patat}, \& {Turatto}}]{2005A&A...436.1021K}
{Kotak}, R., {Meikle}, W.~P.~S., {Pignata}, G., {et~al.} 2005, \aap, 436, 1021

\bibitem[{{Krisciunas} {et~al.}(2000){Krisciunas}, {Hastings}, {Loomis}, {McMillan}, {Rest}, {Riess}, \& {Stubbs}}]{2000ApJ...539..658K}
{Krisciunas}, K., {Hastings}, N.~C., {Loomis}, K., {et~al.} 2000, \apj, 539, 658

\bibitem[{{Krisciunas} {et~al.}(2001){Krisciunas}, {Phillips}, {Stubbs}, {Rest}, {Miknaitis}, {Riess}, {Suntzeff}, {Roth}, {Persson}, \& {Freedman}}]{2001AJ....122.1616K}
{Krisciunas}, K., {Phillips}, M.~M., {Stubbs}, C., {et~al.} 2001, \aj, 122, 1616

\bibitem[{{Krisciunas} {et~al.}(2003){Krisciunas}, {Suntzeff}, {Candia}, {Arenas}, {Espinoza}, {Gonzalez}, {Gonzalez}, {H{\"o}flich}, {Landolt}, {Phillips}, \& {Pizarro}}]{2003AJ....125..166K}
{Krisciunas}, K., {Suntzeff}, N.~B., {Candia}, P., {et~al.} 2003, \aj, 125, 166

\bibitem[{{Krisciunas} {et~al.}(2004){Krisciunas}, {Suntzeff}, {Phillips}, {Candia}, {Prieto}, {Antezana}, {Chassagne}, {Chen}, {Dickinson}, {Eisenhardt}, {Espinoza}, {Garnavich}, {Gonz{\'a}lez}, {Harrison}, {Hamuy}, {Ivanov}, {Krzemi{\'n}ski}, {Kulesa}, {McCarthy}, {Moro-Mart{\'\i}n}, {Muena}, {Noriega-Crespo}, {Persson}, {Pinto}, {Roth}, {Rubenstein}, {Stanford}, {Stringfellow}, {Zapata}, {Porter}, \& {Wischnjewsky}}]{2004AJ....128.3034K}
{Krisciunas}, K., {Suntzeff}, N.~B., {Phillips}, M.~M., {et~al.} 2004, \aj, 128, 3034

\bibitem[{{Lair} {et~al.}(2006){Lair}, {Leising}, {Milne}, \& {Williams}}]{2006AJ....132.2024L}
{Lair}, J.~C., {Leising}, M.~D., {Milne}, P.~A., \& {Williams}, G.~G. 2006, \aj, 132, 2024

\bibitem[{{Lee} {et~al.}(2005){Lee}, {Ponticello}, \& {Li}}]{2005IAUC.8632....2L}
{Lee}, E., {Ponticello}, N.~J., \& {Li}, W. 2005, \iaucirc, 8632, 2

\bibitem[{{Lee} {et~al.}(1999){Lee}, {Park}, {Lee}, {Lee}, {Park}, {Jeon}, \& {Bolte}}]{1999IAUC.7237....3L}
{Lee}, M.~G., {Park}, C., {Lee}, J., {et~al.} 1999, \iaucirc, 7237, 3

\bibitem[{{Leonard}(2007)}]{2007AIPC..937..311L}
{Leonard}, D.~C. 2007, in American Institute of Physics Conference Series, Vol. 937, Supernova 1987A: 20 Years After: Supernovae and Gamma-Ray Bursters, ed. S.~{Immler}, K.~{Weiler}, \& R.~{McCray} (AIP), 311--315

\bibitem[{{Leonard} {et~al.}(2006){Leonard}, {Filippenko}, {Ganeshalingam}, {Serduke}, {Li}, {Swift}, {Gal-Yam}, {Foley}, {Fox}, {Park}, {Hoffman}, \& {Wong}}]{2006Natur.440..505L}
{Leonard}, D.~C., {Filippenko}, A.~V., {Ganeshalingam}, M., {et~al.} 2006, \nat, 440, 505

\bibitem[{{Leonard} {et~al.}(2002){Leonard}, {Filippenko}, {Li}, {Matheson}, {Kirshner}, {Chornock}, {Van Dyk}, {Berlind}, {Calkins}, {Challis}, {Garnavich}, {Jha}, \& {Mahdavi}}]{2002AJ....124.2490L}
{Leonard}, D.~C., {Filippenko}, A.~V., {Li}, W., {et~al.} 2002, \aj, 124, 2490

\bibitem[{{Leonard} {et~al.}(2005){Leonard}, {Li}, {Filippenko}, {Foley}, \& {Chornock}}]{2005ApJ...632..450L}
{Leonard}, D.~C., {Li}, W., {Filippenko}, A.~V., {Foley}, R.~J., \& {Chornock}, R. 2005, \apj, 632, 450

\bibitem[{{Levy} {et~al.}(2021){Levy}, {Mohayaee}, \& {von Hausegger}}]{Levy_2021}
{Levy}, B., {Mohayaee}, R., \& {von Hausegger}, S. 2021, \mnras, 506, 1165

\bibitem[{{Li} {et~al.}(2005){Li}, {Van Dyk}, {Filippenko}, \& {Cuillandre}}]{2005PASP..117..121L}
{Li}, W., {Van Dyk}, S.~D., {Filippenko}, A.~V., \& {Cuillandre}, J.-C. 2005, \pasp, 117, 121

\bibitem[{{Lira} {et~al.}(1998){Lira}, {Suntzeff}, {Phillips}, {Hamuy}, {Maza}, {Schommer}, {Smith}, {Wells}, {Avil{\'e}s}, {Baldwin}, {Elias}, {Gonz{\'a}lez}, {Layden}, {Navarrete}, {Ugarte}, {Walker}, {Williger}, {Baganoff}, {Crotts}, {Rich}, {Tyson}, {Dey}, {Guhathakurta}, {Hibbard}, {Kim}, {Rehner}, {Siciliano}, {Roth}, {Seitzer}, \& {Williams}}]{1998AJ....115..234L}
{Lira}, P., {Suntzeff}, N.~B., {Phillips}, M.~M., {et~al.} 1998, \aj, 115, 234

\bibitem[{{Lochner} {et~al.}(2016){Lochner}, {McEwen}, {Peiris}, {Lahav}, \& {Winter}}]{Lochner2016}
{Lochner}, M., {McEwen}, J.~D., {Peiris}, H.~V., {Lahav}, O., \& {Winter}, M.~K. 2016, \apjs, 225, 31

\bibitem[{{L{\'o}pez} {et~al.}(1991){L{\'o}pez}, {Dominguez}, {G{\'\i}mez}, {Isern}, {Mampaso}, {Ruiz-Lapuente}, \& {S{\'a}nchez}}]{1991ESOC...37..721L}
{L{\'o}pez}, R., {Dominguez}, I., {G{\'\i}mez}, G., {et~al.} 1991, in European Southern Observatory Conference and Workshop Proceedings, Vol.~37, European Southern Observatory Conference and Workshop Proceedings, 721

\bibitem[{{LSST Science Collaboration} {et~al.}(2009){LSST Science Collaboration}, {Abell}, {Allison}, {Anderson}, {Andrew}, {Angel}, {Armus}, {Arnett}, {Asztalos}, {Axelrod}, {Bailey}, {Ballantyne}, {Bankert}, {Barkhouse}, {Barr}, {Barrientos}, {Barth}, {Bartlett}, {Becker}, {Becla}, {Beers}, {Bernstein}, {Biswas}, {Blanton}, {Bloom}, {Bochanski}, {Boeshaar}, {Borne}, {Bradac}, {Brandt}, {Bridge}, {Brown}, {Brunner}, {Bullock}, {Burgasser}, {Burge}, {Burke}, {Cargile}, {Chandrasekharan}, {Chartas}, {Chesley}, {Chu}, {Cinabro}, {Claire}, {Claver}, {Clowe}, {Connolly}, {Cook}, {Cooke}, {Cooray}, {Covey}, {Culliton}, {de Jong}, {de Vries}, {Debattista}, {Delgado}, {Dell'Antonio}, {Dhital}, {Di Stefano}, {Dickinson}, {Dilday}, {Djorgovski}, {Dobler}, {Donalek}, {Dubois-Felsmann}, {Durech}, {Eliasdottir}, {Eracleous}, {Eyer}, {Falco}, {Fan}, {Fassnacht}, {Ferguson}, {Fernandez}, {Fields}, {Finkbeiner}, {Figueroa}, {Fox}, {Francke}, {Frank}, {Frieman}, {Fromenteau}, {Furqan}, {Galaz}, {Gal-Yam}, {Garnavich},
  {Gawiser}, {Geary}, {Gee}, {Gibson}, {Gilmore}, {Grace}, {Green}, {Gressler}, {Grillmair}, {Habib}, {Haggerty}, {Hamuy}, {Harris}, {Hawley}, {Heavens}, {Hebb}, {Henry}, {Hileman}, {Hilton}, {Hoadley}, {Holberg}, {Holman}, {Howell}, {Infante}, {Ivezic}, {Jacoby}, {Jain}, {R}, {Jedicke}, {Jee}, {Garrett Jernigan}, {Jha}, {Johnston}, {Jones}, {Juric}, {Kaasalainen}, {Styliani}, {Kafka}, {Kahn}, {Kaib}, {Kalirai}, {Kantor}, {Kasliwal}, {Keeton}, {Kessler}, {Knezevic}, {Kowalski}, {Krabbendam}, {Krughoff}, {Kulkarni}, {Kuhlman}, {Lacy}, {Lepine}, {Liang}, {Lien}, {Lira}, {Long}, {Lorenz}, {Lotz}, {Lupton}, {Lutz}, {Macri}, {Mahabal}, {Mandelbaum}, {Marshall}, {May}, {McGehee}, {Meadows}, {Meert}, {Milani}, {Miller}, {Miller}, {Mills}, {Minniti}, {Monet}, {Mukadam}, {Nakar}, {Neill}, {Newman}, {Nikolaev}, {Nordby}, {O'Connor}, {Oguri}, {Oliver}, {Olivier}, {Olsen}, {Olsen}, {Olszewski}, {Oluseyi}, {Padilla}, {Parker}, {Pepper}, {Peterson}, {Petry}, {Pinto}, {Pizagno}, {Popescu}, {Prsa}, {Radcka}, {Raddick},
  {Rasmussen}, {Rau}, {Rho}, {Rhoads}, {Richards}, {Ridgway}, {Robertson}, {Roskar}, {Saha}, {Sarajedini}, {Scannapieco}, {Schalk}, {Schindler}, {Schmidt}, {Schmidt}, {Schneider}, {Schumacher}, {Scranton}, {Sebag}, {Seppala}, {Shemmer}, {Simon}, {Sivertz}, {Smith}, {Allyn Smith}, {Smith}, {Spitz}, {Stanford}, {Stassun}, {Strader}, {Strauss}, {Stubbs}, {Sweeney}, {Szalay}, {Szkody}, {Takada}, {Thorman}, {Trilling}, {Trimble}, {Tyson}, {Van Berg}, {Vanden Berk}, {VanderPlas}, {Verde}, {Vrsnak}, {Walkowicz}, {Wandelt}, {Wang}, {Wang}, {Warner}, {Wechsler}, {West}, {Wiecha}, {Williams}, {Willman}, {Wittman}, {Wolff}, {Wood-Vasey}, {Wozniak}, {Young}, {Zentner}, \& {Zhan}}]{LSST}
{LSST Science Collaboration}, {Abell}, P.~A., {Allison}, J., {et~al.} 2009, arXiv e-prints, arXiv:0912.0201

\bibitem[{{Lu} {et~al.}(2023){Lu}, {Hsiao}, {Phillips}, {Burns}, {Ashall}, {Morrell}, {Ng}, {Kumar}, {Shahbandeh}, {Hoeflich}, {Baron}, {Uddin}, {Stritzinger}, {Suntzeff}, {Baltay}, {Davis}, {Diamond}, {Folatelli}, {F{\"o}rster}, {Gagn{\'e}}, {Galbany}, {Gall}, {Gonz{\'a}lez-Gait{\'a}n}, {Holmbo}, {Kirshner}, {Krisciunas}, {Marion}, {Perlmutter}, {Pessi}, {Piro}, {Rabinowitz}, {Ryder}, \& {Sand}}]{Lu2023}
{Lu}, J., {Hsiao}, E.~Y., {Phillips}, M.~M., {et~al.} 2023, \apj, 948, 27

\bibitem[{{Madison} {et~al.}(2006){Madison}, {Ponticello}, {Li}, {Newton}, {Cox}, {Puckett}, {Gonzalez}, {Tokimasa}, {Naito}, \& {Yamaoka}}]{2006IAUC.8691....2M}
{Madison}, D.~R., {Ponticello}, N.~J., {Li}, W., {et~al.} 2006, \iaucirc, 8691, 2

\bibitem[{{Maguire} {et~al.}(2010){Maguire}, {Di Carlo}, {Smartt}, {Pastorello}, {Tsvetkov}, {Benetti}, {Spiro}, {Arkharov}, {Beccari}, {Botticella}, {Cappellaro}, {Cristallo}, {Dolci}, {Elias-Rosa}, {Fiaschi}, {Gorshanov}, {Harutyunyan}, {Larionov}, {Navasardyan}, {Pietrinferni}, {Raimondo}, {di Rico}, {Valenti}, {Valentini}, \& {Zampieri}}]{2010MNRAS.404..981M}
{Maguire}, K., {Di Carlo}, E., {Smartt}, S.~J., {et~al.} 2010, \mnras, 404, 981

\bibitem[{{Maguire} {et~al.}(2012){Maguire}, {Jerkstrand}, {Smartt}, {Fransson}, {Pastorello}, {Benetti}, {Valenti}, {Bufano}, \& {Leloudas}}]{2012MNRAS.420.3451M}
{Maguire}, K., {Jerkstrand}, A., {Smartt}, S.~J., {et~al.} 2012, \mnras, 420, 3451

\bibitem[{{Maguire} {et~al.}(2014){Maguire}, {Sullivan}, {Pan}, {Gal-Yam}, {Hook}, {Howell}, {Nugent}, {Mazzali}, {Chotard}, {Clubb}, {Filippenko}, {Kasliwal}, {Kandrashoff}, {Poznanski}, {Saunders}, {Silverman}, {Walker}, \& {Xu}}]{2014arXiv1408.1430M}
{Maguire}, K., {Sullivan}, M., {Pan}, Y.~C., {et~al.} 2014, \mnras, 444, 3258

\bibitem[{Marshall \& Severson(2022)}]{1995PAZh...21..678T}
Marshall, J. \& Severson, S. 2022, Journal of Undergraduate Research in Physics [\eprint[arXiv]{2212.06942}]

\bibitem[{{Matheson} {et~al.}(2008){Matheson}, {Kirshner}, {Challis}, {Jha}, {Garnavich}, {Berlind}, {Calkins}, {Blondin}, {Balog}, {Bragg}, {Caldwell}, {Dendy Concannon}, {Falco}, {Graves}, {Huchra}, {Kuraszkiewicz}, {Mader}, {Mahdavi}, {Phelps}, {Rines}, {Song}, \& {Wilkes}}]{2008AJ....135.1598M}
{Matheson}, T., {Kirshner}, R.~P., {Challis}, P., {et~al.} 2008, \aj, 135, 1598

\bibitem[{{Mattei} {et~al.}(1986){Mattei}, {Hurst}, {Lubbock}, {Ariail}, {Chaple}, {Scovil}, {Dyck}, {Dartmouth}, \& {Ducoty}}]{1986IAUC.4188....1M}
{Mattei}, J.~A., {Hurst}, G.~M., {Lubbock}, S., {et~al.} 1986, \iaucirc, 4188, 1

\bibitem[{{Mauerhan} {et~al.}(2017){Mauerhan}, {Van Dyk}, {Johansson}, {Hu}, {Fox}, {Wang}, {Graham}, {Filippenko}, \& {Shivvers}}]{2017ApJ...834..118M}
{Mauerhan}, J.~C., {Van Dyk}, S.~D., {Johansson}, J., {et~al.} 2017, \apj, 834, 118

\bibitem[{{Mazzali} {et~al.}(1993){Mazzali}, {Lucy}, {Danziger}, {Gouiffes}, {Cappellaro}, \& {Turatto}}]{1993A&A...269..423M}
{Mazzali}, P.~A., {Lucy}, L.~B., {Danziger}, I.~J., {et~al.} 1993, \aap, 269, 423

\bibitem[{{Mazzali} {et~al.}(2014){Mazzali}, {Sullivan}, {Hachinger}, {Ellis}, {Nugent}, {Howell}, {Gal-Yam}, {Maguire}, {Cooke}, {Thomas}, {Nomoto}, \& {Walker}}]{2014MNRAS.439.1959M}
{Mazzali}, P.~A., {Sullivan}, M., {Hachinger}, S., {et~al.} 2014, \mnras, 439, 1959

\bibitem[{{McClelland} {et~al.}(2013){McClelland}, {Garnavich}, {Milne}, {Shappee}, \& {Pogge}}]{2013ApJ...767..119M}
{McClelland}, C.~M., {Garnavich}, P.~M., {Milne}, P.~A., {Shappee}, B.~J., \& {Pogge}, R.~W. 2013, \apj, 767, 119

\bibitem[{{Millard} \& {Richardson}(2015)}]{Millard2015}
{Millard}, K. \& {Richardson}, M. 2015, Remote Sensing, 7, 8489

\bibitem[{{Minkowski}(1941)}]{minkowski}
{Minkowski}, R. 1941, \pasp, 53, 224

\bibitem[{{M{\"o}ller} \& {de Boissi{\`e}re}(2020)}]{Moller2020}
{M{\"o}ller}, A. \& {de Boissi{\`e}re}, T. 2020, \mnras, 491, 4277

\bibitem[{Monge(1784)}]{Monge1784}
Monge, G. 1784, Histoire de l’Académie Royale des Sciences (1781), 666--704

\bibitem[{{Mukai} \& {Ishida}(1999)}]{1999IAUC.7205....2M}
{Mukai}, K. \& {Ishida}, M. 1999, \iaucirc, 7205, 2

\bibitem[{{Munari} {et~al.}(2013){Munari}, {Henden}, {Belligoli}, {Castellani}, {Cherini}, {Righetti}, \& {Vagnozzi}}]{2013NewA...20...30M}
{Munari}, U., {Henden}, A., {Belligoli}, R., {et~al.} 2013, \na, 20, 30

\bibitem[{{Navasardyan} {et~al.}(2006){Navasardyan}, {Benetti}, {Harutyunyan}, {Bufano}, {Elias-Rosa}, {Zampieri}, {Turatto}, \& {Cappellaro}}]{2006IAUC.8667....2N}
{Navasardyan}, H., {Benetti}, S., {Harutyunyan}, A., {et~al.} 2006, \iaucirc, 8667, 2

\bibitem[{{Navasardyan} {et~al.}(2009){Navasardyan}, {Cappellaro}, \& {Benetti}}]{2009CBET.1918....2N}
{Navasardyan}, H., {Cappellaro}, E., \& {Benetti}, S. 2009, Central Bureau Electronic Telegrams, 1918, 2

\bibitem[{{Newton} {et~al.}(2009){Newton}, {Puckett}, \& {Orff}}]{2009CBET.1694....1N}
{Newton}, J., {Puckett}, T., \& {Orff}, T. 2009, Central Bureau Electronic Telegrams, 1694, 1

\bibitem[{{Nikakhtar} {et~al.}(2023){Nikakhtar}, {Padmanabhan}, {L{\'e}vy}, {Sheth}, \& {Mohayaee}}]{Nikakhtar_2023}
{Nikakhtar}, F., {Padmanabhan}, N., {L{\'e}vy}, B., {Sheth}, R.~K., \& {Mohayaee}, R. 2023, \prd, 108, 083534

\bibitem[{{Nikakhtar} {et~al.}(2022){Nikakhtar}, {Sheth}, {L{\'e}vy}, \& {Mohayaee}}]{Nikakhtar_2022}
{Nikakhtar}, F., {Sheth}, R.~K., {L{\'e}vy}, B., \& {Mohayaee}, R. 2022, \prl, 129, 251101

\bibitem[{{Nugent} {et~al.}(2002){Nugent}, {Kim}, \& {Perlmutter}}]{Nugent2002}
{Nugent}, P., {Kim}, A., \& {Perlmutter}, S. 2002, \pasp, 114, 803

\bibitem[{{Olivares} {et~al.}(2010){Olivares}, {Hamuy}, {Pignata}, {Maza}, {Bersten}, {Phillips}, {Suntzeff}, {Filippenko}, {Morrel}, {Kirshner}, \& {Matheson}}]{Olivares2010}
{Olivares}, F., {Hamuy}, M., {Pignata}, G., {et~al.} 2010, \apj, 715, 833

\bibitem[{{{\"O}stman} {et~al.}(2011){{\"O}stman}, {Nordin}, {Goobar}, {Amanullah}, {Smith}, {Sollerman}, {Stanishev}, {Stritzinger}, {Bassett}, {Davis}, {Edmondson}, {Frieman}, {Garnavich}, {Lampeitl}, {Leloudas}, {Marriner}, {Nichol}, {Romer}, {Sako}, {Schneider}, \& {Zheng}}]{2011A&A...526A..28O}
{{\"O}stman}, L., {Nordin}, J., {Goobar}, A., {et~al.} 2011, \aap, 526, A28

\bibitem[{{Pan} {et~al.}(2015){Pan}, {Foley}, {Kromer}, {Fox}, {Zheng}, {Challis}, {Clubb}, {Filippenko}, {Folatelli}, {Graham}, {Hillebrandt}, {Kirshner}, {Lee}, {Pakmor}, {Patat}, {Phillips}, {Pignata}, {R{\"o}pke}, {Seitenzahl}, {Silverman}, {Simon}, {Sternberg}, {Stritzinger}, {Taubenberger}, {Vinko}, \& {Wheeler}}]{2015arXiv150402396P}
{Pan}, Y.~C., {Foley}, R.~J., {Kromer}, M., {et~al.} 2015, \mnras, 452, 4307

\bibitem[{{Panagia} {et~al.}(2006){Panagia}, {Van Dyk}, {Weiler}, {Sramek}, {Stockdale}, \& {Murata}}]{2006ApJ...646..369P}
{Panagia}, N., {Van Dyk}, S.~D., {Weiler}, K.~W., {et~al.} 2006, \apj, 646, 369

\bibitem[{{Pastorello} {et~al.}(2007{\natexlab{a}}){Pastorello}, {Mazzali}, {Pignata}, {Benetti}, {Cappellaro}, {Filippenko}, {Li}, {Meikle}, {Arkharov}, {Blanc}, {Bufano}, {Derekas}, {Dolci}, {Elias-Rosa}, {Foley}, {Ganeshalingam}, {Harutyunyan}, {Kiss}, {Kotak}, {Larionov}, {Lucey}, {Napoleone}, {Navasardyan}, {Patat}, {Rich}, {Ryder}, {Salvo}, {Schmidt}, {Stanishev}, {Sz{\'e}kely}, {Taubenberger}, {Temporin}, {Turatto}, \& {Hillebrandt}}]{2007MNRAS.377.1531P}
{Pastorello}, A., {Mazzali}, P.~A., {Pignata}, G., {et~al.} 2007{\natexlab{a}}, \mnras, 377, 1531

\bibitem[{{Pastorello} {et~al.}(2006){Pastorello}, {Sauer}, {Taubenberger}, {Mazzali}, {Nomoto}, {Kawabata}, {Benetti}, {Elias-Rosa}, {Harutyunyan}, {Navasardyan}, {Zampieri}, {Iijima}, {Botticella}, {di Rico}, {Del Principe}, {Dolci}, {Gagliardi}, {Ragni}, \& {Valentini}}]{2006MNRAS.370.1752P}
{Pastorello}, A., {Sauer}, D., {Taubenberger}, S., {et~al.} 2006, \mnras, 370, 1752

\bibitem[{{Pastorello} {et~al.}(2007{\natexlab{b}}){Pastorello}, {Taubenberger}, {Elias-Rosa}, {Mazzali}, {Pignata}, {Cappellaro}, {Garavini}, {Nobili}, {Anupama}, {Bayliss}, {Benetti}, {Bufano}, {Chakradhari}, {Kotak}, {Goobar}, {Navasardyan}, {Patat}, {Sahu}, {Salvo}, {Schmidt}, {Stanishev}, {Turatto}, \& {Hillebrandt}}]{2007MNRAS.376.1301P}
{Pastorello}, A., {Taubenberger}, S., {Elias-Rosa}, N., {et~al.} 2007{\natexlab{b}}, \mnras, 376, 1301

\bibitem[{{Pastorello} {et~al.}(2009){Pastorello}, {Valenti}, {Zampieri}, {Navasardyan}, {Taubenberger}, {Smartt}, {Arkharov}, {B{\"a}rnbantner}, {Barwig}, {Benetti}, {Birtwhistle}, {Botticella}, {Cappellaro}, {Del Principe}, {di Mille}, {di Rico}, {Dolci}, {Elias-Rosa}, {Efimova}, {Fiedler}, {Harutyunyan}, {H{\"o}flich}, {Kloehr}, {Larionov}, {Lorenzi}, {Maund}, {Napoleone}, {Ragni}, {Richmond}, {Ries}, {Spiro}, {Temporin}, {Turatto}, \& {Wheeler}}]{2009MNRAS.394.2266P}
{Pastorello}, A., {Valenti}, S., {Zampieri}, L., {et~al.} 2009, \mnras, 394, 2266

\bibitem[{{Pereira} {et~al.}(2013){Pereira}, {Thomas}, {Aldering}, {Antilogus}, {Baltay}, {Benitez-Herrera}, {Bongard}, {Buton}, {Canto}, {Cellier-Holzem}, {Chen}, {Childress}, {Chotard}, {Copin}, {Fakhouri}, {Fink}, {Fouchez}, {Gangler}, {Guy}, {Hillebrandt}, {Hsiao}, {Kerschhaggl}, {Kowalski}, {Kromer}, {Nordin}, {Nugent}, {Paech}, {Pain}, {P{\'e}contal}, {Perlmutter}, {Rabinowitz}, {Rigault}, {Runge}, {Saunders}, {Smadja}, {Tao}, {Taubenberger}, {Tilquin}, \& {Wu}}]{2013A&A...554A..27P}
{Pereira}, R., {Thomas}, R.~C., {Aldering}, G., {et~al.} 2013, \aap, 554, A27

\bibitem[{{Perlmutter} {et~al.}(1999){Perlmutter}, {Aldering}, {Goldhaber}, {Knop}, {Nugent}, {Castro}, {Deustua}, {Fabbro}, {Goobar}, {Groom}, {Hook}, {Kim}, {Kim}, {Lee}, {Nunes}, {Pain}, {Pennypacker}, {Quimby}, {Lidman}, {Ellis}, {Irwin}, {McMahon}, {Ruiz-Lapuente}, {Walton}, {Schaefer}, {Boyle}, {Filippenko}, {Matheson}, {Fruchter}, {Panagia}, {Newberg}, {Couch}, \& {Project}}]{Perlmutter1999}
{Perlmutter}, S., {Aldering}, G., {Goldhaber}, G., {et~al.} 1999, \apj, 517, 565

\bibitem[{Peyré \& Cuturi(2020)}]{Peyre2020}
Peyré, G. \& Cuturi, M. 2020, Computational Optimal Transport

\bibitem[{{Pignata} {et~al.}(2008{\natexlab{a}}){Pignata}, {Benetti}, {Mazzali}, {Kotak}, {Patat}, {Meikle}, {Stehle}, {Leibundgut}, {Suntzeff}, {Buson}, {Cappellaro}, {Clocchiatti}, {Hamuy}, {Maza}, {Mendez}, {Ruiz-Lapuente}, {Salvo}, {Schmidt}, {Turatto}, \& {Hillebrandt}}]{2008MNRAS.388..971P}
{Pignata}, G., {Benetti}, S., {Mazzali}, P.~A., {et~al.} 2008{\natexlab{a}}, \mnras, 388, 971

\bibitem[{{Pignata} {et~al.}(2008{\natexlab{b}}){Pignata}, {Maza}, {Hamuy}, {Antezana}, {Gonzalez}, {Gonzalez}, {Lopez}, {Silva}, {Folatelli}, {Iturra}, {Cartier}, {Forster}, {Reichart}, {Ivarsen}, {Crain}, {Foster}, {Nysewander}, \& {Lacluyze}}]{2008CBET.1506....1P}
{Pignata}, G., {Maza}, J., {Hamuy}, M., {et~al.} 2008{\natexlab{b}}, Central Bureau Electronic Telegrams, 1506, 1

\bibitem[{{Pignata} {et~al.}(2004){Pignata}, {Patat}, {Benetti}, {Blinnikov}, {Hillebrandt}, {Kotak}, {Leibundgut}, {Mazzali}, {Meikle}, {Qiu}, {Ruiz-Lapuente}, {Smartt}, {Sorokina}, {Stritzinger}, {Stehle}, {Turatto}, {Marsh}, {Martin-Luis}, {McBride}, {Mendez}, {Morales-Rueda}, {Narbutis}, \& {Street}}]{2004MNRAS.355..178P}
{Pignata}, G., {Patat}, F., {Benetti}, S., {et~al.} 2004, \mnras, 355, 178

\bibitem[{{Ponticello} {et~al.}(2005){Ponticello}, {Lee}, \& {Li}}]{2005IAUC.8608....2P}
{Ponticello}, N., {Lee}, E., \& {Li}, W. 2005, \iaucirc, 8608, 2

\bibitem[{{Prasad} \& {Li}(2007)}]{2007CBET..914....1P}
{Prasad}, R. \& {Li}, W. 2007, Central Bureau Electronic Telegrams, 914, 1

\bibitem[{{Puckett} {et~al.}(2007{\natexlab{a}}){Puckett}, {Crowley}, \& {Orff}}]{2007CBET..966....1P}
{Puckett}, T., {Crowley}, T., \& {Orff}, T. 2007{\natexlab{a}}, Central Bureau Electronic Telegrams, 966, 1

\bibitem[{{Puckett} {et~al.}(2008){Puckett}, {Gagliano}, \& {Orff}}]{2008CBET.1243....2P}
{Puckett}, T., {Gagliano}, R., \& {Orff}, T. 2008, Central Bureau Electronic Telegrams, 1243, 2

\bibitem[{{Puckett} {et~al.}(2007{\natexlab{b}}){Puckett}, {Gagliano}, \& {Sehgal}}]{2007CBET..803....1P}
{Puckett}, T., {Gagliano}, R., \& {Sehgal}, A. 2007{\natexlab{b}}, Central Bureau Electronic Telegrams, 803, 1

\bibitem[{{Puckett} {et~al.}(2006){Puckett}, {Langoussis}, {Newton}, {Briggs}, {Colesanti}, {Jacques}, {Pimentel}, {Napoleao}, {Khandrika}, {Li}, {Boles}, \& {Sehgal}}]{2006IAUC.8716....1P}
{Puckett}, T., {Langoussis}, A., {Newton}, J., {et~al.} 2006, \iaucirc, 8716, 1

\bibitem[{{Rajala} {et~al.}(2004){Rajala}, {Fox}, \& {Gal-Yam}}]{2004IAUC.8386....4R}
{Rajala}, A., {Fox}, D.~B., \& {Gal-Yam}, A. 2004, \iaucirc, 8386, 4

\bibitem[{{Rawson} \& {Hultgren}(2022)}]{Rawson_2022}
{Rawson}, M. \& {Hultgren}, J. 2022, Proceedings of the 30th European Signal Processing Conference, EUSIPCO, 2022., arXiv:2202.05354

\bibitem[{{Rex} {et~al.}(2008){Rex}, {Li}, \& {Filippenko}}]{2008CBET.1437....1R}
{Rex}, J., {Li}, W., \& {Filippenko}, A.~V. 2008, Central Bureau Electronic Telegrams, 1437, 1

\bibitem[{{Richards} {et~al.}(2012){Richards}, {Homrighausen}, {Freeman}, {Schafer}, \& {Poznanski}}]{Richards2012}
{Richards}, J.~W., {Homrighausen}, D., {Freeman}, P.~E., {Schafer}, C.~M., \& {Poznanski}, D. 2012, \mnras, 419, 1121

\bibitem[{{Richardson} {et~al.}(2001){Richardson}, {Thomas}, {Casebeer}, {Blankenship}, {Ratowt}, {Baron}, \& {Branch}}]{2001AAS...199.8408R}
{Richardson}, D., {Thomas}, R.~C., {Casebeer}, D., {et~al.} 2001, in American Astronomical Society Meeting Abstracts, Vol. 199, American Astronomical Society Meeting Abstracts, 84.08

\bibitem[{{Richmond} {et~al.}(1995){Richmond}, {Treffers}, {Filippenko}, {van Dyk}, {Paik}, {Peng}, {Marschall}, {Laaksonen}, {Macintosh}, \& {McLean}}]{1995AJ....109.2121R}
{Richmond}, M.~W., {Treffers}, R.~R., {Filippenko}, A.~V., {et~al.} 1995, \aj, 109, 2121

\bibitem[{{Riess} {et~al.}(1998){Riess}, {Filippenko}, {Challis}, {Clocchiatti}, {Diercks}, {Garnavich}, {Gilliland}, {Hogan}, {Jha}, {Kirshner}, {Leibundgut}, {Phillips}, {Reiss}, {Schmidt}, {Schommer}, {Smith}, {Spyromilio}, {Stubbs}, {Suntzeff}, \& {Tonry}}]{Riess1998}
{Riess}, A.~G., {Filippenko}, A.~V., {Challis}, P., {et~al.} 1998, \aj, 116, 1009

\bibitem[{{Riess} {et~al.}(1999){Riess}, {Kirshner}, {Schmidt}, {Jha}, {Challis}, {Garnavich}, {Esin}, {Carpenter}, {Grashius}, {Schild}, {Berlind}, {Huchra}, {Prosser}, {Falco}, {Benson}, {Brice{\~n}o}, {Brown}, {Caldwell}, {dell'Antonio}, {Filippenko}, {Goodman}, {Grogin}, {Groner}, {Hughes}, {Green}, {Jansen}, {Kleyna}, {Luu}, {Macri}, {McLeod}, {McLeod}, {McNamara}, {McLean}, {Milone}, {Mohr}, {Moraru}, {Peng}, {Peters}, {Prestwich}, {Stanek}, {Szentgyorgyi}, \& {Zhao}}]{1999AJ....117..707R}
{Riess}, A.~G., {Kirshner}, R.~P., {Schmidt}, B.~P., {et~al.} 1999, \aj, 117, 707

\bibitem[{{Riess} {et~al.}(2005){Riess}, {Li}, {Stetson}, {Filippenko}, {Jha}, {Kirshner}, {Challis}, {Garnavich}, \& {Chornock}}]{2005ApJ...627..579R}
{Riess}, A.~G., {Li}, W., {Stetson}, P.~B., {et~al.} 2005, \apj, 627, 579

\bibitem[{{Rodr{\'\i}guez} {et~al.}(2019){Rodr{\'\i}guez}, {Pignata}, {Hamuy}, {Clocchiatti}, {Phillips}, {Krisciunas}, {Morrell}, {Folatelli}, {Roth}, {Castell{\'o}n}, {Jang}, {Apostolovski}, {L{\'o}pez}, {Marchi}, {Ram{\'\i}rez}, \& {S{\'a}nchez}}]{Osmar_2019}
{Rodr{\'\i}guez}, {\'O}., {Pignata}, G., {Hamuy}, M., {et~al.} 2019, \mnras, 483, 5459

\bibitem[{{Rostopchin} \& {Westfall}(2008)}]{2008CBET.1273....2R}
{Rostopchin}, S. \& {Westfall}, A. 2008, Central Bureau Electronic Telegrams, 1273, 2

\bibitem[{{Rubin} {et~al.}(2016){Rubin}, {Gal-Yam}, {De Cia}, {Horesh}, {Khazov}, {Ofek}, {Kulkarni}, {Arcavi}, {Manulis}, {Yaron}, {Vreeswijk}, {Kasliwal}, {Ben-Ami}, {Perley}, {Cao}, {Cenko}, {Rebbapragada}, {Wo{\'z}niak}, {Filippenko}, {Clubb}, {Nugent}, {Pan}, {Badenes}, {Howell}, {Valenti}, {Sand}, {Sollerman}, {Johansson}, {Leonard}, {Horst}, {Armen}, {Fedrow}, {Quimby}, {Mazzali}, {Pian}, {Sternberg}, {Matheson}, {Sullivan}, {Maguire}, \& {Lazarevic}}]{2016ApJ...820...33R}
{Rubin}, A., {Gal-Yam}, A., {De Cia}, A., {et~al.} 2016, \apj, 820, 33

\bibitem[{{Russell} {et~al.}(2003){Russell}, {Lynch}, {Kim}, {Hammel}, \& {Perry}}]{2003IAUC.8186....3R}
{Russell}, R.~W., {Lynch}, D.~K., {Kim}, D.~L., {Hammel}, H.~B., \& {Perry}, R.~B. 2003, \iaucirc, 8186, 3

\bibitem[{{Sahu} {et~al.}(2006){Sahu}, {Anupama}, {Srividya}, \& {Muneer}}]{2006MNRAS.372.1315S}
{Sahu}, D.~K., {Anupama}, G.~C., {Srividya}, S., \& {Muneer}, S. 2006, \mnras, 372, 1315

\bibitem[{{Sahu} {et~al.}(2008){Sahu}, {Tanaka}, {Anupama}, {Kawabata}, {Maeda}, {Tominaga}, {Nomoto}, {Mazzali}, \& {Prabhu}}]{2008ApJ...680..580S}
{Sahu}, D.~K., {Tanaka}, M., {Anupama}, G.~C., {et~al.} 2008, \apj, 680, 580

\bibitem[{{Sako} {et~al.}(2018){Sako}, {Bassett}, {Becker}, {Brown}, {Campbell}, {Wolf}, {Cinabro}, {D'Andrea}, {Dawson}, {DeJongh}, {Depoy}, {Dilday}, {Doi}, {Filippenko}, {Fischer}, {Foley}, {Frieman}, {Galbany}, {Garnavich}, {Goobar}, {Gupta}, {Hill}, {Hayden}, {Hlozek}, {Holtzman}, {Hopp}, {Jha}, {Kessler}, {Kollatschny}, {Leloudas}, {Marriner}, {Marshall}, {Miquel}, {Morokuma}, {Mosher}, {Nichol}, {Nordin}, {Olmstead}, {{\"O}stman}, {Prieto}, {Richmond}, {Romani}, {Sollerman}, {Stritzinger}, {Schneider}, {Smith}, {Wheeler}, {Yasuda}, \& {Zheng}}]{2014arXiv1401.3317S}
{Sako}, M., {Bassett}, B., {Becker}, A.~C., {et~al.} 2018, \pasp, 130, 064002

\bibitem[{{Sauer} {et~al.}(2008){Sauer}, {Mazzali}, {Blondin}, {Stehle}, {Benetti}, {Challis}, {Filippenko}, {Kirshner}, {Li}, \& {Matheson}}]{2008MNRAS.391.1605S}
{Sauer}, D.~N., {Mazzali}, P.~A., {Blondin}, S., {et~al.} 2008, \mnras, 391, 1605

\bibitem[{{Schmidt} {et~al.}(1993){Schmidt}, {Kirshner}, {Schild}, {Leibundgut}, {Jeffery}, {Willner}, {Peletier}, {Zabludoff}, {Phillips}, {Suntzeff}, {Hamuy}, {Wells}, {Smith}, {Baldwin}, {Weller}, {Navarette}, {Gonzalez}, {Filippenko}, {Shields}, {Steidel}, {Perlmutter}, {Pennypacker}, {Smith}, {Porter}, {Boroson}, {Stathakis}, {Cannon}, {Peters}, {Horine}, {Freeman}, {Womble}, {Stone}, {Marschall}, {Phillips}, {Saha}, \& {Bond}}]{1993AJ....105.2236S}
{Schmidt}, B.~P., {Kirshner}, R.~P., {Schild}, R., {et~al.} 1993, \aj, 105, 2236

\bibitem[{{Silverman} {et~al.}(2012{\natexlab{a}}){Silverman}, {Foley}, {Filippenko}, {Ganeshalingam}, {Barth}, {Chornock}, {Griffith}, {Kong}, {Lee}, {Leonard}, {Matheson}, {Miller}, {Steele}, {Barris}, {Bloom}, {Cobb}, {Coil}, {Desroches}, {Gates}, {Ho}, {Jha}, {Kandrashoff}, {Li}, {Mandel}, {Modjaz}, {Moore}, {Mostardi}, {Papenkova}, {Park}, {Perley}, {Poznanski}, {Reuter}, {Scala}, {Serduke}, {Shields}, {Swift}, {Tonry}, {Van Dyk}, {Wang}, \& {Wong}}]{2012MNRAS.425.1789S}
{Silverman}, J.~M., {Foley}, R.~J., {Filippenko}, A.~V., {et~al.} 2012{\natexlab{a}}, \mnras, 425, 1789

\bibitem[{{Silverman} {et~al.}(2012{\natexlab{b}}){Silverman}, {Ganeshalingam}, {Cenko}, {Filippenko}, {Li}, {Barth}, {Carson}, {Childress}, {Clubb}, {Cucchiara}, {Graham}, {Marion}, {Nguyen}, {Pei}, {Tucker}, {Vinko}, {Wheeler}, \& {Worseck}}]{2012ApJ...756L...7S}
{Silverman}, J.~M., {Ganeshalingam}, M., {Cenko}, S.~B., {et~al.} 2012{\natexlab{b}}, \apjl, 756, L7

\bibitem[{{Silverman} {et~al.}(2013{\natexlab{a}}){Silverman}, {Ganeshalingam}, \& {Filippenko}}]{2013MNRAS.430.1030S}
{Silverman}, J.~M., {Ganeshalingam}, M., \& {Filippenko}, A.~V. 2013{\natexlab{a}}, \mnras, 430, 1030

\bibitem[{{Silverman} {et~al.}(2009){Silverman}, {Kandrashoff}, \& {Filippenko}}]{2009CBET.1969....2S}
{Silverman}, J.~M., {Kandrashoff}, M.~T., \& {Filippenko}, A.~V. 2009, Central Bureau Electronic Telegrams, 1969, 2

\bibitem[{{Silverman} {et~al.}(2013{\natexlab{b}}){Silverman}, {Nugent}, {Gal-Yam}, {Sullivan}, {Howell}, {Filippenko}, {Arcavi}, {Ben-Ami}, {Bloom}, {Cenko}, {Cao}, {Chornock}, {Clubb}, {Coil}, {Foley}, {Graham}, {Griffith}, {Horesh}, {Kasliwal}, {Kulkarni}, {Leonard}, {Li}, {Matheson}, {Miller}, {Modjaz}, {Ofek}, {Pan}, {Perley}, {Poznanski}, {Quimby}, {Steele}, {Sternberg}, {Xu}, \& {Yaron}}]{2013ApJS..207....3S}
{Silverman}, J.~M., {Nugent}, P.~E., {Gal-Yam}, A., {et~al.} 2013{\natexlab{b}}, \apjs, 207, 3

\bibitem[{{Silverman} {et~al.}(2017){Silverman}, {Pickett}, {Wheeler}, {Filippenko}, {Vink{\'o}}, {Marion}, {Cenko}, {Chornock}, {Clubb}, {Foley}, {Graham}, {Kelly}, {Matheson}, \& {Shields}}]{2017MNRAS.467..369S}
{Silverman}, J.~M., {Pickett}, S., {Wheeler}, J.~C., {et~al.} 2017, \mnras, 467, 369

\bibitem[{{Simon} {et~al.}(2009){Simon}, {Gal-Yam}, {Gnat}, {Quimby}, {Ganeshalingam}, {Silverman}, {Blondin}, {Li}, {Filippenko}, {Wheeler}, {Kirshner}, {Patat}, {Nugent}, {Foley}, {Vogt}, {Butler}, {Peek}, {Rosolowsky}, {Herczeg}, {Sauer}, \& {Mazzali}}]{2009ApJ...702.1157S}
{Simon}, J.~D., {Gal-Yam}, A., {Gnat}, O., {et~al.} 2009, \apj, 702, 1157

\bibitem[{{Simon} {et~al.}(2007){Simon}, {Gal-Yam}, {Penprase}, {Li}, {Quimby}, {Silverman}, {Allende Prieto}, {Wheeler}, {Filippenko}, {Martinez}, {Beeler}, \& {Patat}}]{2007ApJ...671L..25S}
{Simon}, J.~D., {Gal-Yam}, A., {Penprase}, B.~E., {et~al.} 2007, \apjl, 671, L25

\bibitem[{{Smartt}(2009)}]{2009ARA&A..47...63S}
{Smartt}, S.~J. 2009, \araa, 47, 63

\bibitem[{{Smartt} {et~al.}(2015){Smartt}, {Valenti}, {Fraser}, {Inserra}, {Young}, {Sullivan}, {Pastorello}, {Benetti}, {Gal-Yam}, {Knapic}, {Molinaro}, {Smareglia}, {Smith}, {Taubenberger}, {Yaron}, {Anderson}, {Ashall}, {Balland}, {Baltay}, {Barbarino}, {Bauer}, {Baumont}, {Bersier}, {Blagorodnova}, {Bongard}, {Botticella}, {Bufano}, {Bulla}, {Cappellaro}, {Campbell}, {Cellier-Holzem}, {Chen}, {Childress}, {Clocchiatti}, {Contreras}, {Dall'Ora}, {Danziger}, {de Jaeger}, {De Cia}, {Della Valle}, {Dennefeld}, {Elias-Rosa}, {Elman}, {Feindt}, {Fleury}, {Gall}, {Gonzalez-Gaitan}, {Galbany}, {Morales Garoffolo}, {Greggio}, {Guillou}, {Hachinger}, {Hadjiyska}, {Hage}, {Hillebrandt}, {Hodgkin}, {Hsiao}, {James}, {Jerkstrand}, {Kangas}, {Kankare}, {Kotak}, {Kromer}, {Kuncarayakti}, {Leloudas}, {Lundqvist}, {Lyman}, {Hook}, {Maguire}, {Manulis}, {Margheim}, {Mattila}, {Maund}, {Mazzali}, {McCrum}, {McKinnon}, {Moreno-Raya}, {Nicholl}, {Nugent}, {Pain}, {Pignata}, {Phillips}, {Polshaw}, {Pumo}, {Rabinowitz},
  {Reilly}, {Romero-Ca{\~n}izales}, {Scalzo}, {Schmidt}, {Schulze}, {Sim}, {Sollerman}, {Taddia}, {Tartaglia}, {Terreran}, {Tomasella}, {Turatto}, {Walker}, {Walton}, {Wyrzykowski}, {Yuan}, \& {Zampieri}}]{2015A&A...579A..40S}
{Smartt}, S.~J., {Valenti}, S., {Fraser}, M., {et~al.} 2015, \aap, 579, A40

\bibitem[{{Smitka} {et~al.}(2015){Smitka}, {Brown}, {Suntzeff}, {Zhang}, {Zhai}, {Wang}, {Mo}, \& {Zhang}}]{2015ApJ...813...30S}
{Smitka}, M.~T., {Brown}, P.~J., {Suntzeff}, N.~B., {et~al.} 2015, \apj, 813, 30

\bibitem[{{Srivastav} {et~al.}(2017){Srivastav}, {Anupama}, {Sahu}, \& {Ravikumar}}]{2017MNRAS.466.2436S}
{Srivastav}, S., {Anupama}, G.~C., {Sahu}, D.~K., \& {Ravikumar}, C.~D. 2017, \mnras, 466, 2436

\bibitem[{{Stahl} {et~al.}(2020){Stahl}, {Zheng}, {de Jaeger}, {Brink}, {Filippenko}, {Silverman}, {Cenko}, {Clubb}, {Graham}, {Halevi}, {Kelly}, {Kleiser}, {Shivvers}, {Yuk}, {Cobb}, {Fox}, {Kandrashoff}, {Kong}, {Mauerhan}, {Wang}, \& {Wang}}]{2020MNRAS.492.4325S}
{Stahl}, B.~E., {Zheng}, W., {de Jaeger}, T., {et~al.} 2020, \mnras, 492, 4325

\bibitem[{{Stahl} {et~al.}(2019){Stahl}, {Zheng}, {de Jaeger}, {Filippenko}, {Bigley}, {Blanchard}, {Blanchard}, {Brink}, {Cargill}, {Casper}, {Channa}, {Choi}, {Choksi}, {Chu}, {Clubb}, {Cohen}, {Ellison}, {Falcon}, {Fazeli}, {Fuller}, {Ganeshalingam}, {Gates}, {Gould}, {Halevi}, {Hayakawa}, {Hestenes}, {Jeffers}, {Joubert}, {Kandrashoff}, {Kim}, {Kim}, {Kislak}, {Kleiser}, {Kong}, {de Kouchkovsky}, {Krishnan}, {Kumar}, {Leja}, {Leonard}, {Li}, {Li}, {Lu}, {Mason}, {Molloy}, {Pina}, {Rex}, {Ross}, {Stegman}, {Tang}, {Thrasher}, {Wang}, {Wilkins}, {Yuk}, {Yunus}, \& {Zhang}}]{2019MNRAS.490.3882S}
{Stahl}, B.~E., {Zheng}, W., {de Jaeger}, T., {et~al.} 2019, \mnras, 490, 3882

\bibitem[{{Stanishev} {et~al.}(2007){Stanishev}, {Goobar}, {Benetti}, {Kotak}, {Pignata}, {Navasardyan}, {Mazzali}, {Amanullah}, {Garavini}, {Nobili}, {Qiu}, {Elias-Rosa}, {Ruiz-Lapuente}, {Mendez}, {Meikle}, {Patat}, {Pastorello}, {Altavilla}, {Gustafsson}, {Harutyunyan}, {Iijima}, {Jakobsson}, {Kichizhieva}, {Lundqvist}, {Mattila}, {Melinder}, {Pavlenko}, {Pavlyuk}, {Sollerman}, {Tsvetkov}, {Turatto}, \& {Hillebrandt}}]{2007A&A...469..645S}
{Stanishev}, V., {Goobar}, A., {Benetti}, S., {et~al.} 2007, \aap, 469, 645

\bibitem[{{Stritzinger} {et~al.}(2005){Stritzinger}, {Suntzeff}, {Hamuy}, {Challis}, {Demarco}, {Germany}, \& {Soderberg}}]{Stritzinger_2005}
{Stritzinger}, M., {Suntzeff}, N.~B., {Hamuy}, M., {et~al.} 2005, \pasp, 117, 810

\bibitem[{{Strolger} {et~al.}(2003){Strolger}, {GOODS Treasury Team}, \& {Hubble Higher-Z Supernova Team}}]{2003IAUC.8150....4S}
{Strolger}, L.~G., {GOODS Treasury Team}, \& {Hubble Higher-Z Supernova Team}. 2003, \iaucirc, 8150, 4

\bibitem[{{Sugerman} {et~al.}(2006){Sugerman}, {Ercolano}, {Barlow}, {Tielens}, {Clayton}, {Zijlstra}, {Meixner}, {Speck}, {Gledhill}, {Panagia}, {Cohen}, {Gordon}, {Meyer}, {Fabbri}, {Bowey}, {Welch}, {Regan}, \& {Kennicutt}}]{2006Sci...313..196S}
{Sugerman}, B. E.~K., {Ercolano}, B., {Barlow}, M.~J., {et~al.} 2006, Science, 313, 196

\bibitem[{{Szab{\'o}} {et~al.}(2003){Szab{\'o}}, {S{\'a}rneczky}, {Vink{\'o}}, {Cs{\'a}k}, {M{\'e}sz{\'a}ros}, {Sz{\'e}kely}, \& {Bebesi}}]{2003A&A...408..915S}
{Szab{\'o}}, G.~M., {S{\'a}rneczky}, K., {Vink{\'o}}, J., {et~al.} 2003, \aap, 408, 915

\bibitem[{{Szalai} \& {Vink{\'o}}(2013)}]{2013A&A...549A..79S}
{Szalai}, T. \& {Vink{\'o}}, J. 2013, \aap, 549, A79

\bibitem[{{Szalai} {et~al.}(2019){Szalai}, {Zs{\'\i}ros}, {Fox}, {Pejcha}, \& {M{\"u}ller}}]{2018arXiv180302571S}
{Szalai}, T., {Zs{\'\i}ros}, S., {Fox}, O.~D., {Pejcha}, O., \& {M{\"u}ller}, T. 2019, \apjs, 241, 38

\bibitem[{{Taddia}(2012)}]{2012IAUS..279..403T}
{Taddia}, F. 2012, in Death of Massive Stars: Supernovae and Gamma-Ray Bursts, ed. P.~{Roming}, N.~{Kawai}, \& E.~{Pian}, Vol. 279, 403--404

\bibitem[{{Taddia} {et~al.}(2012{\natexlab{a}}){Taddia}, {Stritzinger}, {Phillips}, {Burns}, {Heinrich-Josties}, {Morrell}, {Sollerman}, {Valenti}, {Anderson}, {Boldt}, {Campillay}, {Castellon}, {Contreras}, {Folatelli}, {Freedman}, {Hamuy}, {Krzeminski}, {Leloudas}, {Maeda}, {Persson}, {Roth}, \& {Suntzeff}}]{2012A&A...545L...7T}
{Taddia}, F., {Stritzinger}, M.~D., {Phillips}, M.~M., {et~al.} 2012{\natexlab{a}}, \aap, 545, L7

\bibitem[{{Taddia} {et~al.}(2012{\natexlab{b}}){Taddia}, {Stritzinger}, {Sollerman}, {Phillips}, {Anderson}, {Ergon}, {Folatelli}, {Fransson}, {Freedman}, {Hamuy}, {Morrell}, {Pastorello}, {Persson}, \& {Gonzalez}}]{2012A&A...537A.140T}
{Taddia}, F., {Stritzinger}, M.~D., {Sollerman}, J., {et~al.} 2012{\natexlab{b}}, \aap, 537, A140

\bibitem[{{Taubenberger} {et~al.}(2011){Taubenberger}, {Benetti}, {Childress}, {Pakmor}, {Hachinger}, {Mazzali}, {Stanishev}, {Elias-Rosa}, {Agnoletto}, {Bufano}, {Ergon}, {Harutyunyan}, {Inserra}, {Kankare}, {Kromer}, {Navasardyan}, {Nicolas}, {Pastorello}, {Prosperi}, {Salgado}, {Sollerman}, {Stritzinger}, {Turatto}, {Valenti}, \& {Hillebrandt}}]{2011MNRAS.412.2735T}
{Taubenberger}, S., {Benetti}, S., {Childress}, M., {et~al.} 2011, \mnras, 412, 2735

\bibitem[{{Taubenberger} {et~al.}(2015){Taubenberger}, {Elias-Rosa}, {Kerzendorf}, {Hachinger}, {Spyromilio}, {Fransson}, {Kromer}, {Ruiter}, {Seitenzahl}, {Benetti}, {Cappellaro}, {Pastorello}, {Turatto}, \& {Marchetti}}]{2015MNRAS.448L..48T}
{Taubenberger}, S., {Elias-Rosa}, N., {Kerzendorf}, W.~E., {et~al.} 2015, \mnras, 448, L48

\bibitem[{{Taubenberger} {et~al.}(2013){Taubenberger}, {Kromer}, {Hachinger}, {Mazzali}, {Benetti}, {Nugent}, {Scalzo}, {Pakmor}, {Stanishev}, {Spyromilio}, {Bufano}, {Sim}, {Leibundgut}, \& {Hillebrandt}}]{2013MNRAS.432.3117T}
{Taubenberger}, S., {Kromer}, M., {Hachinger}, S., {et~al.} 2013, \mnras, 432, 3117

\bibitem[{{Thomas} {et~al.}(2007){Thomas}, {Aldering}, {Antilogus}, {Aragon}, {Bailey}, {Baltay}, {Baron}, {Bauer}, {Buton}, {Bongard}, {Copin}, {Gangler}, {Gilles}, {Kessler}, {Loken}, {Nugent}, {Pain}, {Parrent}, {P{\'e}contal}, {Pereira}, {Perlmutter}, {Rabinowitz}, {Rigaudier}, {Runge}, {Scalzo}, {Smadja}, {Wang}, {Weaver}, \& {Nearby Supernova Factory}}]{2007ApJ...654L..53T}
{Thomas}, R.~C., {Aldering}, G., {Antilogus}, P., {et~al.} 2007, \apjl, 654, L53

\bibitem[{{Tinyanont} {et~al.}(2016){Tinyanont}, {Kasliwal}, {Fox}, {Lau}, {Smith}, {Williams}, {Jencson}, {Perley}, {Dykhoff}, {Gehrz}, {Johansson}, {Van Dyk}, {Masci}, {Cody}, \& {Prince}}]{2016ApJ...833..231T}
{Tinyanont}, S., {Kasliwal}, M.~M., {Fox}, O.~D., {et~al.} 2016, \apj, 833, 231

\bibitem[{{Turatto} {et~al.}(1990){Turatto}, {Cappellaro}, {Barbon}, {della Valle}, {Ortolani}, \& {Rosino}}]{1990AJ....100..771T}
{Turatto}, M., {Cappellaro}, E., {Barbon}, R., {et~al.} 1990, \aj, 100, 771

\bibitem[{{Valenti} {et~al.}(2016){Valenti}, {Howell}, {Stritzinger}, {Graham}, {Hosseinzadeh}, {Arcavi}, {Bildsten}, {Jerkstrand}, {McCully}, {Pastorello}, {Piro}, {Sand}, {Smartt}, {Terreran}, {Baltay}, {Benetti}, {Brown}, {Filippenko}, {Fraser}, {Rabinowitz}, {Sullivan}, \& {Yuan}}]{2016MNRAS.459.3939V}
{Valenti}, S., {Howell}, D.~A., {Stritzinger}, M.~D., {et~al.} 2016, \mnras, 459, 3939

\bibitem[{{Valenti} {et~al.}(2014){Valenti}, {Sand}, {Pastorello}, {Graham}, {Howell}, {Parrent}, {Tomasella}, {Ochner}, {Fraser}, {Benetti}, {Yuan}, {Smartt}, {Maund}, {Arcavi}, {Gal-Yam}, {Inserra}, \& {Young}}]{2014MNRAS.438L.101V}
{Valenti}, S., {Sand}, D., {Pastorello}, A., {et~al.} 2014, \mnras, 438, L101

\bibitem[{{Van Dyk} {et~al.}(2003){Van Dyk}, {Li}, \& {Filippenko}}]{2003PASP..115.1289V}
{Van Dyk}, S.~D., {Li}, W., \& {Filippenko}, A.~V. 2003, \pasp, 115, 1289

\bibitem[{Villani(2009)}]{Villani2009}
Villani, C. 2009, Optimal Transport: Old and New, Grundlehren der mathematischen Wissenschaften (Berlin: Springer)

\bibitem[{{Villar} {et~al.}(2019){Villar}, {Berger}, {Miller}, {Chornock}, {Rest}, {Jones}, {Drout}, {Foley}, {Kirshner}, {Lunnan}, {Magnier}, {Milisavljevic}, {Sanders}, \& {Scolnic}}]{villar_2019}
{Villar}, V.~A., {Berger}, E., {Miller}, G., {et~al.} 2019, \apj, 884, 83

\bibitem[{{Vincenzi} {et~al.}(2019){Vincenzi}, {Sullivan}, {Firth}, {Guti{\'e}rrez}, {Frohmaier}, {Smith}, {Angus}, \& {Nichol}}]{Vincenzi2019}
{Vincenzi}, M., {Sullivan}, M., {Firth}, R.~E., {et~al.} 2019, \mnras, 489, 5802

\bibitem[{{Vink{\'o}} {et~al.}(2003){Vink{\'o}}, {B{\'\i}r{\'o}}, {Cs{\'a}k}, {Csizmadia}, {Derekas}, {Fur{\'e}sz}, {Heiner}, {S{\'a}rneczky}, {Sipocz}, {Szab{\'o}}, {Szab{\'o}}, {Szil{\'a}di}, \& {Szatm{\'a}ry}}]{2003A&A...397..115V}
{Vink{\'o}}, J., {B{\'\i}r{\'o}}, I.~B., {Cs{\'a}k}, B., {et~al.} 2003, \aap, 397, 115

\bibitem[{{Vink{\'o}} {et~al.}(2018){Vink{\'o}}, {Ordasi}, {Szalai}, {S{\'a}rneczky}, {B{\'a}nyai}, {B{\'\i}r{\'o}}, {Borkovits}, {Heged{\"u}s}, {Hodos{\'a}n}, {Kelemen}, {Klagyivik}, {Kriskovics}, {Kun}, {Marion}, {Marschalk{\'o}}, {Moln{\'a}r}, {Nagy}, {P{\'a}l}, {Silverman}, {Szak{\'a}ts}, {Szegedi-Elek}, {Sz{\'e}kely}, {Szing}, {Vida}, \& {Wheeler}}]{2018PASP..130f4101V}
{Vink{\'o}}, J., {Ordasi}, A., {Szalai}, T., {et~al.} 2018, \pasp, 130, 064101

\bibitem[{{Vink{\'o}} {et~al.}(2009){Vink{\'o}}, {S{\'a}rneczky}, {Balog}, {Immler}, {Sugerman}, {Brown}, {Misselt}, {Szab{\'o}}, {Csizmadia}, {Kun}, {Klagyivik}, {Foley}, {Filippenko}, {Cs{\'a}k}, \& {Kiss}}]{2009ApJ...695..619V}
{Vink{\'o}}, J., {S{\'a}rneczky}, K., {Balog}, Z., {et~al.} 2009, \apj, 695, 619

\bibitem[{{Vink{\'o}} {et~al.}(2006){Vink{\'o}}, {Tak{\'a}ts}, {S{\'a}rneczky}, {Szab{\'o}}, {M{\'e}sz{\'a}ros}, {Csorv{\'a}si}, {Szalai}, {G{\'a}sp{\'a}r}, {P{\'a}l}, {Csizmadia}, {K{\'o}sp{\'a}l}, {R{\'a}cz}, {Kun}, {Cs{\'a}k}, {F{\"u}r{\'e}sz}, {DeBond}, {Grunhut}, {Thomson}, {Mochnacki}, \& {Koktay}}]{2006MNRAS.369.1780V}
{Vink{\'o}}, J., {Tak{\'a}ts}, K., {S{\'a}rneczky}, K., {et~al.} 2006, \mnras, 369, 1780

\bibitem[{{Walker} {et~al.}(1994){Walker}, {Dahle}, {Midtskogen}, {Hurst}, \& {Sekino}}]{1994IAUC.5950....1W}
{Walker}, A., {Dahle}, H., {Midtskogen}, O., {Hurst}, G.~M., \& {Sekino}, Y. 1994, \iaucirc, 5950, 1

\bibitem[{{Walker} {et~al.}(2015){Walker}, {Baltay}, {Campillay}, {Citrenbaum}, {Contreras}, {Ellman}, {Feindt}, {Gonz{\'a}lez}, {Graham}, {Hadjiyska}, {Hsiao}, {Krisciunas}, {McKinnon}, {Ment}, {Morrell}, {Nugent}, {Phillips}, {Rabinowitz}, {Rostami}, {Ser{\'o}n}, {Stritzinger}, {Sullivan}, \& {Tucker}}]{2015ApJS..219...13W}
{Walker}, E.~S., {Baltay}, C., {Campillay}, A., {et~al.} 2015, \apjs, 219, 13

\bibitem[{{Wang} {et~al.}(2003){Wang}, {Baade}, {H{\"o}flich}, {Khokhlov}, {Wheeler}, {Kasen}, {Nugent}, {Perlmutter}, {Fransson}, \& {Lundqvist}}]{2003ApJ...591.1110W}
{Wang}, L., {Baade}, D., {H{\"o}flich}, P., {et~al.} 2003, \apj, 591, 1110

\bibitem[{{Wang} {et~al.}(2009){Wang}, {Li}, {Filippenko}, {Foley}, {Kirshner}, {Modjaz}, {Bloom}, {Brown}, {Carter}, {Friedman}, {Gal-Yam}, {Ganeshalingam}, {Hicken}, {Krisciunas}, {Milne}, {Silverman}, {Suntzeff}, {Wood-Vasey}, {Cenko}, {Challis}, {Fox}, {Kirkman}, {Li}, {Li}, {Malkan}, {Moore}, {Reitzel}, {Rich}, {Serduke}, {Shang}, {Steele}, {Swift}, {Tao}, {Wong}, \& {Zhang}}]{2009ApJ...697..380W}
{Wang}, X., {Li}, W., {Filippenko}, A.~V., {et~al.} 2009, \apj, 697, 380

\bibitem[{{Wang} {et~al.}(2008){Wang}, {Li}, {Filippenko}, {Krisciunas}, {Suntzeff}, {Li}, {Zhang}, {Deng}, {Foley}, {Ganeshalingam}, {Li}, {Lou}, {Qiu}, {Shang}, {Silverman}, {Zhang}, \& {Zhang}}]{2008ApJ...675..626W}
{Wang}, X., {Li}, W., {Filippenko}, A.~V., {et~al.} 2008, \apj, 675, 626

\bibitem[{{Wheeler} \& {Harkness}(1990)}]{W&H1990}
{Wheeler}, J.~C. \& {Harkness}, R.~P. 1990, Reports on Progress in Physics, 53, 1467

\bibitem[{{Wheeler} \& {Levreault}(1985)}]{wheler}
{Wheeler}, J.~C. \& {Levreault}, R. 1985, \apjl, 294, L17

\bibitem[{Whitaker \& Hayes(2018)}]{Whitaker2018}
Whitaker, D.~A. \& Hayes, K. 2018, Chemometrics and Intelligent Laboratory Systems, 179, 82

\bibitem[{{Yamanaka} {et~al.}(2014){Yamanaka}, {Maeda}, {Kawabata}, {Tanaka}, {Takaki}, {Ueno}, {Masumoto}, {Kawabata}, {Itoh}, {Moritani}, {Akitaya}, {Arai}, {Honda}, {Nishiyama}, {Kabashima}, {Matsumoto}, {Nogami}, \& {Yoshida}}]{2014ApJ...782L..35Y}
{Yamanaka}, M., {Maeda}, K., {Kawabata}, M., {et~al.} 2014, \apjl, 782, L35

\bibitem[{{Yamanaka} {et~al.}(2009){Yamanaka}, {Naito}, {Kinugasa}, {Takanashi}, {Tanaka}, {Kawabata}, {Ozaki}, {Narusawa}, \& {Sadakane Kozo}}]{2009PASJ...61..713Y}
{Yamanaka}, M., {Naito}, H., {Kinugasa}, K., {et~al.} 2009, \pasj, 61, 713

\bibitem[{{Yaron} \& {Gal-Yam}(2012)}]{2012PASP..124..668Y}
{Yaron}, O. \& {Gal-Yam}, A. 2012, \pasp, 124, 668

\bibitem[{{Yuan} {et~al.}(2016){Yuan}, {Jerkstrand}, {Valenti}, {Sollerman}, {Seitenzahl}, {Pastorello}, {Schulze}, {Chen}, {Childress}, {Fraser}, {Fremling}, {Kotak}, {Ruiter}, {Schmidt}, {Smartt}, {Taddia}, {Terreran}, {Tucker}, {Barbarino}, {Benetti}, {Elias-Rosa}, {Gal-Yam}, {Howell}, {Inserra}, {Kankare}, {Lee}, {Li}, {Maguire}, {Margheim}, {Mehner}, {Ochner}, {Sullivan}, {Tomasella}, \& {Young}}]{2016MNRAS.461.2003Y}
{Yuan}, F., {Jerkstrand}, A., {Valenti}, S., {et~al.} 2016, \mnras, 461, 2003

\bibitem[{{Yuan} {et~al.}(2008){Yuan}, {Quimby}, {Chamarro}, {Sisson}, {Akerlof}, {Wheeler}, {Caldwell}, \& {Riley}}]{2008CBET.1314....1Y}
{Yuan}, F., {Quimby}, R., {Chamarro}, D., {et~al.} 2008, Central Bureau Electronic Telegrams, 1314, 1

\bibitem[{{Zanotta} {et~al.}(1986){Zanotta}, {Morris}, \& {Merlin}}]{1986IAUC.4260....4Z}
{Zanotta}, M.~V., {Morris}, C.~S., \& {Merlin}, J.~C. 1986, \iaucirc, 4260, 4

\bibitem[{{Zhai} {et~al.}(2016){Zhai}, {Zhang}, {Wang}, {Zhang}, {Liu}, {Brown}, {Huang}, {Zhao}, {Chang}, {Yi}, {Wang}, {Xin}, {Wang}, {Lun}, {Zhang}, {Fan}, {Zheng}, \& {Bai}}]{2016AJ....151..125Z}
{Zhai}, Q., {Zhang}, J.-J., {Wang}, X.-F., {et~al.} 2016, \aj, 151, 125

\bibitem[{{Zhang} {et~al.}(2014{\natexlab{a}}){Zhang}, {Wang}, {Mazzali}, {Bai}, {Zhang}, {Bersier}, {Huang}, {Fan}, {Mo}, {Wang}, {Yi}, {Wang}, {Xin}, {Liangchang}, {Zhang}, {Lun}, {Wang}, {He}, \& {Walker}}]{2014ApJ...797....5Z}
{Zhang}, J., {Wang}, X., {Mazzali}, P.~A., {et~al.} 2014{\natexlab{a}}, \apj, 797, 5

\bibitem[{Zhang {et~al.}(2021)Zhang, Zhong, \& Ma}]{Zhang2021}
Zhang, J., Zhong, W., \& Ma, P. 2021, A Review on Modern Computational Optimal Transport Methods with Applications in Biomedical Research

\bibitem[{{Zhang} {et~al.}(2014{\natexlab{b}}){Zhang}, {Wang}, {Bai}, {Zhang}, {Wang}, {Liu}, {Zhao}, \& {Chen}}]{2014AJ....148....1Z}
{Zhang}, J.-J., {Wang}, X.-F., {Bai}, J.-M., {et~al.} 2014{\natexlab{b}}, \aj, 148, 1

\bibitem[{{Zhang} {et~al.}(2006){Zhang}, {Wang}, {Li}, {Zhou}, {Ma}, {Jiang}, \& {Chen}}]{2006AJ....131.2245Z}
{Zhang}, T., {Wang}, X., {Li}, W., {et~al.} 2006, \aj, 131, 2245

\bibitem[{{Zheng} {et~al.}(2017){Zheng}, {Filippenko}, {Mauerhan}, {Graham}, {Yuk}, {Hosseinzadeh}, {Silverman}, {Rui}, {Arbour}, {Foley}, {Abolfathi}, {Abramson}, {Arcavi}, {Barth}, {Bennert}, {Brandel}, {Cooper}, {Cosens}, {Fillingham}, {Fulton}, {Halevi}, {Howell}, {Hsyu}, {Kelly}, {Kumar}, {Li}, {Li}, {Malkan}, {Manzano-King}, {McCully}, {Nugent}, {Pan}, {Pei}, {Scott}, {Sexton}, {Shivvers}, {Stahl}, {Treu}, {Valenti}, {Vogler}, {Walsh}, \& {Wang}}]{2017ApJ...841...64Z}
{Zheng}, W., {Filippenko}, A.~V., {Mauerhan}, J., {et~al.} 2017, \apj, 841, 64

\bibitem[{{Zheng} {et~al.}(2013){Zheng}, {Silverman}, {Filippenko}, {Kasen}, {Nugent}, {Graham}, {Wang}, {Valenti}, {Ciabattari}, {Kelly}, {Fox}, {Shivvers}, {Clubb}, {Cenko}, {Balam}, {Howell}, {Hsiao}, {Li}, {Marion}, {Sand}, {Vinko}, {Wheeler}, \& {Zhang}}]{2013ApJ...778L..15Z}
{Zheng}, W., {Silverman}, J.~M., {Filippenko}, A.~V., {et~al.} 2013, \apjl, 778, L15

\end{thebibliography}
%


\begin{appendix}

\section{Data cleaning}
\label{sec:Data_Cleaning}

Before interpolating the spectra, it is important to remove contaminants, including cosmic rays and features coming from sources such as H II regions or atmospheric telluric lines.

To automate the removal of cosmic ray interference from the spectral data, we developed a Python-based application called the Cosmic Ray Deleter (CRD)\footnote{\url{https://github.com/mramirezz/Cosmic-Ray-Deleter}}. The CRD can identify and eliminate spikes using an algorithm founded on the method proposed by \citet{Whitaker2018}. The z-scores tell us how far a value is from the average in units of standard deviation and is defined as follows:

\begin{equation}
    z_i=\frac{x_i-\mu}{\sigma},
    \label{eq:z_scores}
\end{equation}

\noindent where $\mu$ is the mean and $\sigma$ is the standard deviation of the population $x_i$. In this work, we use a robust statistics approach using the median $M$ and the median absolute deviation ($MAD$) rather than the $\mu$ and $\sigma$ as proposed by \citet{Whitaker2018}. The authors employed a modified z-scores outlier detection technique to identify spikes:

\begin{equation}
    z_i=0.6745 \frac{\nabla x_i-M}{MAD}.
    \label{eq:modified_z_scores}
\end{equation}Here $MAD=median (|x-M|)$, $\nabla x_i= x_i-x_{i-1}$, and the value 0.6745 corresponds to the 0.75th quartile of a normal distribution. We take the absolute value of the z-scores and select a threshold value of 3.5 as is recommended in \citet{Hoaglin2013Volume1H}. 

Spikes are corrected by calculating the mean values within a 2m+1 window surrounding them, where m represents an adjustable input value that determines the window size. By default, m=3, but we may alter this value and the threshold to better accommodate the data.

Additionally, we have incorporated into the CRD the functionality to interactively remove cosmic rays. This allows users to input the specific wavelengths between which a spike is observed. This feature proves beneficial when dealing with particularly noisy spectra that contain numerous spikes, which may not necessarily be attributed to cosmic rays. Once the cosmic rays are detected it is removed connecting the two edges of the window with a straight line. 

We visually inspected each spectrum and identified telluric lines using a reference spectrum. These lines were then removed through linear interpolation between the edges of each telluric feature. 

\section{Standard stars}\label{sec:Standars_stars}

For the calibration of $F_0$ in Eq.~\ref{eq:Syn_photometry} we compare the synthetic photometry of three stars reported in  \citet{Stritzinger_2005}, specifically HR0718, HR4468, and HR4963 with the observed photometry reported in \citet{Cousins_1980,Cousins_1984} for the \textit{BVRI} bands and \citet{Fukugita_1996} for the \textit{g'r'i'} bands. 
For computing the synthetic photometry we use the \textit{BVRI} and \textit{g'r'i'}  bands reported in  \citet{Bessell_1998} and  \citet{Fukugita_1996}, respectively.

For each band we also computed the root mean square, mean, and relative error $\sigma$ of the values obtained for the three stars, which are reported in Table \ref{tab:K_values_for_standars2}.


\begin{table}
\caption{Mean $F_0$ value and relative error $\sigma$ for each band.}             
\label{tab:K_values_for_standars2}      
\centering                          
\begin{tabular}{c c c }        
\hline\hline                 
 Band    &  $F_0 \thinspace[erg \thinspace cm^{-2} s^{-1} A^{-1}]$  &  $\sigma$ \\    
\hline                        
            \textit{B} & $6.460\times10^{-9}$ & $0.002 $\\
           \textit{V} & $3.675\times10^{-9}$ & $0.002$ \\
          \textit{ R} & $2.232\times10^{-9}$ & $0.001$ \\
            \textit{I} & $1.177\times10^{-9}$ & $0.005$ \\
            \textit{g} & $4.791\times10^{-9}$ & $0.001$ \\
           \textit{r} & $2.818\times10^{-9}$ & $0.005$ \\
            \textit{i}& $1.907\times10^{-9}$ & $0.014$ \\
\hline                                   
\end{tabular}
\end{table}
\FloatBarrier

\section{Extra figures and tables}
\begin{figure*}[!htb]
    \centering
    \includegraphics[width=\textwidth,height=10cm]{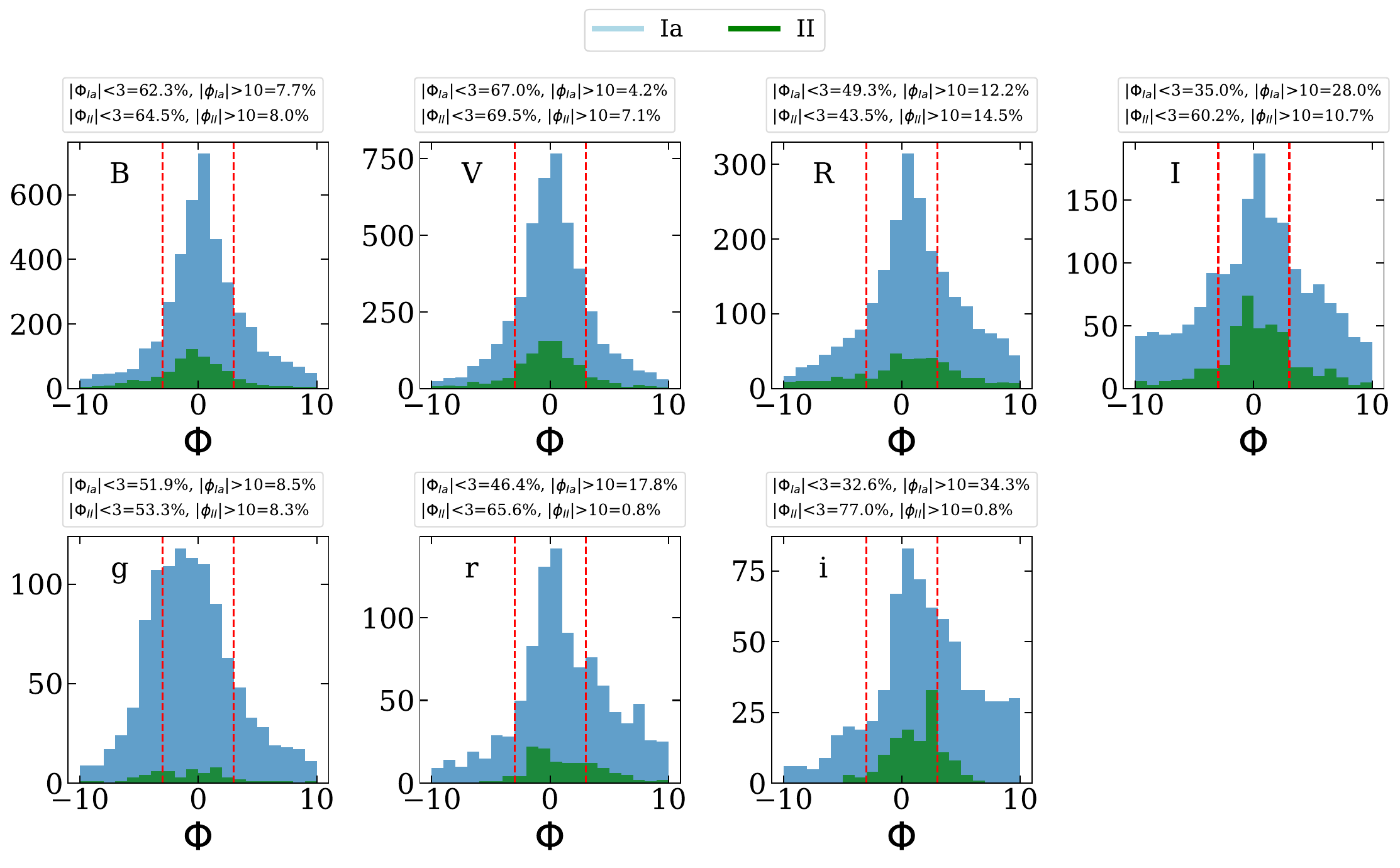}
    \caption{Relative photometric residuals $\Phi$ for each of the \textit{BVRIgri} filters. This residuals are weighted by the error in the photometry. The red dotted line denotes the zones where $|\Phi|<3.0$. \tred{The legend also includes the fraction of relative residuals that fall outside the range shown in the histogram ($|\Phi| > 10.0$).}}
    \label{fig:weighted_Ia}
\end{figure*}

\begin{figure*}[!htb]
    \centering
    \includegraphics[width=\textwidth,height=10cm]{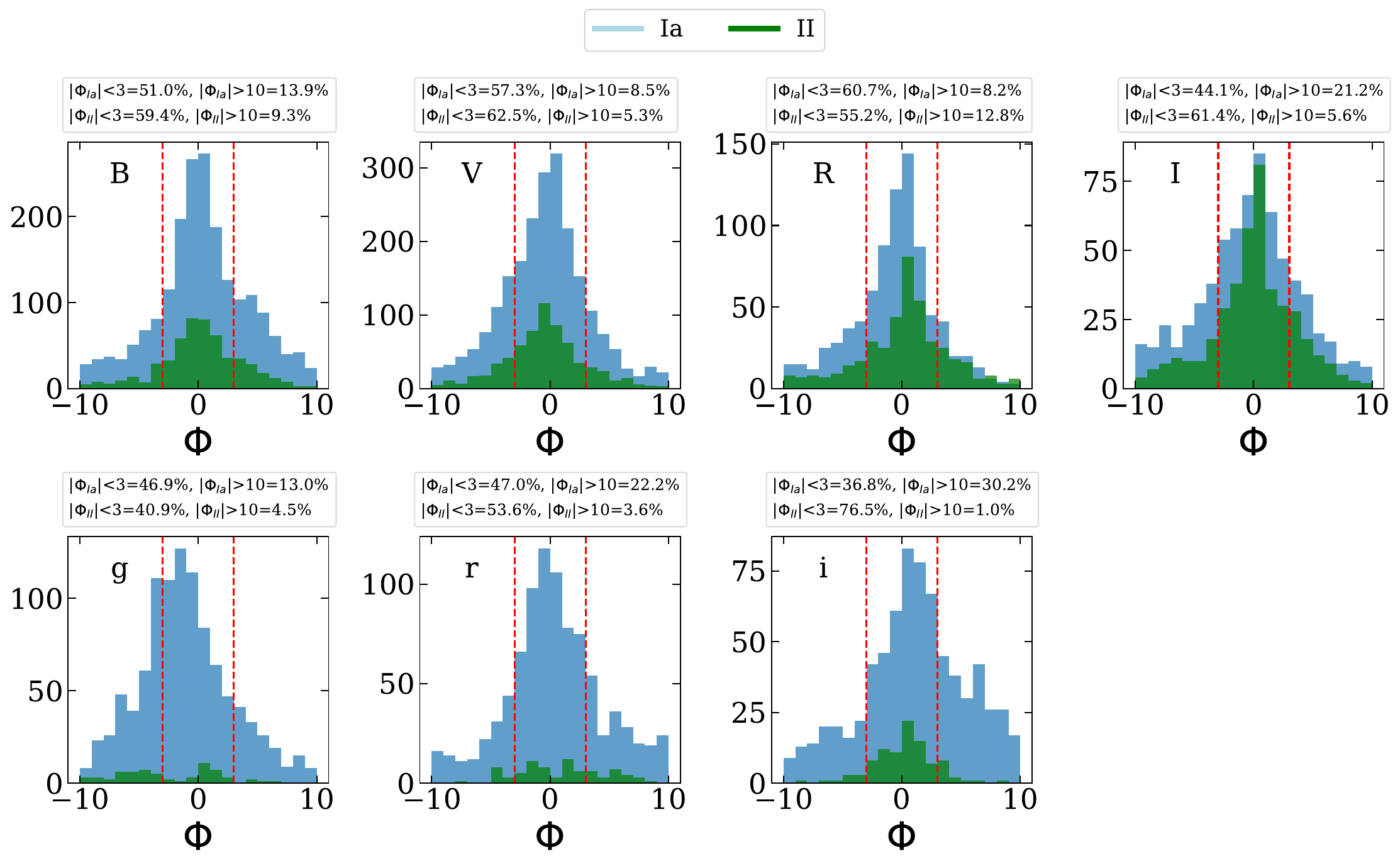}
    \caption{\tred{Same as Fig.~\ref{fig:weighted_Ia}, but with the relative residuals computed  using the leave-one-out cross-validation test.}}
    \label{fig:weighted_Ia_oneout}
\end{figure*}

The following figures display the weighted relative photometric residuals $\Phi$, as referenced in Sect. \ref{subsec:OT_observed_spectra}. Fig.  \ref{fig:weighted_Ia} corresponds to the standard test, while Fig. \ref{fig:weighted_Ia_oneout} shows the results using the leave-one-out cross-validation test. Table \ref{tab:SN included} presents the final golden sample as referenced in Sect. \ref{sec:Data_sample}

\FloatBarrier


\begin{table*}[htbp]
\small
\caption{SNe included in this work.}
\centering

\label{tab:SN included}
\begin{tabular}{llllll}
\hline
    SN Name & Type &  N° spec & Phase Range$^{(a)}$ & Photometry data & References$^{(b)}$ \\
\hline
ASASSN-14lp &   Ia &        8 &   [-13,61] &       \textit{UBVugri} &                                                       (1), (2) \\
   SN2006le &   Ia &       19 &   [-11,35] &         \textit{UBVRI} &                        (79), (38), (1), (31), (14), (17), (65) \\
   SN2007hj &   Ia &       17 &    [-4,70] &           \textit{BV} &                        (85), (1), (17), (14), (89), (31), (56) \\
   SN2007cq &   Ia &        7 &    [-7,37] &           \textit{BVR} &                         (31), (4), (14), (17), (38), (1), (65) \\
   SN2007ci &   Ia &       14 &   [-10,25] &          \textit{BVRI} &                              (14), (87), (17), (31), (1), (38) \\
   SN2007ca &   Ia &       10 &     [0,32] &        \textit{BVugri} &                  (31), (59), (14), (56), (38), (17), (1), (65) \\
   SN2007bm &   Ia &       14 &   [-10,39] &          \textit{BVRI} &                   (14), (4), (56), (1), (69), (38), (17), (59) \\
   SN2007bd &   Ia &       12 &    [-9,13] &        \textit{BVugri} &                  (38), (59), (86), (69), (17), (56), (14), (1) \\
   SN2007au &   Ia &        5 &    [-6,15] &          \textit{BVRI} &                                    (14), (17), (1), (31), (38) \\
   SN2007af &   Ia &       40 &   [-12,68] &     \textit{UBVRIugri} &            (31), (59), (56), (84), (17), (83), (4), (38), (14) \\
    SN2007S &   Ia &       26 &    [-7,32] &       \textit{UBVugri} &                   (14), (4), (56), (38), (59), (1), (17), (65) \\
    SN2007F &   Ia &       14 &    [-8,53] &         \textit{BVRI}&                              (38), (1), (69), (88), (14), (17) \\
    SN2007A &   Ia &        9 &    [-5,12] &           \textit{BVr} &                              (56), (14), (85), (1), (17), (59) \\
   SN2006oa &   Ia &        7 &    [-12,0] &       \textit{UBVgriz} &                              (66), (38), (3), (14), (17), (80) \\
   SN2006lf &   Ia &       19 &    [-9,45] &          \textit{BVRI} &                        (1), (14), (79), (31), (17), (38), (65) \\
   SN2006kf &   Ia &        7 &    [-3,28] &           \textit{Vgr} &                        (38), (14), (17), (1), (56), (59), (78) \\
   SN2007on &   Ia &       21 &    [-5,76] &       \textit{UBVugri} &                                     (4), (56), (1), (59), (29) \\
   SN2006gz &   Ia &       21 &    [-14,9] &        \textit{UBVri} &                                         (17), (16), (14), (77) \\
   SN2006gr &   Ia &       18 &   [-11,49] &         \textit{UBVRI} &                        (14), (38), (1), (31), (76), (17), (65) \\
   SN2006et &   Ia &        9 &     [1,43] &        \textit{BVugri} &                              (59), (56), (38), (14), (1), (17) \\
   SN2006cp &   Ia &        8 &    [-11,8] &         \textit{UBVRI} &                        (31), (38), (17), (1), (14), (73), (65) \\
   SN2006ax &   Ia &       15 &   [-11,32] &       \textit{UBVugri} &                  (38), (56), (59), (14), (17), (1), (72), (65) \\
    SN2006X &   Ia &       50 &   [-14,71] &     \textit{UBVRIugri} &            (38), (31), (65), (81), (59), (82), (17), (83), (4) \\
    SN2006S &   Ia &       19 &    [-6,20] &            \textit{BV} &                                          (38), (1), (14), (17) \\
   LSQ12gdj &   Ia &       11 &   [-13,48] &       \textit{UBVRIri} &                                   (6), (7), (3), (4), (8), (9) \\
   SN2005mz &   Ia &        3 &    [-8,11] &           \textit{UBV} &                                          (17), (38), (14), (4) \\
   SN2005ki &   Ia &        5 &     [0,45] &     \textit{UBVRIugri} &                  (14), (56), (1), (69), (17), (59), (70), (38) \\
   SN2005kc &   Ia &        8 &     [0,13] &       \textit{UBVugri} &                              (59), (56), (17), (38), (1), (14) \\
   SN2005hk &   Ia &       36 &   [-12,71] &    \textit{UBVRIugriz} &            (67), (14), (65), (4), (16), (68), (17), (69), (66) \\
   SN2005hc &   Ia &        6 &    [-3,13] &        \textit{Vugriz} &                  (17), (66), (38), (67), (56), (14), (1), (59) \\
   SN2007le &   Ia &       48 &   [-12,65] &      \textit{BVRIugri} &      (65), (90), (14), (56), (85), (83), (31), (59), (1), (17) \\
   SN2007qe &   Ia &       13 &   [-11,32] &          \textit{BVRI} &                              (38), (14), (1), (31), (17), (65) \\
   SN2005eq &   Ia &       17 &    [-6,55] &          \textit{BVRr} &                  (64), (56), (31), (59), (17), (38), (14), (1) \\
   SN2011by &   Ia &       13 &   [-13,49] &         \textit{UBVRI} &                    (29), (65), (1), (3), (4), (2), (108), (69) \\
  SN2017hbi &   Ia &        5 &   [-14,59] &         \textit{BVRIi} &                                                 (2), (1), (69) \\
  SN2017erp &   Ia &       12 &   [-12,42] &       \textit  {UBVRI} &                                            (4), (2), (1), (69) \\
  SN2017drh &   Ia &        5 &     [0,72] &        \textit  {BVRI} &                                          (132), (1), (2), (69) \\
  SN2016coj &   Ia &       19 &   [-11,19] &     \textit {BVRIgriz} &                         (132), (3), (133), (4), (2), (1), (69) \\
   SN2015bp &   Ia &       11 &    [-7,30] &         \textit  {UBV} &                                                (4), (3), (131) \\
   SN2014dg &   Ia &        3 &    [-7,17] &          \textit {UBV} &                                                  (1), (4), (2) \\
    SN2014J &   Ia &       16 &   [-11,59] &     \textit{UBVRIgriz} &              (115), (1), (128), (29), (129), (130), (120), (3) \\
   SN2013gq &   Ia &        5 &    [-4,12] &          \textit {BVR} &                                       (9), (2), (1), (3), (69) \\
   SN2013dy &   Ia &       32 &   [-13,62] &     \textit{UBVRIgriz} &         (55), (115), (125), (126), (2), (118), (4), (127), (1) \\
   SN2012ht &   Ia &       17 &   [-13,77] &     \textit{UBVRIugri} &             (123), (1), (29), (7), (3), (4), (2), (124), (115) \\
   SN2012fr &   Ia &       46 &   [-15,75] &    \textit {UBVRIugri} &    (118), (1), (119), (120), (2), (29), (121), (3), (122), (4) \\
   SN2012cg &   Ia &       19 &   [-17,45] &       \textit  {UBVRI} &            (115), (3), (116), (83), (29), (1), (4), (2), (117) \\
   SN2011fe &   Ia &       49 &   [-17,77] &      \textit {UBVRIgr} &            (83), (109), (3), (110), (111), (112), (113), (114) \\
   SN2009kq &   Ia &        3 &     [4,27] &         \textit  {VRr} &                              (65), (1), (85), (2), (107), (69) \\
    SN2008J &   Ia &        3 &    [-9,12] &       \textit   {Vgri} &                              (95), (96), (3), (29), (97), (98) \\
   SN2009ig &   Ia &       13 &   [-15,64] &       \textit  {UBVRI} &            (83), (105), (65), (106), (4), (3), (99), (2), (94) \\
   SN2009dc &   Ia &       21 &    [-9,80] &       \textit {BVRIri} &                  (29), (59), (65), (4), (3), (2), (103), (104) \\
   SN2009ad &   Ia &        4 &    [-1,32] &          \textit  {Vr} &                                        (56), (85), (102), (65) \\
    SN2009Y &   Ia &       12 &    [-8,64] &         \textit  {UBV} &                                (65), (4), (1), (56), (85), (2) \\
   SN2008s1 &   Ia &       17 &    [-7,32] &          \textit  {BV} &                               (4), (94), (17), (1), (14), (65) \\
   SN2008hv &   Ia &        9 &    [-7,70] &       \textit{UBVugri} &               (65), (4), (1), (56), (2), (59), (3), (85), (94) \\
   SN2008fp &   Ia &       10 &    [-4,62] &         \textit {BVgr} &                                          (56), (3), (93), (59) \\
   SN2008ec &   Ia &        5 &     [0,43] &        \textit {UBVRI} &                                           (1), (31), (92), (4) \\
   SN2008bf &   Ia &       23 &   [-10,41] &          \textit {BVR} &                              (31), (56), (38), (1), (14), (17) \\
   SN2008bc &   Ia &        7 &    [-5,18] &       \textit {BVugri} &                                                     (59), (56) \\
   SN2008ar &   Ia &       15 &    [-9,48] &        \textit{BVRIri} &                        (91), (14), (56), (85), (1), (17), (31) \\
    SN2008Z &   Ia &       13 &    [-8,56] &        \textit{UBVRri} &                       (85), (1), (101), (14), (17), (31), (65) \\
    SN2008Q &   Ia &        6 &    [-8,-1] &       \textit  {UBVRr} &                  (4), (14), (99), (1), (31), (85), (100), (17) \\
   SN2005eu &   Ia &        6 &    [-9,23] &       \textit   {BVRI} &                              (38), (1), (31), (14), (17), (65) \\
    SN2006D &   Ia &       19 &    [-7,68] &       \textit   {BVgr} &            (31), (59), (74), (38), (56), (17), (14), (1), (75) \\

\end{tabular}

\end{table*}

\addtocounter{table}{-1}

\begin{table*}[htbp]
\small
\caption{Continued.}
\centering

\label{tab:SN included2}
\begin{tabular}{llllll}
\hline
    SN Name & Type &  N° spec & Phase Range$^{(a)}$ & Photometry data & References$^{(b)}$ \\
\hline
SN2005el &   Ia &       14 &    [-7,59] &      \textit {BVRugri} &                  (63), (56), (31), (59), (38), (14), (1), (17) \\
   SN2002ha &   Ia &        9 &    [-5,38] &       \textit   {BVRI} &                                    (14), (17), (1), (38), (31) \\
   SN2002dp &   Ia &        9 &    [-7,42] &       \textit    {BVR} &                                    (14), (17), (1), (38), (31) \\
   SN2002dj &   Ia &       10 &   [-11,13] &       \textit   {BVRI} &                        (38), (14), (48), (1), (16), (17), (31) \\
   SN2002cs &   Ia &        7 &   [-10,30] &       \textit   {BVRI} &                               (17), (31), (3), (14), (47), (1) \\
   SN2005de &   Ia &        4 &    [-2,39] &        \textit  {BVRI} &                                           (3), (1), (31), (47) \\
   SN2002cf &   Ia &        3 &    [-8,13] &         \textit {BVRI} &                                          (17), (31), (14), (1) \\
   SN2002cd &   Ia &       13 &    [-9,51] &       \textit  {UBVRI} &                                    (1), (17), (31), (38), (14) \\
   SN2002bo &   Ia &       40 &   [-13,72] &        \textit  {UBVI} &                        (25), (17), (45), (14), (46), (31), (1) \\
   SN2001ep &   Ia &       24 &    [-4,66] &        \textit  {BVRI} &                              (14), (42), (17), (1), (38), (31) \\
   SN2001en &   Ia &       10 &     [2,34] &        \textit  {BVRI} &                                    (1), (14), (31), (17), (38) \\
   SN2001el &   Ia &       10 &    [-7,54] &        \textit  {BVRI} &                                    (1), (39), (40), (16), (41) \\
   SN2001cp &   Ia &        9 &    [-8,27] &         \textit {BVRI} &                               (14), (17), (1), (3), (31), (38) \\
   SN2001ck &   Ia &        3 &     [4,33] &        \textit  {BVRI} &                                          (1), (17), (14), (31) \\
   SN2001bf &   Ia &       11 &     [0,72] &       \textit   {BVRI} &                              (14), (17), (29), (1), (38), (31) \\
    SN2001V &   Ia &       69 &   [-12,75] &       \textit  {UBVRI} &                        (43), (17), (14), (1), (38), (44), (31) \\
   SN2000fa &   Ia &       13 &   [-12,41] &       \textit  {UBVRI} &                                    (1), (14), (30), (17), (31) \\
   SN2000dn &   Ia &        5 &    [-9,15] &       \textit   {BVRI} &                                          (1), (17), (31), (14) \\
   SN2000dk &   Ia &       11 &    [-6,58] &        \textit   {BVR} &                                    (14), (17), (1), (30), (31) \\
   SN2000cu &   Ia &        3 &     [7,39] &        \textit   {BVR} &                                          (1), (17), (31), (14) \\
   SN2000cn &   Ia &        7 &   [-10,21] &     \textit    {UBVRI} &                                    (14), (17), (31), (1), (30) \\
   SN1999dk &   Ia &        4 &   [-10,42] &        \textit  {BVRI} &                                    (1), (36), (21), (31), (37) \\
   SN1999cp &   Ia &        3 &   [-12,13] &        \textit  {BVRI} &                              (14), (34), (1), (31), (35), (17) \\
   SN1998dm &   Ia &       13 &   [-12,48] &        \textit  {BVRI} &                              (30), (17), (31), (33), (14), (1) \\
   SN1998dh &   Ia &       12 &   [-10,70] &        \textit  {BVRI} &                              (1), (30), (31), (17), (14), (32) \\
   SN1998aq &   Ia &       41 &   [-11,77] &        \textit {UBVRI} &                        (16), (27), (1), (17), (28), (29), (14) \\
    SN1994S &   Ia &        4 &     [-4,0] &         \textit  {BVR} &                              (14), (15), (16), (17), (18), (1) \\
    SN1994D &   Ia &       47 &   [-11,59] &        \textit {UBVRI} &            (19), (20), (21), (1), (22), (23), (24), (25), (26) \\
    SN1990N &   Ia &        3 &    [-1,18] &        \textit {UBVRI} &                                    (1), (10), (11), (12), (13) \\
   SN2002er &   Ia &       14 &   [-12,22] &        \textit {UBVRI} &                  (1), (49), (50), (31), (17), (51), (16), (14) \\
   SN2002cr &   Ia &        8 &   [-10,31] &        \textit {UBVRI} &                                    (14), (1), (17), (31), (38) \\
  iPTF14bdn &   Ia &        7 &   [-19,17] &         \textit  {UBV} &                                                  (3), (4), (5) \\
    SN2005M &   Ia &       24 &    [-5,68] &      \textit{BVRIugri} &                  (31), (59), (71), (56), (17), (38), (14), (1) \\
   SN2004eo &   Ia &       20 &   [-12,70] &    \textit  {BVRIugri} &                              (31), (59), (16), (56), (61), (1) \\
   SN2004ey &   Ia &        5 &    [-9,51] &      \textit  {BVugri} &                                          (59), (56), (31), (1) \\
   SN2004ef &   Ia &       25 &    [-9,58] &      \textit{BVRIugri} &                  (38), (31), (59), (56), (60), (1), (14), (17) \\
   SN2004dt &   Ia &       25 &   [-10,83] &        \textit{BVRIri} &             (56), (57), (1), (53), (3), (16), (17), (58), (14) \\
   SN2004bv &   Ia &        4 &     [6,62] &       \textit   {BVRI} &                                                (1), (31), (55) \\
   SN2004fz &   Ia &        4 &    [-9,20] &       \textit   {BVRI} &                                          (1), (17), (31), (14) \\
   SN2004gs &   Ia &       11 &    [-5,30] &       \textit     {Vr} &                        (14), (31), (56), (59), (62), (1), (17) \\
   SN2003du &   Ia &       56 &   [-12,71] &        \textit {UBVRI} &            (1), (53), (38), (54), (17), (31), (14), (16), (29) \\
   SN2003fa &   Ia &       21 &   [-12,39] &        \textit {UBVRI} &                                    (14), (1), (31), (17), (38) \\
   SN2003cg &   Ia &       22 &    [-8,15] &         \textit {UBVI} &                  (16), (14), (52), (17), (29), (1), (31), (38) \\
   SN2005cf &   Ia &       70 &   [-14,75] &        \textit {UBVRI} &         (134), (38), (1), (135), (16), (31), (136), (137), (4) \\
    SN2003Y &   Ia &        5 &    [-7,20] &        \textit  {BVRI} &                                          (17), (31), (14), (1) \\
    SN2003W &   Ia &        8 &    [-9,47] &         \textit{UBVRI} &                                    (14), (1), (31), (17), (38) \\
   \hline
   SN2014cy &   II &        6 & [-120,-50] &     \textit  {BVRIgri} &                                    (9), (175), (1), (3), (145) \\
   SN2006au &   II &        8 &  [-61,-19] &      \textit  {BVugri} &                                              (185), (3), (187) \\
   SN2005ay &   II &        8 & [-106,-28] &      \textit   {UBVRI} &                                  (138), (3), (166), (1), (141) \\
   SN2013ej &   II &       23 &  [-91,334] &   \textit {UBVRIugriz} & (169), (4), (170), (171), (172), (3), (145), (9), (173), (174) \\
    SN2006V &   II &        7 &   [-24,18] &       \textit {BVugri} &                                       (3), (185), (186), (187) \\
   SN2013am &   II &       13 & [-106,-27] &      \textit{UBVRIgri} &               (170), (188), (189), (1), (4), (145), (3), (159) \\
   SN2013ai &   II &        8 &   [-11,37] &       \textit {UBVgri} &                                    (9), (4), (175), (3), (170) \\
   SN2005cs &   II &       20 &  [-127,36] &       \textit {UBVRIz} &         (138), (139), (140), (141), (1), (4), (3), (142), (16) \\
    SN1990E &   II &        4 &   [27,175] &        \textit  {BVRI} &              (3), (159), (176), (177), (16), (1), (178), (144) \\
    SN1986L &   II &        9 & [-107,-74] &         \textit   {BV} &                       (143), (146), (147), (144), (148), (149) \\
   SN1999em &   II &       12 & [-111,-40] &        \textit {UBVRI} &                         (138), (143), (150), (1), (151), (141) \\
   SN1999gi &   II &        9 &  [-120,16] &         \textit {BVRI} &                                        (138), (165), (16), (1) \\
   SN2002gw &   II &        9 &  [-87,-10] &         \textit  {BVI} &                                       (149), (143), (3), (144) \\
   SN2004et &   II &       26 & [-102,231] &        \textit {UBVRI} &            (138), (16), (190), (191), (192), (193), (1), (194) \\
   SN2006be &   II &        4 &  [-72,-39] &         \textit {BVri} &                                       (144), (143), (1), (166) \\
   SN2007il &   II &        5 &  [-80,-25] &         \textit   {VI} &                                (1), (144), (143), (195), (145) \\
    SN2003T &   II &        3 &  [-85,-48] &         \textit  {BVI} &                                            (143), (144), (149) \\

\end{tabular}

\end{table*}

\addtocounter{table}{-1}

\begin{table*}[htbp]
\small
\centering

\label{tab:model_comparison}
\begin{tabular}{llllll}
\hline
    SN Name & Type &  N° spec & Phase Range$^{(a)}$ & Photometry data & References$^{(b)}$ \\
\hline
SN2007od &   II &       20 & [-115,-72] &      \textit {UBVRIri} &             (166), (179), (145), (143), (144), (180), (1), (4) \\
   SN2003bn &   II &        7 & [-103,-16] &       \textit    {BVI} &                                            (143), (144), (149) \\
   SN2007aa &   II &        9 &    [-73,3] &        \textit  {BVri} &                    (144), (4), (166), (143), (167), (3), (168) \\
   SN2002bx &   II &        9 &  [-49,-17] &        \textit {UBVRI} &                                              (138), (1), (166) \\
   SN2008aw &   II &       23 &    [-72,6] &         \textit{UBVRI} &                                  (4), (1), (143), (144), (145) \\
   SN2008bj &   II &        8 & [-103,-36] &       \textit  {UBVri} &                                                   (182), (166) \\
   SN2008bx &   II &        6 &   [-74,-4] &        \textit{BVRIri} &                                              (1), (145), (166) \\
   SN2003gd &   II &        4 &   [-23,99] &        \textit  {BVRI} &      (152), (138), (153), (154), (1), (144), (143), (155), (3) \\
   SN2003hn &   II &        5 &   [-66,32] &        \textit {UBVRI} &                          (143), (156), (3), (144), (47), (149) \\
    SN2004A &   II &        3 &   [-10,69] &        \textit  {BVRI} &                           (3), (162), (159), (163), (164), (1) \\
   SN2004dj &   II &       11 &  [36,1125] &        \textit {UBVRI} &              (3), (157), (16), (158), (1), (159), (160), (161) \\
   SN2002hx &   II &        6 &   [-48,47] &          \textit {BVI} &                                            (143), (144), (149) \\
   SN2009js &   II &        3 &  [-94,-50] &         \textit {BVRI} &                                  (1), (3), (145), (183), (184) \\
   SN2009dd &   II &       10 & [-105,114] &      \textit {UBVRIri} &                                              (181), (3), (166) \\

\end{tabular}
    \begin{tablenotes}\footnotesize
        \item $^{(a)}$ Phase range of the given SN.
        \item $^{(b )}$ \emph{References:} (1) \citet{2012MNRAS.425.1789S}; (2) \citet{2020MNRAS.492.4325S}; (3) \citet{2012PASP..124..668Y}; (4) \citet{2014Ap&SS.354...89B}; (5) \citet{2015ApJ...813...30S}; (6) \citet{2015ApJS..219...13W}; (7) \citet{2015A&A...579A..40S}; (8) \citet{2012ATel.4566....1C}; (9) \citet{2016arXiv160708526C}; (10) \citet{1992ApJ...397..304J}; (11) \citet{1998AJ....115..234L}; (12) \citet{1998AJ....115.1096G}; (13) \citet{1993A&A...269..423M}; (14) \citet{2008AJ....135.1598M}; (15) \citet{1999AJ....117..707R}; (16) \citet{2001AAS...199.8408R}; (17) \citet{2012AJ....143..126B}; (18) \citet{1991ESOC...37..721L}; (19) \citet{1995AJ....109.2121R}; (20) \citet{1995PAZh...21..678T}; (21) \citet{2004MNRAS.349.1344A}; (22) \citet{1994IAUC.5952....3H}; (23) \citet{1994IAUC.5976....3A}; (24) \citet{1994IAUC.5950....1W}; (25) \citet{2006ApJ...646..369P}; (26) \citet{1997ASIC..486.....R}; (27) \citet{2005ApJ...627..579R}; (28) \citet{2003AJ....126.1489B}; (29) \citet{2016ApJ...821..119C}; (30) \citet{2006AJ....131..527J}; (31) \citet{2010ApJS..190..418G}; (32) \citet{1998IAUC.6978....4H}; (33) \citet{1998IAUC.6993....3B}; (34) \citet{2000ApJ...539..658K}; (35) \citet{1999IAUC.7205....2M}; (36) \citet{2001AJ....122.1616K}; (37) \citet{1999IAUC.7237....3L}; (38) \citet{2009ApJ...700..331H}; (39) \citet{2003AJ....125..166K}; (40) \citet{2003ApJ...591.1110W}; (41) \citet{2008ApJ...686..117F}; (42) \citet{2008MNRAS.391.1605S}; (43) \citet{2003A&A...397..115V}; (44) \citet{2006AJ....132.2024L}; (45) \citet{2003A&A...408..915S}; (46) \citet{2004AJ....128.3034K}; (47) \citet{2008A&A...488..383H}; (48) \citet{2008MNRAS.388..971P}; (49) \citet{2004MNRAS.355..178P}; (50) \citet{2003A&A...401..479C}; (51) \citet{2005A&A...436.1021K}; (52) \citet{2006MNRAS.369.1880E}; (53) \citet{2005ApJ...632..450L}; (54) \citet{2007A&A...469..645S}; (55) \citet{2018arXiv180302571S}; (56) \citet{2013ApJ...773...53F}; (57) \citet{2004IAUC.8386....4R}; (58) \citet{2007A&A...475..585A}; (59) \citet{2010AJ....139..519C}; (60) \citet{2004IAUC.8399....3A}; (61) \citet{2007MNRAS.377.1531P}; (62) \citet{2004IAUC.8453....3F}; (63) \citet{2005IAUC.8604....3G}; (64) \citet{2005IAUC.8608....2P}; (65) \citet{2015ApJS..220....9F}; (66) \citet{2014arXiv1401.3317S}; (67) \citet{2008AJ....136.2306H}; (68) \citet{2008ApJ...680..580S}; (69) \citet{2019MNRAS.490.3882S}; (70) \citet{2005IAUC.8632....2L}; (71) \citet{2005IAUC.8470....3G}; (72) \citet{2006IAUC.8691....2M}; (73) \citet{2006IAUC.8716....1P}; (74) \citet{2002SPIE.4836...61A}; (75) \citet{2007ApJ...654L..53T}; (76) \citet{2006IAUC.8754....1F}; (77) \citet{2007ApJ...669L..17H}; (78) \citet{2006IAUC.8766....3D}; (79) \citet{2006IAUC.8770....2F}; (80) \citet{2011A&A...526A..28O}; (81) \citet{2008ApJ...675..626W}; (82) \citet{2009PASJ...61..713Y}; (83) \citet{2017MNRAS.466.3442J}; (84) \citet{2007ApJ...671L..25S}; (85) \citet{2012ApJS..200...12H}; (86) \citet{2007CBET..914....1P}; (87) \citet{2007CBET..966....1P}; (88) \citet{2007CBET..803....1P}; (89) \citet{2007CBET.1048....2B}; (90) \citet{2009ApJ...702.1157S}; (91) \citet{2008CBET.1273....2R}; (92) \citet{2008CBET.1437....1R}; (93) \citet{2008CBET.1506....1P}; (94) \citet{2012ApJ...749...18B}; (95) \citet{2013AJ....146....2F}; (96) \citet{2012A&A...545L...7T}; (97) \citet{2013ApJS..207....3S}; (98) \citet{2010ATel.2665....1F}; (99) \citet{2013ApJ...767..119M}; (100) \citet{2008CBET.1228....2C}; (101) \citet{2008CBET.1243....2P}; (102) \citet{2009CBET.1694....1N}; (103) \citet{2013MNRAS.432.3117T}; (104) \citet{2011MNRAS.412.2735T}; (105) \citet{2012ApJ...744...38F}; (106) \citet{2009CBET.1918....2N}; (107) \citet{2009CBET.2005....1G}; (108) \citet{2013MNRAS.430.1030S}; (109) \citet{2014MNRAS.439.1959M}; (110) \citet{2014arXiv1408.1430M}; (111) \citet{2012ApJ...750..164C}; (112) \citet{2015MNRAS.448L..48T}; (113) \citet{2013NewA...20...30M}; (114) \citet{2013A&A...554A..27P}; (115) \citet{2018PASP..130f4101V}; (116) \citet{2012ApJ...756L...7S}; (117) \citet{2016ApJ...819...31G}; (118) \citet{2017MNRAS.472.3437G}; (119) \citet{2009ApJ...696..870D}; (120) \citet{2015arXiv150702501C}; (121) \citet{2013ApJ...770...29C}; (122) \citet{2014AJ....148....1Z}; (123) \citet{2018arXiv180906381B}; (124) \citet{2014ApJ...782L..35Y}; (125) \citet{2016AJ....151..125Z}; (126) \citet{2015arXiv150402396P}; (127) \citet{2013ApJ...778L..15Z}; (128) \citet{2014ApJ...784L..12G}; (129) \citet{2014MNRAS.443.2887F}; (130) \citet{2015arXiv151006596G}; (131) \citet{2017MNRAS.466.2436S}; (132) \citet{2017arXiv171102474F}; (133) \citet{2017ApJ...841...64Z}; (134) \citet{2007AIPC..937..311L}; (135) \citet{2007MNRAS.376.1301P}; (136) \citet{2009ApJ...697..380W}; (137) \citet{2009ApJ...700.1456B}; (138) \citet{2014MNRAS.442..844F}; (139) \citet{2008ApJ...675..644D}; (140) \citet{2006MNRAS.370.1752P}; (141) \citet{2008ApJ...685L.117G}; (142) \citet{2009MNRAS.394.2266P}; (143) \citet{2014ApJ...786...67A}; (144) \citet{2017ApJ...850...89G}; (145) \citet{2019MNRAS.490.2799D}; (146) \citet{1986IAUC.4188....1M}; (147) \citet{1986IAUC.4260....4Z}; (148) \citet{1990AJ....100..771T}; (149) \citet{2016AJ....151...33G}; (150) \citet{2001ApJ...558..615H}; (151) \citet{2003MNRAS.338..939E}; (152) \citet{2005MNRAS.359..906H}; (153) \citet{2003PASP..115.1289V}; (154) \citet{2003IAUC.8150....4S}; (155) \citet{2006Sci...313..196S}; (156) \citet{2003IAUC.8186....3R}; (157) \citet{2006MNRAS.369.1780V}; (158) \citet{2006Natur.440..505L}; (159) \citet{2017MNRAS.467..369S}; (160) \citet{2006AJ....131.2245Z}; (161) \citet{2009ApJ...695..619V}; (162) \citet{2006MNRAS.369.1303H}; (163) \citet{2013A&A...549A..79S}; (164) \citet{2004IAUC.8265....2B}; (165) \citet{2002AJ....124.2490L}; (166) \citet{2017ApJS..233....6H}; (167) \citet{2007CBET..848....1D}; (168) \citet{2012MNRAS.420.3451M}; (169) \citet{2017ApJ...834..118M}; (170) \citet{2016ApJ...833..231T}; (171) \citet{2016ApJ...817...22C}; (172) \citet{2016ApJ...822....6D}; (173) \citet{2016MNRAS.461.2003Y}; (174) \citet{2014MNRAS.438L.101V}; (175) \citet{2016MNRAS.459.3939V}; (176) \citet{1993AJ....105.2236S}; (177) \citet{1994A&A...285..147B}; (178) \citet{2000AJ....120..367G}; (179) \citet{2010ApJ...715..541A}; (180) \citet{2011MNRAS.417..261I}; (181) \citet{2013A&A...555A.142I}; (182) \citet{2008CBET.1314....1Y}; (183) \citet{2013ApJ...767..166G}; (184) \citet{2009CBET.1969....2S}; (185) \citet{2012IAUS..279..403T}; (186) \citet{2006IAUC.8667....2N}; (187) \citet{2012A&A...537A.140T}; (188) \citet{2016ApJ...820...33R}; (189) \citet{2014ApJ...797....5Z}; (190) \citet{2006MNRAS.372.1315S}; (191) \citet{2005PASP..117..121L}; (192) \citet{2010MNRAS.404..981M}; (193) \citet{2009ApJ...704..306K}; (194) \citet{2011MNRAS.418.1285F}; (195) \citet{2007CBET.1062....2B}

    \end{tablenotes}

\end{table*}

\end{appendix}
\end{document}